\documentclass[12pt,preprint]{aastex}

\usepackage{graphicx,amssymb,amsmath}
\begin{document}

\title{Evidence for Warped Disks of Young Stars in the Galactic Center}

\newcommand{\commentOut}[1]{}
\newcommand{\commentOn}[2]{{\emph #1}{\em {(\footnotesize Comment: #2)}}}

\author{
H.~Bartko\altaffilmark{a,}\altaffilmark{*}, 
F.~Martins\altaffilmark{b}, 
T.~K.~Fritz\altaffilmark{a}, 
R.~Genzel\altaffilmark{a,c}, 
Y.~Levin\altaffilmark{d}, %
H.~B.~Perets\altaffilmark{f}, %
T.~Paumard\altaffilmark{e}, 
S.~Nayakshin\altaffilmark{g}, %
O.~Gerhard\altaffilmark{a}, %
T.~Alexander\altaffilmark{f}, %
K.~Dodds-Eden\altaffilmark{a}, 
F.~Eisenhauer\altaffilmark{a}, 
S.~Gillessen\altaffilmark{a}, 
L.~Mascetti\altaffilmark{a}, 
T.~Ott\altaffilmark{a}, 
G.~Perrin\altaffilmark{e}, %
O.~Pfuhl\altaffilmark{a}, 
M.~J.~Reid\altaffilmark{h}, 
D.~Rouan\altaffilmark{e}, %
A.~Sternberg\altaffilmark{i},
S.~Trippe\altaffilmark{a}
}
 \altaffiltext{a} {Max-Planck-Institute for Extraterrestrial Physics, Garching, Germany}
 \altaffiltext{b} {GRAAL-CNRS, Universit Montpelier II, Montpelier, France}
 \altaffiltext{c} {Department of Physics, University of California, Berkeley, USA}
 \altaffiltext{d} {Leiden University, Leiden Observatory and Lorentz Institute, , NL-2300 RA Leiden, the Netherlands}

 \altaffiltext{e} {LESIA, Observatoire de Paris, CNRS, UPMC, Université Paris Direrot, Meudon, France}
 \altaffiltext{f} {Faculty of Physics, Weizmann Institute of Science, Rehovot 76100, Israel}
 \altaffiltext{g} {Department of Physics \& Astronomy, University of Leicester, Leicester, UK}
 \altaffiltext{h} {Harvard-Smithsonian Center for Astrophysics, 60 Garden Street, Cambridge, USA}
 \altaffiltext{i} {School of Physics and Astronomy, Tel Aviv University, Tel Aviv 69978, Israel}
 \altaffiltext{*} {correspondence: H.~Bartko, hbartko@mpe.mpg.de}
 

\begin{abstract}

The central parsec around the super-massive black hole in the Galactic Center hosts more than 100 young and massive stars. Outside the central cusp ($R\sim1''$) the majority of these O and Wolf-Rayet (WR) stars reside in a main clockwise system, plus a second, less prominent disk or streamer system at large angles with respect to the main system. Here we present the results from new observations of the Galactic Center with the AO-assisted near-infrared imager NACO and the integral field spectrograph SINFONI on the ESO/VLT.
These include the detection of 27 new reliably measured WR/O stars in the central 12'' and improved measurements of 63 previously detected stars, with proper motion uncertainties reduced by a factor of four compared to our earlier work. 
Based on the sample of 90 well measured WR/O stars, we develop a detailed statistical analysis of their orbital properties and orientations. 
We show that half of the WR/O stars are compatible with being members of a clockwise rotating system. The rotation axis of this system shows a strong transition from the inner to the outer regions as a function of the projected distance from Sgr~A*.

The main clockwise system either is either a strongly warped single disk with a thickness of about $10^{\circ}$, or consists of a series of streamers with significant radial variation in their orbital planes. 11 out of 61 clockwise moving stars have an angular separation of more than $30^{\circ}$ from the local angular momentum direction of the clockwise system. The mean eccentricity of the clockwise system is $0.36\pm0.06$.
The distribution of the counter-clockwise WR/O star is not isotropic at the 98\% confidence level. 
It is compatible with a coherent structure such as stellar filaments, streams, small clusters or possibly a disk in a dissolving state: 10 out of 29 counter-clockwise moving WR/O stars have an angular separation of more than $30^{\circ}$ from the local angular momentum direction of the counter-clockwise system. 
The observed disk warp and the steep surface density distribution favor in situ star formation in gaseous accretion disks as the origin of the young massive stars. 

\end{abstract}


\section{Introduction}

The Galactic Center (GC) is a uniquely accessible laboratory for studying the properties and evolution of galactic nuclei \citep[for reviews see e.g.][]{Genzel1987,Morris1996,Mezger1996,Alexander2005}. At a distance of about 8~kpc \citep{Eisenhauer2003a,Gillessen2008,Ghez2008,Gronewegen2008,Trippe2008}, processes in the Galactic Center can be studied at much higher resolution compared to any other galactic nucleus.
Stellar orbits show that the gravitational potential to a scale of a few light hours is dominated by a concentrated mass of about $4\times10^6M_{\odot}$. It is associated with the compact radio source Sgr~A*, which must be a massive black hole, beyond any reasonable doubt \citep{Schoedel2002,Ghez2005,Gillessen2008}. We will adopt a distance and a mass of Sgr~A*  of $R_{0}= 8$~kpc and $M_{\mathrm{Sgr~A*}}=4.0 \times 10^6$~$M_{\odot}$ for all analyses presented in this paper.
The evolution and the star-formation history in the central pc of the Galaxy may also be used as a probe of star formation processes near supermassive black holes in general, also relevant to other galactic nuclei \citep{Collin2008,Levin2007}. 

The central parsec of the Galaxy contains about a hundred massive young stars. The majority are O-type supergiants and Wolf-Rayet (WR) stars \citep{Forrest1987,Allen1990,Krabbe1991,Navarro1994,Krabbe1995,Blum1995,Tamblyn1996,Navarro1997,Genzel2003,Ghez2003,Paumard2006,Martins2007} with an estimated age of about $6\times10^6$ years. \citet{Genzel1996,Genzel2000,Genzel2003,Levin2003,Beloborodov2006,Paumard2006} inferred that most of the dynamical properties of the WR/O stars (located at projected distances to SgrA* between 0.8'' and 12'') are compatible with belonging to either of two moderately thick counter-rotating stellar disks. However, \citet{Tanner2006} were only able to assign 15 out of 30 early-type stars near the Galactic Center as disk members. \citet{Lu2006,Lu2008} confirm one stellar disk but do not observe a significant number of stars in the other one.

The existence of these young massive stars indicates that star formation must have recently taken place at or near the Galactic Center within the last few million years. 
This is surprising, since regular star
formation processes are likely to be suppressed by the tidal forces
from the massive black hole. Many scenarios have been suggested for the origin of
these stars (see \citet{Alexander2005,Paumard2006,Paumard2008} for
recent reviews). These include
in situ star formation through gravitational fragmentation of gas in
disk(s) formed from infalling molecular cloud(s); transport of
stars from far out by an infalling young stellar cluster, or through
disruption of binary stars on highly elliptical orbits by the massive black hole; and
rejuvenation of old stars due to stellar collisions and tidal
stripping.  The young stars observed in the inner $R\sim1''$ are less
massive B-stars (so called the 'S-stars') and are likely to originate
from a different scenario then the O and WR stars
(e.g. \citet{Perets2007}, but see
\citet{Levin2007,Loeckmann_etal2008}). Here we discuss only our
observations of the O and WR stars outside the central 0.8'' \citep[other
 populations of young stars in the GC are discussed elsewhere; ][]{Gillessen2008,Martins2009}, and interpret them in the context
of the two leading formation scenarios, the {\it
infalling cluster} \citep{Gerhard2001,McMillan2003,Portegiesetal2003,Kim2003,Kim2004,Guerkan2005} and the {\it in situ} formation \citep{Levin2003,Genzel2003,Goodman2003,Milosljevic2004,Nayakshin2005,Paumard2006} scenarios.
The {\it infalling cluster} and the {\it in situ formation} scenarios can be distinguished by different phase space distributions of the stars \citep[see also][]{Paumard2006,Lu2008}. Key observables are the number of disks, the fractions of disk and isotropic stars, the disk orientation, thickness, eccentricity and warp as well as the radial density of the stars and the stellar mass function.

In the following we present the results of new observations of the Galactic Center with the adaptive optics (AO) assisted near-infrared imager NACO and the integral field spectrograph SINFONI on the ESO/VLT. 
These include the detection of 27 new reliably measured
WR/O stars in the central 12'' and improved measurements of previously
detected stars, with proper motion uncertainties reduced by a factor
of four compared to our earlier work. Based on a sample of 90 well
measured WR/O stars, we develop a detailed statistical analysis of
their orbital properties and orientations.
To this end, we use a Monte-Carlo technique to simulate observations of a large number of isotropic stars with the same measurement uncertainties as present in the data. From these simulated measurements, we determine the probability of finding 
coherent dynamical structures against isotropic stars.
We find strong evidence for
the existence of a warped disk in the distribution of the clockwise 
rotating stars and a non-random structure among the counter-clockwise
rotating stars, which is possibly an additional disk.
We then analyze the properties of the stellar disks using both the 3D velocity information and the stellar positions. 
We discuss the implications of our observational results for models for the
origin of the O and WR stellar population in the GC.

This paper is structured as follows: First, we describe our observations, the data selection criteria and present the properties of our data set in section \ref{sec:data}. Thereafter, in section \ref{sec:MC_simulation}, we describe our simulations of the observations of isotropically distributed stars and disk stars. In section \ref{sec:analysis} we introduce our analysis method to search for features in the star distribution. In section 5 we present our results, including a thorough study of the significance of the counter-clockwise system, the determination of the orbital properties of the disk stars and a comparison to previous work. After a discussion of our results in section 6 we summarize our conclusions in section 7.

\section{Data} \label{sec:data}

\subsection{Observations}

\begin{figure}[t!]
\begin{center}
\includegraphics[totalheight=10cm]{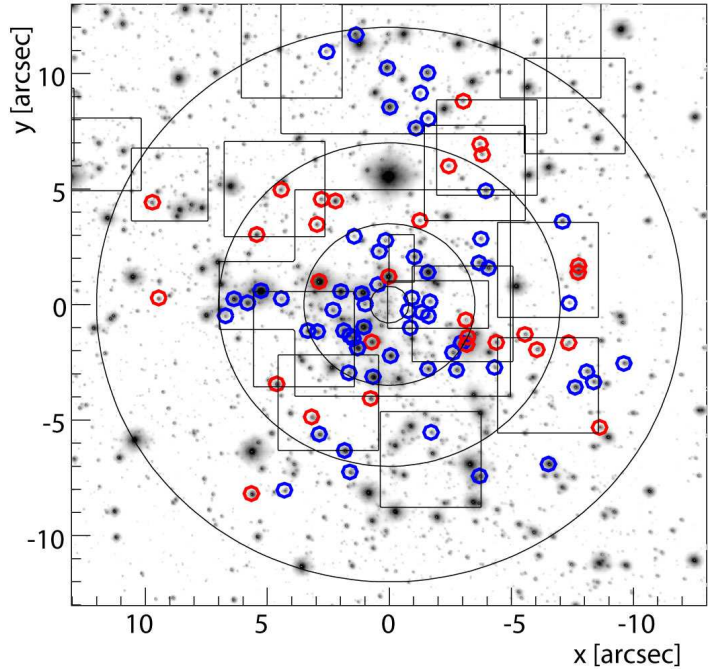}
\caption{\it \small The sample of 90 WR/O stars ($m_K<14$ and $\Delta(v_z)\leq 100$~km/s) in the central 0.5~pc of our Galaxy: Blue circles indicate clockwise orbits (61 WR/O stars) and red circles indicate counter-clockwise orbits (29 WR/O stars). The black circles show projected distances of 0.8'', 3.5'', 7'' and 12'' from Sgr~A*. Squares indicate the exposed fields with SINFONI in the 25~mas/pixel and 100~mas/pixel scale. The whole inner 0.5~pc region is contained in lower resolution (250~mas/pixel scale) SINFONI observations \citep{Paumard2006}.
}
\label{fig:stars_used}
\end{center}
\end{figure}

The data set previously analyzed by \citet{Paumard2006} contained 63 reliably (labeled ``quality 2'') measured WR/O stars in the innermost 12'' and several candidates. 
In 2006-2008 we carried out new observations with the integral field spectrograph SINFONI \citep{Eisenhauer2003,Bonnet2004} at the ESO/VLT. We covered two regions west and north of Sgr A*
with the AO scale (25~mas/pixel) resulting in a final K-band full width at half maximum (FWHM) of typically about 100~mas.
In addition, we observed sixteen $4.2'' \times 4.2''$ fields with the 100~mas/pixel scale resulting in typical K-band FWHMs of about 200~mas. For some of the fields we used the laser guide star facility \citep{Rabien2003,Bonaccini2006}. 
The location of the observed fields is indicated by black squares in figure \ref{fig:stars_used}. The details of the observations and the data analysis will be presented by \citet{Martins2009}. These observations resulted in the reliable detection of 27 new WR/O stars near the Galactic Center. 25 out of the 27 new stars are located at projected distances between 5'' and 12''. Four of the new stars were listed by \citet{Paumard2006} as early type candidates (quality 0 and 1). We determined the radial velocities of these new stars by fitting the observed spectra with template spectra \citep{Martins2007}. We also updated the radial velocities given by \citet{Paumard2006} for all stars in the re-observed SINFONI fields.
In addition, for all these early-type stars we derived proper motions from the NAOS/CONICA \citep{Rousset2003,Hartung2003} imaging data set of the Galactic Center covering six epochs from May 2002 to March 2007 in the 27~mas/pixel scale \citep{Trippe2008}.
Table \ref{tab_param} summarizes the K magnitudes, positions, proper motions and radial velocities of the 90 WR/O stars used in our analysis.

\begin{table*}[t!]
\caption{K magnitudes, positions, proper motions and radial velocities of the 90 WR/O stars used in our analysis. {\it The full machine-readable table is available online}. \label{tab_param}}
\begin{center}
\begin{tabular}{ccccccccccc}
\hline 
\hline
Star    & m$_{\rm K}$\tablenotemark{a}  & x(\arcsec)\tablenotemark{b} & y(\arcsec)\tablenotemark{b} & d$_{\rm SgrA*}$(\arcsec) & v$_{\rm RA}$\tablenotemark{c} & $\sigma$(v$_{\rm RA}$) & v$_{\rm DEC}$ & $\sigma$(v$_{\rm DEC}$) & v$_{\rm r}$ & $\sigma$(v$_{\rm r}$) \\
\hline   
1  &  13.7 & -0.776  &  -0.2771 &  0.822 &  83.2  &  0.7  &  -57.1  &  0.9  &  -75 &  26 \\
\hline
\end{tabular}
\tablenotetext{a}{The uncertainty on the observed magnitude is 0.1 mag.}
\tablenotetext{b}{Positions are relative to SgrA*, typical position uncertainties are as low as 0.0002''.}
\tablenotetext{c}{All velocities are in km/s assuming $R_0=8$~kpc.}
\end{center}
\end{table*}

\subsection{Data Selection Criteria}


In this work we focus on the analysis of the dynamics of WR/O stars and B supergiants.
The brightest of the so-called S-stars within the central 0.8'', S2, is an early B dwarf (B0 –- B2.5 V) \citep{Martins2008a} with $m_K=14.0$ \citep{Paumard2006}. We define $m_K=14.0$ as the border between O and B dwarfs and only include early-type stars with $m_K<14$ in our analysis. Hence we exclude B dwarfs --- S-stars type stars --- from the present analysis.  
O/WR and B supergiants all have initial masses larger than $\sim15 M_{\odot}$, while B dwarfs are less massive.
The S-stars have different dynamical properties than the disk stars. They have isotropic orbits with large eccentricities \citep{Eisenhauer2005,Ghez2005,Gillessen2008}.
Our analysis shows that a large fraction of the B dwarfs in the region of the disks have different kinematics than the WR/O stars. A thorough analysis of the properties of B dwarfs in the region of the WR/O stars ($R>0.8''$) will be presented elsewhere \citep{Martins2009}.  

In order to perform a reliable analysis of the stellar dynamics, we require the observation of a high quality early-type spectrum, such that the error of the radial velocity is 100~km/s or smaller; $\Delta(v_z)\leq 100$~km/s.  

After applying these cuts our sample comprises 90 WR/O stars. There are about 10 candidate WR/O stars with either a too large radial velocity uncertainty or for which no reliable proper motions could be determined due to crowding. No WR/O star is reliably measured in our data set in the central 0.8'' and beyond 12''. 
Figure \ref{fig:stars_used} shows the locations of these stars relative to Sgr~A*, and indicates whether these stars are on clockwise or counter-clockwise orbits. 
We only covered a relatively small area beyond 12'' by deep integral-field spectroscopic observations as indicated in figure \ref{fig:stars_used}. The large ``frame'' between 15'' and 20'' around Sgr~A* shown in figure 1 of \citet{Paumard2006} did not contain any WR/O star, but also had a poor signal to noise ratio.

\subsection{Properties of the Data Set}

\begin{figure}[t!]
\begin{center}
\includegraphics[totalheight=7cm]{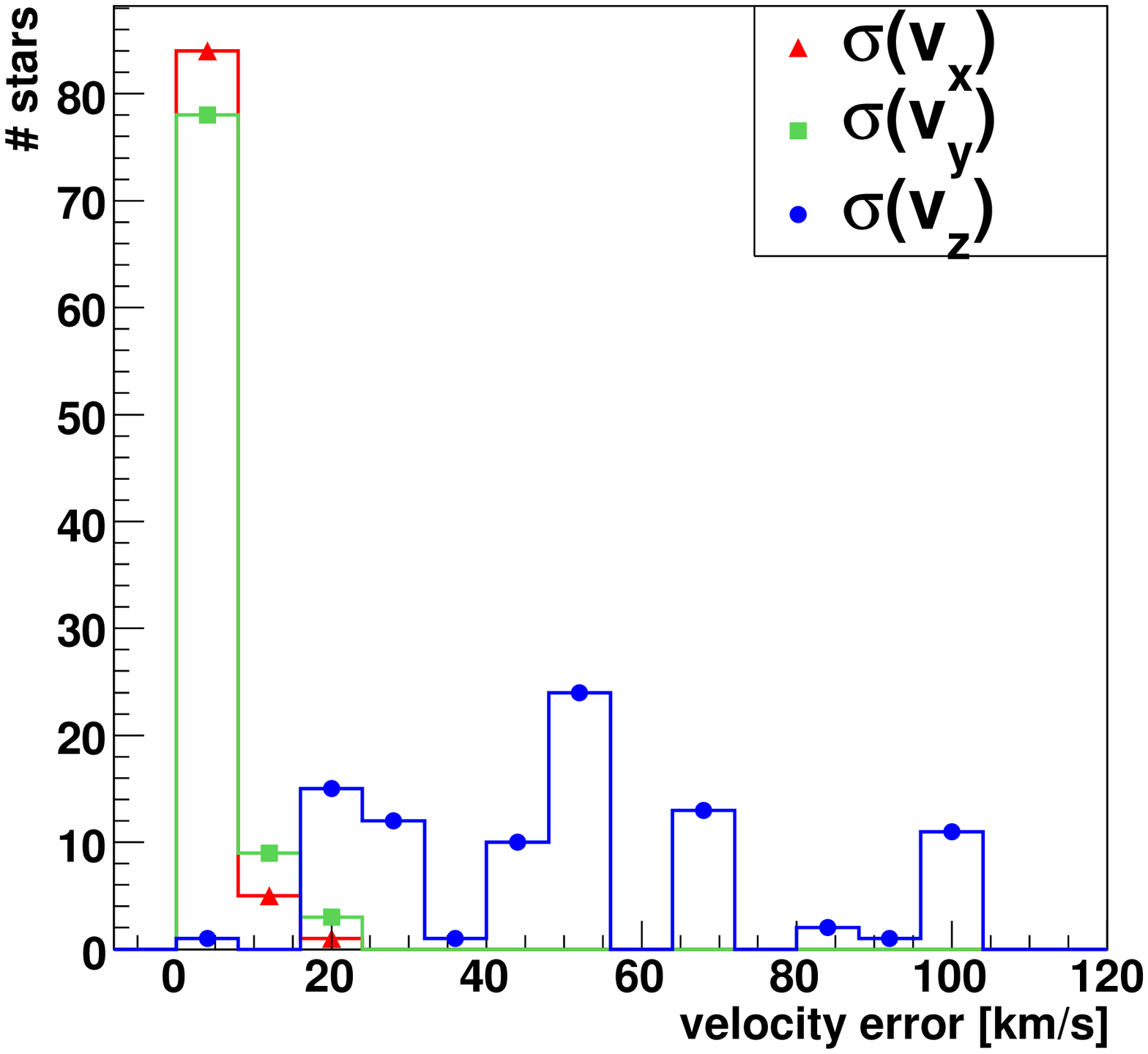}
\includegraphics[totalheight=7cm]{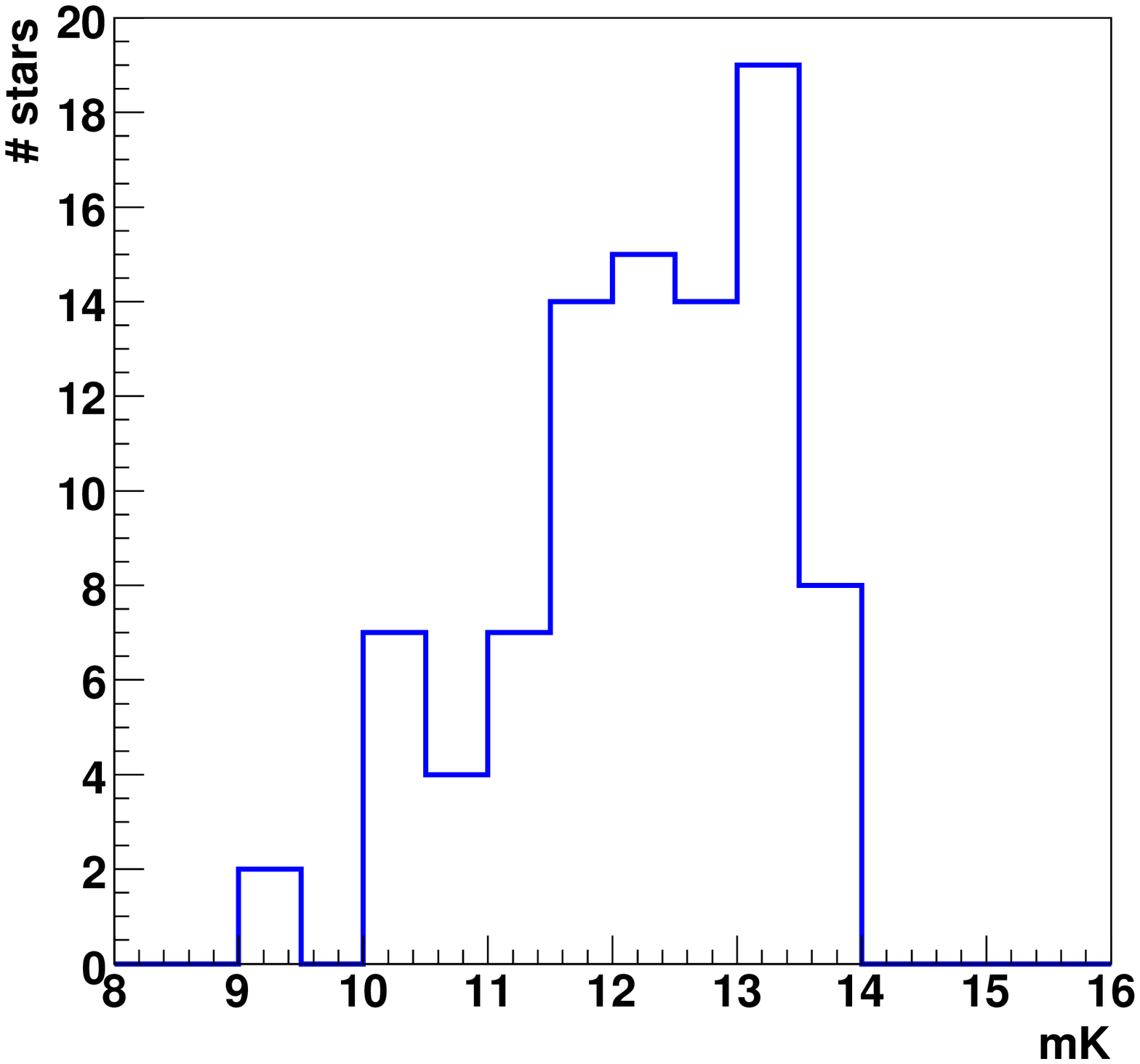}
\caption{\it \small The sample of 90 WR/O stars in the central 0.5~pc of our Galaxy: (Left) Distribution of the statistical velocity errors. The distributions of velocity uncertainties in the $x$ and $y$ directions have both a mean of 5~km/s and an RMS of 3~km/s. The distribution of radial velocity uncertainties has a mean of 51~km/s and an RMS of 25~km/s. (Right) Distribution of K band magnitudes.
}
\label{fig:velocity_erros_disto}
\end{center}
\end{figure}

\begin{figure}[t!]
\begin{center}
\includegraphics[totalheight=7cm]{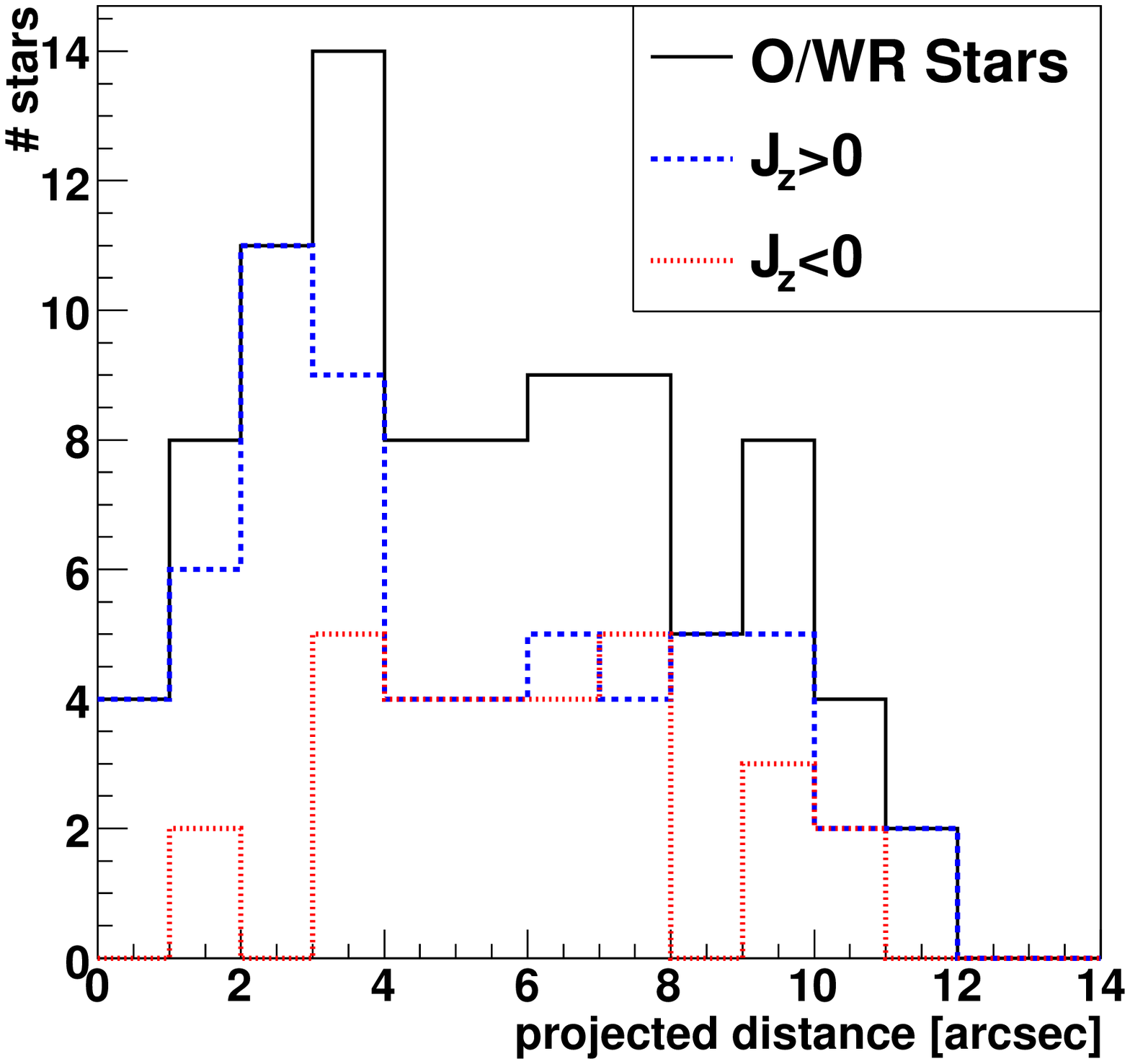}
\includegraphics[totalheight=7cm]{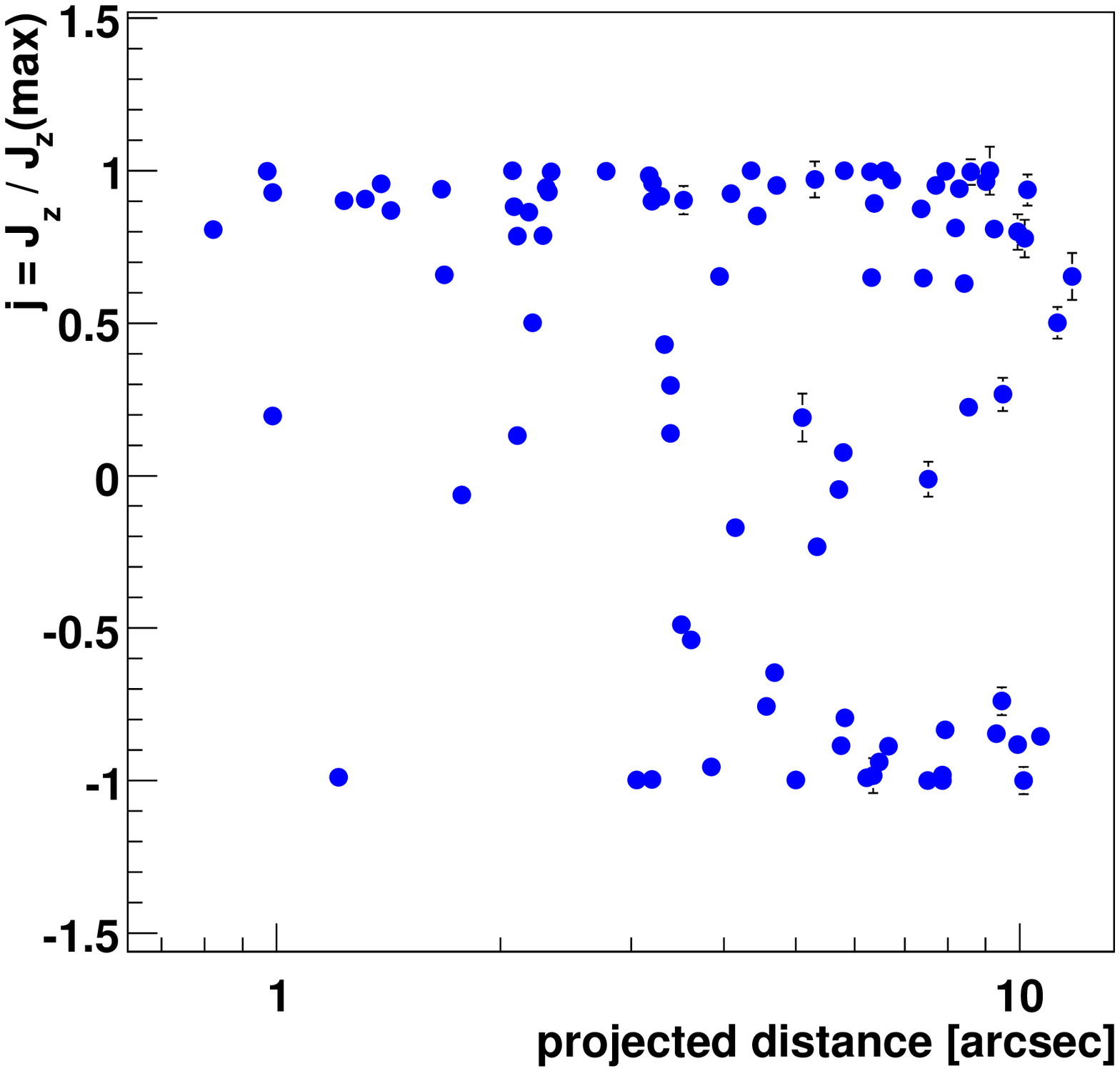}
\caption{\it \small The sample of 90 WR/O stars in the central 0.5~pc of our Galaxy: (Left) Distribution of projected distances to Sgr~A* for all WR/O stars (black full histogram), clockwise moving WR/O stars (blue dashed histogram) and counter-clockwise moving WR/O stars (red dotted histogram). (Right) projected and normalized angular momentum on the sky $j=J_{z} / J_{z,\mathrm{max}}$ as a function of projected distance to Sgr~A*. The error bars for most of the stars are smaller than the markers. Clockwise orbits correspond to $j>0$ and counter-clockwise orbits correspond to $j<0$.
}
\label{fig:jz}
\end{center}
\end{figure}
%


Figure \ref{fig:velocity_erros_disto} (left panel) presents the distribution of the statistical velocity uncertainties in the $x,y$ and $z$ directions.
The distributions of velocity uncertainties in the $x$ and $y$ directions both have a mean of 5~km/s and an RMS of 3~km/s. The systematic uncertainty of the astrometric reference frame is 6.4~km/s \citep{Trippe2008}. The uncertainty of the distance to Sgr~A*, which we used to convert the measured proper motions on the sky to velocities in km/s, introduces a systematic error of about 6\% in $v_x$ and $v_y$.
The mass of Sgr~A* and its distance are tightly correlated: $M_{\mathrm{Sgr~A*}} = (3.95\pm0.06)\times 10^6(R_0/8\ \mathrm{kpc})^{2.19}M_{\odot}$ \citep{Gillessen2008}. In order to evaluate the systematic errors introduced by the choise of $R_0=8$~kpc and $M_{\mathrm{Sgr~A*}}=4 \times 10^6 M_{\odot}$ we ran all analysis steps also with $R_0=7.5$~kpc, $M_{\mathrm{Sgr~A*}}=3.5 \times 10^6 M_{\odot}$ and $R_0=8.5$~kpc, $M_{\mathrm{Sgr~A*}}=4.5 \times 10^6 M_{\odot}$.

The combined statistical and systematic proper motion uncertainties in the \citet{Paumard2006} analysis had a mean of 35~km/s and an RMS of 9~km/s. The reduced errors of the present proper motion data are due to the larger data set, a correction of the geometric image distortion and smaller systematic uncertainties of the movement of the coordinate system \citep{Trippe2008}. 
The distribution of radial velocity uncertainties has a mean of 51~km/s and an RMS of 25~km/s, about the same as for the \citet{Paumard2006} data set (mean 46~km/s and RMS 26~km/s) but for 90 instead of 63 high quality stars.
Figure \ref{fig:velocity_erros_disto} (right panel) shows the distribution of observed K band magnitudes. 

For some stars with low projected distances from Sgr~A*, significant accelerations are measured, which allowed \citet{Gillessen2008} to determine full orbital solutions (see below figure \ref{fig:sky_zoom_Sstars}). Moreover, upper limits to accelerations translate into lower limits for the absolute value of the $z$-coordinate \citep[see][]{Trippe2008,Lu2008}. However, a timeline of 5~years of NACO data has not yet been sufficient to constrain the $z$-coordinate for more than a handful of stars. We have chosen not to include these limits as priors in our analysis in order to have a data set with uniform errors independent of projected distance to Sgr~A*. Instead we compare below the statistical results of the entire data set with the properties of the subset of stars with measured orbits.

Figure \ref{fig:jz} (left panel) shows the distribution of projected distances to Sgr~A* for the data set of 90 WR/O stars. 
The sample of WR/O stars is not complete, i.e. the probability to detect a star with a given magnitude in an image (photometric completeness) and to identify its spectrum as early-type (spectroscopic completeness) is below one. The completeness depends on the respective field as well as on the apparent magnitude of the star \citep[see e.g.][]{Martins2009}. O stars in the magnitude interval $m_{K} = 13-14$ typically have a combined photometric and spectroscopic completeness of 75\%. 
To obtain the most reliable radial distribution a correction for completeness will eventually have to be applied. However, considering the emphasis of the current paper on the angular momentum distribution this correction is unnecessary.
Figure \ref{fig:jz} (right panel) shows, for each of the early-type stars, the projected and normalized angular momentum on the sky $j=J_{z} / J_{z,\mathrm{max}}$ \citep{Genzel2003,Paumard2006} as a function of projected distance to Sgr~A*. There are 61 WR/O stars on clockwise orbits ($j>0$) and 29 WR/O stars on counter-clockwise orbits ($j<0$). Most of the WR/O stars with projected distances below 3'' are on projected clockwise tangential orbits ($j\simeq1$). For larger projected distances there are two concentrations of stars with $j\simeq1$ and $j\simeq-1$ (projected clockwise and counter-clockwise) tangential orbits. The ``diagonal feature'' observed by \citet{Paumard2006} looks less pronounced in our data. Our data rather suggest two systems of stars with $j\simeq1$ and $j\simeq-1$ and a background of random stars.

\begin{table*}[t!]
\caption{Numbers and fractions of spectral subclasses of the 90 selected WR/O stars. }
\begin{center}
\begin{tabular}{l|r|r|r|r}
\hline 
\hline
& \multicolumn{2}{c}{clockwise} & \multicolumn{2}{c}{counter-clockwise} \\
\hline
Type & Number & fraction & number & fraction\\
\hline
  OB       & 42 & 0.69 &  17 & 0.58 \\
  Ofpe/WN9 &  5 & 0.08 &   4 & 0.14 \\
  WN       &  8 & 0.13 &   2 & 0.07 \\
  WC       &  6 & 0.10 &   6 & 0.21 \\
\hline
  sum      & 61 & 1.00 &  29 & 1.00 \\
\hline
\hline
\end{tabular} 
\end{center}
\label{tab:spectral_types}
\end{table*} 

Table~\ref{tab:spectral_types} lists the numbers and
relative fractions of the different sub-types of the selected 90 WR/O stars. 
The fraction of different spectral sub-types is strikingly similar for the clockwise and the counter-clockwise rotating stars. 
This resemblance
in content of massive stars strongly suggests that the clockwise and counter-clockwise stars are, within measurement accuracies of $\sim2$~Myr, coeval \citep{Genzel2003,Paumard2006}. Moreover, there is no significant difference in the fraction of spectral sub-types with distance to Sgr~A*.

\section{MC Simulation of Signal and Background} \label{sec:MC_simulation}

To assess the probability of the observed features in the stellar distribution being compatible with an isotropic star distribution, we simulated measurements of isotropically distributed stars and disks. For these simulations we assumed bound orbits, which are described by the following orbital elements (see appendix \ref{sec:coordinate} for a visual representation of the chosen coordinate system and the orbital elements):

\begin{enumerate}
\item{$\Omega, i$: Longitude of the ascending node, inclination \\ 
  $\Leftrightarrow \theta_J, \phi_J$: Orientation of the orbital plane / the orbital angular momentum vector}
\item{$a$: semi-major axis}
\item{$\epsilon$: eccentricity}
\item{$\omega$: argument of periapsis}
\item{$\tau$: time difference between periastron and the current position of the star.}
\end{enumerate}

To simulate an isotropic cusp stellar distribution we used the following algorithm:

\begin{enumerate}
\item{Generate the direction of the angular momentum vector $\vec{J}/|\vec{J}|$ uniformly distributed on a sphere. Calculate $\Omega$ and $i$.}
\item{Generate $\omega$ uniformly in the interval $[0,2\pi]$ and $\tau$ uniformly in the interval $[0,T]$.}
\item{Generate $\epsilon$ according to the distribution $\mathrm{d}N_{\mathrm{stars}}/\mathrm{d}\epsilon \propto \epsilon$ \citep{Binney1987} for isotropic orbits. Note that the clockwise disk of \citet{Paumard2006} has a different distribution of eccentricities.}
\item{Generate $a$ according to the probability density $\mathrm{d}N_{\mathrm{stars}}/\mathrm{d}a \propto a^{-\beta+1}$ with $\beta=2$ in the interval $[0.2'' \leq a \leq 40'']$. We motivate this choice as follows: For a power-law cusp the radial star number density is related to the surface number density $\Sigma(r_{\mathrm{disk}}) = \mathrm{d}N_{\mathrm{stars}}/ \mathrm{d} A_{\mathrm{disk}} \propto r_{\mathrm{disk}}^{-\beta}$ \citep{Schoedel2003,Alexander2005}. $r_{\mathrm{disk}}$ denotes the 3D distance in the plane of the disk. Each
projection (e.g going from 3D density to 2D (surface) density)
decreases the power of the radial coordinate by 1.
%
%
\citet{Paumard2006} measured the surface number density of the early-type stars in the disks as $\Sigma(r_{\mathrm{disk}}) \propto r_{\mathrm{disk}}^{-2.1\pm0.2}$ and in our work we find  $\Sigma(r_{\mathrm{disk}}) \propto r_{\mathrm{disk}}^{-1.95\pm0.25}$, see section \ref{chi2_analysis}.
}
\end{enumerate}


For the simulation of a thick disk we assumed the following distribution of the stellar angular momentum directions:
\begin{equation}
\frac{\mathrm{d}N_{\mathrm{stars}}}{\mathrm{d}S} \propto \exp\left[-\frac{\psi^2}{2\sigma_{\psi}^2}\right] \ . \label{eq:inclin_2D}
\end{equation} 
$S$ is the solid angle in which the stellar angular momentum points, $\psi$ is the angular distance between the generated angular momentum direction of the star $(\phi_J,\theta_J)$ and the disk angular momentum direction $(\phi_{\mathrm{disk}},\theta_{\mathrm{disk}})$. In the following we will call the parameter $\sigma_{\psi}$ the two-dimensional Gaussian sigma thickness.

The generation of $\epsilon$ and $a$ in steps 3 and 4 implies a strong prior to the simulated isotropic stars drawn from a $\beta=2$ cusp distribution.
However, we will only use the simulated isotropic stars to determine expectation values of the mean and RMS of the distribution of stellar angular momenta per solid angle. To first order, these distributions depend only on the direction of the orbital plane (given by $\Omega$ and $i$).
In the simulations we only considered the potential of Sgr~A*.
For simplicity, we neglect in our simulations the mass of the cluster of late-type stars \citep{Trippe2008}, about $2 \times 10^5 M_{\odot}$ enclosed in the innermost 10''. This is justified as we are only interested in the distribution of the stellar angular momenta of the simulated stars, which is, to lowest order, independent of the enclosed mass.

For each simulated star we calculated the true positions and velocities $x,y,z$ and $v_x,v_y,v_z$ from the generated orbital elements. We simulated the measurement uncertainties for the position and velocities by adding random numbers to these positions and velocities following the distribution of errors in the measured data, see figure \ref{fig:velocity_erros_disto}.

\section{Analysis Method to Search for Features in the Star Distribution} \label{sec:analysis}

The prime criterion for distinguishing a disk from an isotropic stellar distribution, which we assume as the null hypothesis, is the presence of a common direction of the angular momentum vectors for all stars. The angular momentum vectors of isotropic stars are uniformly distributed over a sphere. The analysis requires three steps:

\begin{enumerate}
\item{computation of a density map of angular momentum vectors of the $n$ observed stars}
\item{computation of the mean and RMS density maps of angular momentum vectors for $n$ MC simulated isotropic stars.}
\item{computation of a significance map by comparing the density map of step 1 to the mean and RMS expectations derived in step 2.}
\end{enumerate}

\subsection{Computation of a Density Map of Angular Momentum Vectors} \label{sec:z_generation}

The computation of a density map of angular momentum vectors for sets of measured or MC generated stars proceeds in three steps:

\begin{enumerate}
\item{MC generation of 1000 values for the unknown $z$-coordinate for each star, computation of the angular momentum direction}
\item{computation of the total reconstructed angular momentum distribution on a sphere}
\item{computation of the density of reconstructed angular momentum vectors in a fixed aperture.}
\end{enumerate}

\subsubsection{Generation of $z$-Coordinates} \label{sec:z_generation}

The orbit of a star in a given gravitational potential is fully described by six orbital elements. For most stars there are only five measurements: the projected positions on the sky $x$ and $y$,  the proper motion velocities $v_x$ and $v_y$, as well as the radial velocity $v_z$ \footnote{For the definition of the coordinate system used, see appendix \ref{sec:coordinate}}. Under these conditions, the direction of the angular momentum of any given star cannot be uniquely determined. All possible directions of the angular momentum lie on a one-dimensional curve \citep[see e.g.][]{Eisenhauer2005}. Assuming that the star is bound, one can derive an upper limit to the line-of-sight position of the star, which in turn limits the curve of possible angular momentum directions. 

The specific angular momentum (angular momentum per mass) $\vec{J}$ is given by:
\begin{equation}
\vec{J} = \vec{r} \times \vec{v} 
 \ .
\end{equation} 
As the $z$-coordinate is unknown, the angular momentum can only be determined as a function of unknown $z$: $\vec{J}=\overrightarrow{J(z)}$. For a cluster in equilibrium the distribution of the $z$-coordinates is related to the star surface number density distribution $\Sigma(r_{\mathrm{disk}}) = \mathrm{d}N_{\mathrm{stars}}/ \mathrm{d} A_{\mathrm{disk}}$, see section \ref{chi2_analysis} and \citet{Genzel2003,Paumard2006}. 
%
%
Assuming a power law of the surface number density $\mathrm{d}N_{\mathrm{stars}}/ \mathrm{d} A_{\mathrm{disk}} \propto (x^2+y^2)^{-\beta/2}$, 
%
%
this density distribution translates into a density distribution for the $z$-positions of the measured stars \citep{Alexander2005}:
\begin{equation}
\frac{\mathrm{d}N_{\mathrm{stars}}}{\mathrm{d}z}\Bigg|_{x_0, y_0} \propto \frac{\mathrm{d}N_{\mathrm{stars}}}{\mathrm{d}V} \Bigg|_{x_0, y_0} \propto (x_0^2+y_0^2+z^2)^{-(\beta+1)/2} \ , \label{eq:z_density}
%
\end{equation}
where $x_0$,$y_0$ are the measured positions on the sky.
%

\begin{figure}[t!]
\begin{center}
\includegraphics[totalheight=7cm]{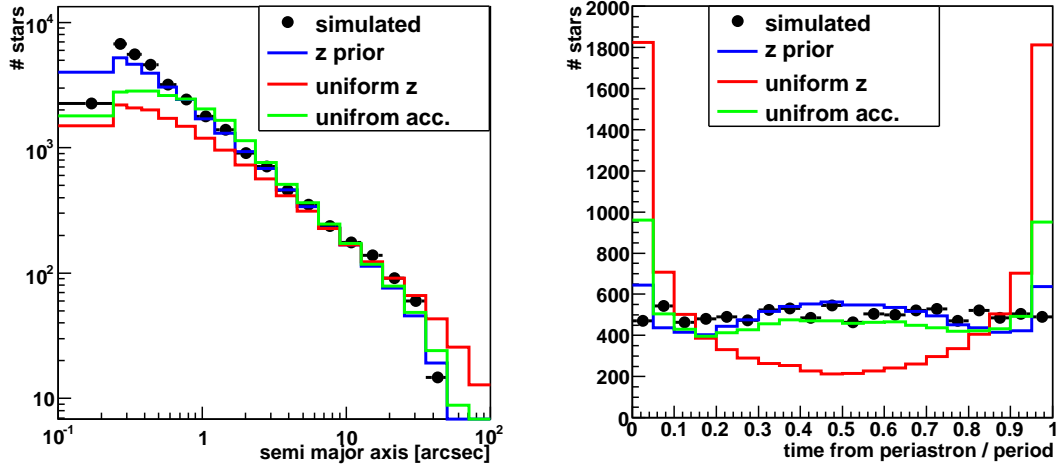}
\caption{\it \small Input (black points) and reconstructed distributions of the semi-major axis and the time difference between periastron and the current position of the star. The uniform $z$ prior is shown by a red histogram, the uniform acceleration prior \citep{Lu2008} is denoted by the green histogram and the $z$-prior used in this work by the blue histogram. 
}
\label{fig:z_prior_comparison}
\end{center}
\end{figure}
%


For each star we generate 1000 $z$-values, assuming $\beta=2$, according to section \ref{chi2_analysis} and \citet{Paumard2006}. The choice of $\beta$ is again a strong prior to our analysis. However, the generation of the $z$-values is based on the same stellar surface density distribution as the generation of the MC simulated stars (see section \ref{sec:MC_simulation}). We also ran our analysis simulations with $\beta$-values of 1.5 and 2.5 and obtained similar results.
The limit $|z|_{\mathrm{max}} \geq |z|$ is given by the requirement that the star must be bound to Sgr~A*.

The way we determine the distances along the line-of-sight is very important as concerns possible biases. 

As a consistency check of our generation method for $z$ we compare the reconstructed orbital element distributions for MC generated isotropic stars with the input distribution into the MC simulation. Moreover, we have computed the distributions of orbital elements for the case of uniform $z$ priors and uniform acceleration priors \citep{Lu2008}. The distributions of eccentricity and the argument of the periapsis match very well the input distribution for all three studied $z$-priors.
Figure \ref{fig:z_prior_comparison} shows the input and reconstructed distributions of the semi-major axis and the time difference between periastron and the current position of the star. The uniform $z$-prior gives strong biases to reconstructed positions near the pericenter of the star. Also the uniform acceleration prior biases the reconstructed star position to the pericenter. Although the reconstructed distribution with our adopted $z$-prior shows small biases, these are smaller than the biases for the other two studied $z$-priors.

\subsubsection{Angular Momentum Distribution on a Sphere}

\begin{figure}[t!]
\begin{center}
\includegraphics[totalheight=7cm]{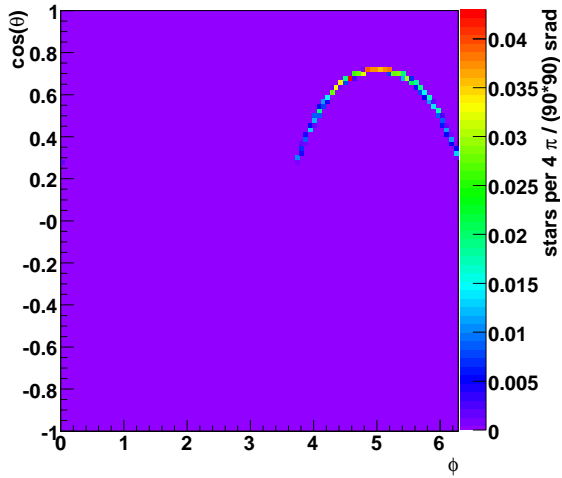}
\caption{\it \small cylindrical equal area projection of the distribution of the direction of the angular momentum vector ($\cos_J$ vs. $\phi_J$) for the star IRS16~CC for 1000 generated $z$-values. The $z$-coordinate is generated according to equation (\ref{eq:z_density}) (assuming a power law stellar surface density with exponent $\beta=2$). Only $z$-values of bound orbits are considered. 
Clockwise orbits have angular momentum directions in the upper hemisphere ($\cos \theta>0$) and counter-clockwise orbits have angular momentum directions in the lower hemisphere ($\cos \theta<0$).
}
\label{fig:IRS16CC}
\end{center}
\end{figure}

Having calculated the angular momentum for the 1000 generated $z$-values, we can compute the distribution of directions of the angular momentum vectors $\overrightarrow{J(z)}$ on a sphere. We parameterize the sphere with the usual spherical coordinates $\phi$ and $\theta$. For our further analysis, we use a cylindrical equal area projection \citep{Gall1885,Peters1983} of this sphere: ($\cos\theta_J$ vs. $\phi_J$). We divide the projected map into a grid of $90 \times 90$ bins in $\phi$ and $\cos\theta$.
As an example, figure \ref{fig:IRS16CC} 
shows a cylindrical equal area projection of the distribution of angular momentum vectors directions $\overrightarrow{J(z)}$ ($\cos\theta_J$ vs. $\phi_J$) for the star IRS16~CC.
The shape of the distribution of reconstructed angular momenta is independent of $M_{\mathrm{Sgr~A*}}$. It determines the maximum $|z|$-value for bound orbits and thus the minimum $|\cos \theta_J|$.


\subsubsection{Computation of Angular Momentum Density Maps}

In a last step, the average density $\rho_{\alpha}(\theta,\phi)$ of the reconstructed angular momentum directions is calculated over an aperture of $\alpha = 15^{\circ}$ radius centered on the sky direction $(\theta,\phi)$:

\begin{equation}
\rho_{\alpha}(\theta,\phi) = \frac{
\sum_{\mathrm{stars}}\sum_{i=1}^{1000}
\left\{
\begin{array}{c c}
1 & \angle [\overrightarrow{J(z)},(\theta,\phi)] \leq \alpha \\
0 & \mathrm{else}
\end{array}
\right\}
}
{ 1000 \cdot 2 \pi (1 -\cos \alpha)} \ .
\end{equation}

This averaging maximizes the signal (mean integrated angular momenta over the aperture for a disk) to noise (RMS integrated angular momenta over the aperture for isotropic stars) ratio for a moderately thick disk with a two-dimensional Gaussian sigma of $10^{\circ}$. We chose a simple aperture in order to have well defined sky regions with a flat weight. A matched filter may yield even higher signal to noise ratios.
Bins of the angular momentum density sky map, which are separated by less than the aperture size, are correlated.

\subsection{Density Maps of Angular Momentum Vectors for Simulated Stars}

\begin{figure}[t!]
\begin{center}
\includegraphics[totalheight=7cm]{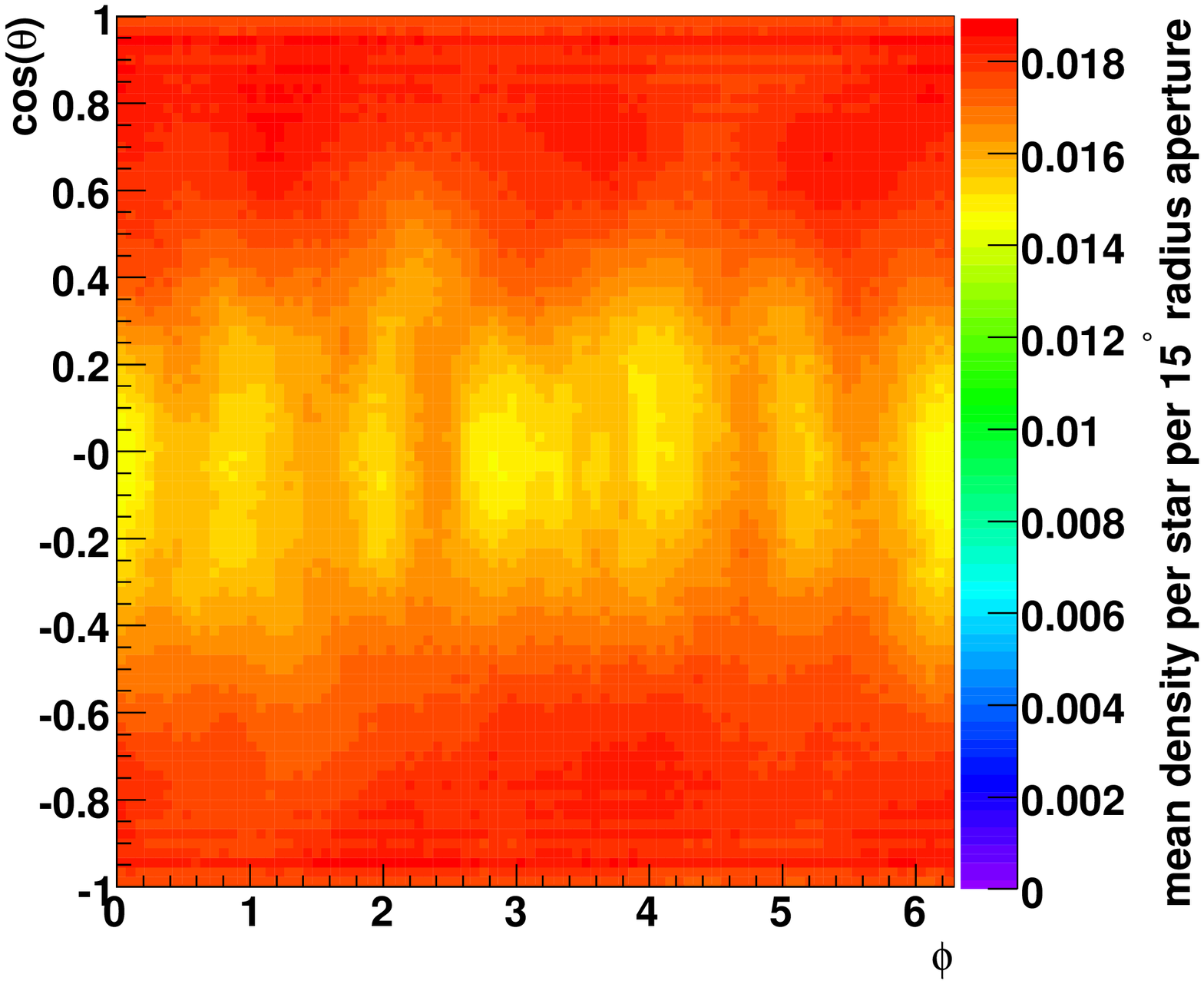}
\includegraphics[totalheight=7cm]{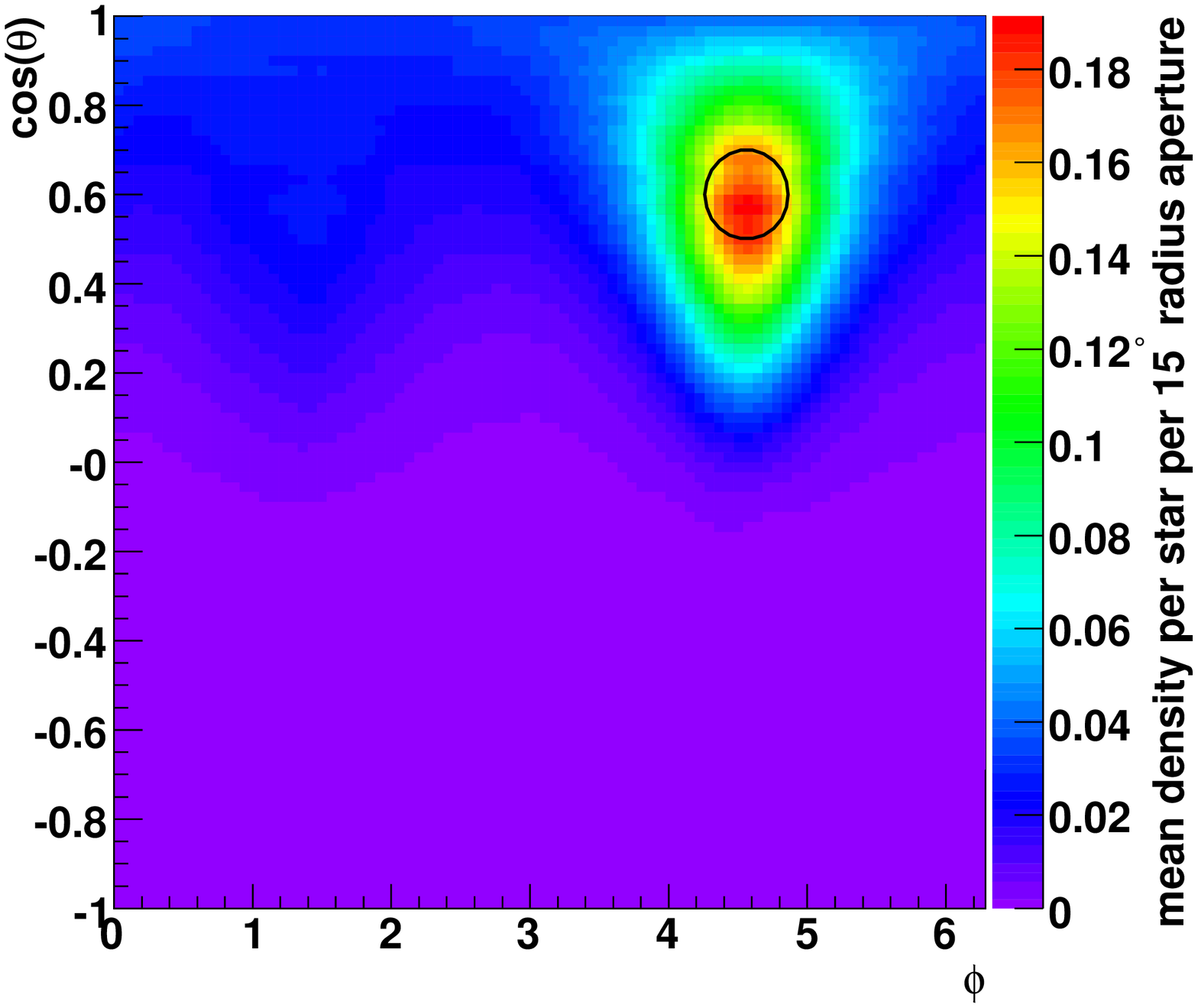}
\caption{\it \small cylindrical equal area projections of the sky distributions of the average density (fixed aperture of $15^{\circ}$ radius) per star of the reconstructed angular momentum directions for isotropic stars with projected angular distances from the Galactic Center between 0.8'' and 12'' (left panel) and for a stellar disk oriented like the clockwise disk in \citet{Paumard2006} in the same radial bin (right panel). The sky distribution of the average density of reconstructed angular momenta is rather flat in the case of isotropic stars. For a disk there is a peak at the simulated disk angular momentum direction. The simulated disk position is marked with the black circle. The plots are based on a simulated data set of $4\times10^5$ isotropic stars and $4\times10^4$ disk stars, see text.}
\label{fig:sky_simul_stars}
\end{center}
\end{figure}

Our null hypothesis is that the WR/O stars are isotropic. In order to be 
able to reject this null hypothesis we have to accurately characterize 
both the average and the fluctuations of the distribution of isotropic 
stars. In order to show that the observed stars are compatible with a
disk of stars we only need to characterize the average distribution of 
disk stars. Therefore, we 
simulated a total of $4\times10^5$ isotropic stars and several disks of $4\times10^4$ stars each with different orientations and thicknesses.
As an example, we selected isotropic stars with projected angular distances from Sgr~A* between 0.8'' and 12''. For these stars we computed sky distributions of reconstructed angular momentum densities for a fixed aperture of $15^{\circ}$ radius. Figure \ref{fig:sky_simul_stars} (left panel) shows a cylindrical equal area projection of the average reconstructed angular momentum density.
The right panel of figure \ref{fig:sky_simul_stars} shows the same distribution but for a disk with a two-dimensional Gaussian sigma thickness of $10^{\circ}$ oriented like the clockwise disk in \citet{Paumard2006} in the same radial interval (right panel).
In case of isotropic stars, the sky distribution of reconstructed angular momenta is rather flat (e.g., $\pm13\%$ variation).  There is a small depletion close to $\cos\theta = 0$ caused by the choice of simulated stars with {\it projected} distances in the interval 0.8''--12''. 
This projection along a cylinder causes a small bias of the actual $z$-coordinate of the star to have a larger absolute value than the generated $z$-coordinates. In our example the distribution of MC generated isotropic stars has a mean of zero and an RMS of 5.7'', while the distribution of the reconstructed z values has an RMS of 5.4''.
This causes the small bias towards face-on reconstructed orbits. 
For a disk there is a peak at the simulated disk angular momentum direction.

In the case of 90 isotropic stars the MC simulations predict an average value of 1.6 stars per $15^{\circ}$ radius aperture centered on the direction of the clockwise disk in \citet{Paumard2006}. This value is similar to the zeroth order expectation of  $90\times4\pi/(2\pi(1-\cos15^{\circ}))=1.53$ stars. The expected RMS is 0.55~stars per $15^{\circ}$ radius aperture.
A simulated thick disk of 90 stars rather yields a reconstructed angular momentum density of 17.2 stars per $15^{\circ}$ radius aperture. Hence a disk can be well distinguished from isotropic stars.

\subsection{Computation of Significance Maps}

\begin{figure}[t!]
\begin{center}
\includegraphics[totalheight=7cm]{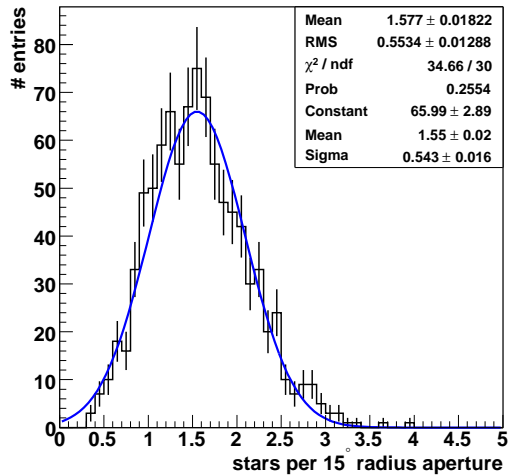}
\caption{\it \small Distribution of reconstructed angular momentum directions per $15^{\circ}$ aperture centered at the position of the clockwise disk of \citet{Paumard2006} for sets of 90 simulated isotropic stars (black histogram). The blue line shows a Gaussian fit.}
\label{fig:distribution_density_15deg_aperture}
\end{center}
\end{figure}

The goal is to test whether the features of a given population of $n$ stars are statistical fluctuations of an isotropic stellar distribution or not. 
We applied the same cuts in projected distance to the MC simulated stars as to the data.
We grouped the simulated isotropic stars in $N$ sets of $n$ stars. 
For each set of simulated stars we computed a sky map of the density of reconstructed angular momentum directions per $15^{\circ}$ aperture, again in a cylindrical equal area projection $\cos\theta$ vs. $\phi$ with a $90 \times 90$ grid of bins. We histogrammed the distribution of $N$ values for the density of reconstructed angular momentum directions per $15^{\circ}$ aperture for each of the $90 \times 90$ bins. These distributions can be well approximated by Gaussian distributions. We produced two sky maps containing the mean and the RMS of these distributions.
As an example, figure \ref{fig:distribution_density_15deg_aperture} shows the distribution of reconstructed angular momentum directions per $15^{\circ}$ aperture centered at the position of the clockwise disk of \citet{Paumard2006} for sets of 90 simulated isotropic stars. This distribution can be approximated by a Gaussian.

We computed significance maps from the sky map of the density of reconstructed angular momentum directions of the observed stars. We define the significance for each bin of the sky map as (see equation 10a of \citet{LiMa1983}):
\begin{equation}
\mathrm{significance} = \frac{\text{observed density - mean density of isotropic stars}}{\text{RMS density of isotropic stars}} \ .
\end{equation}
As the expected distribution of angular momentum densities for isotropic stars is approximately Gaussian, we will call an excess of $x$ times the RMS density over the mean density expected for isotropic stars an excess of $x$ $\sigma$.


\subsection{Definition of $\chi^2$ for Disks} \label{sec:chi2_definition}


\begin{figure}[t!]
\begin{center}
\includegraphics[totalheight=7cm]{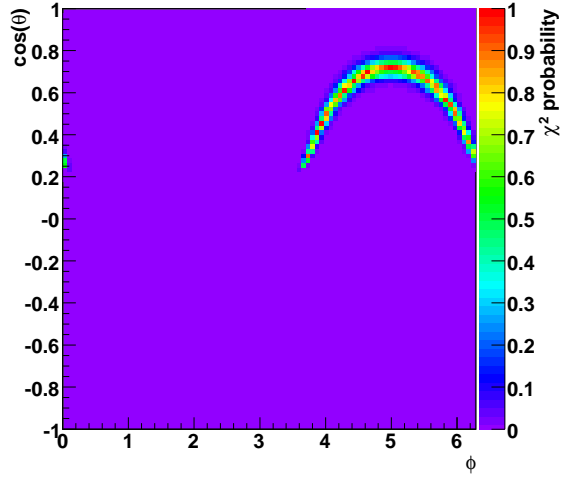}
\caption{\it \small cylindrical equal area projection of the probability that the star IRS16~CC is part of a thin disk with an angular momentum direction given by $(\theta,\phi)$, without assuming any prior on the $z$-position of the star.
}
\label{fig:IRS16CC_chi2}
\end{center}
\end{figure}

With the method described above we can determine the position and significance of possible disk features in the distribution of the early-type stars. The next step is to quantify if a given star is a candidate 
member of the disk. We define a $\chi^2(\theta_{\mathrm{disk}},\phi_{\mathrm{disk}})$ value that a star with measured 3D velocity $(v_{x,\mathrm{m}}, v_{y,\mathrm{m}}, v_{z,\mathrm{m}})$ and 2D position $(x_{\mathrm{m}}, y_{\mathrm{m}})$ has an angular momentum direction $\vec{n} = \vec{J}/|\vec{J}| = (\cos\phi_{\mathrm{disk}} \sin\theta_{\mathrm{disk}}, \sin\phi_{\mathrm{disk}} \sin\theta_{\mathrm{disk}}, \cos\theta_{\mathrm{disk}})$ by minimizing the following function with respect to $(a,\epsilon,\omega,\tau)$:
%
%

\begin{equation}
\chi^2(\theta_{\mathrm{disk}},\phi_{\mathrm{disk}}) = \mathrm{Min}
\left[ \begin{array}{c} \normalsize
 \left( \frac{x_{\mathrm{m}} - x(a,\epsilon,\omega,\tau,\Omega_{\mathrm{disk}},i_{\mathrm{disk}})}{\Delta x_{\mathrm{m}}} \right)^{2} + \left( \frac{y_{\mathrm{m}} - y(a,\epsilon,\omega,\tau,\Omega_{\mathrm{disk}},i_{\mathrm{disk}})}{\Delta y_{\mathrm{m}}} \right)^{2} \\ 
+ \left( \frac{v_{x,\mathrm{m}} - v_x(a,\epsilon,\omega,\tau,\Omega_{\mathrm{disk}},i_{\mathrm{disk}})}{\Delta v_{x,\mathrm{m}}} \right)^{2} + \left( \frac{v_{y,\mathrm{m}} - v_y(a,\epsilon,\omega,\tau,\Omega_{\mathrm{disk}},i_{\mathrm{disk}})}{\Delta v_{y,\mathrm{m}}} \right)^{2}\\
+ \left( \frac{v_{z,\mathrm{m}} - v_z(a,\epsilon,\omega,\tau,\Omega_{\mathrm{disk}},i_{\mathrm{disk}})}{\Delta v_{z,\mathrm{m}}} \right)^{2} 
\end{array} \right] \ . \label{eq_chi2}
\end{equation}
This definition does not use any priors except bound orbits to Sgr~A*. The orbital elements $(a,\epsilon,\omega,\tau,\Omega_{\mathrm{disk}},i_{\mathrm{disk}})$ define the motion of the star (see section \ref{sec:MC_simulation}). $\Omega_{\mathrm{disk}}$ and $i_{\mathrm{disk}}$ are fixed by the disk angular momentum direction: $\Omega_{\mathrm{disk}} = \arctan(-\tan \phi_{\mathrm{disk}})$, $i_{\mathrm{disk}} = \arccos(-\sin{\phi_{\mathrm{disk}}}\sin{\theta_{\mathrm{disk}}})$. The above defined $\chi^2$ has only observational errors in the denominators, and thus it has no bias to any models or priors.
Contrary to the $\chi^2$ definition of \citet{Levin2003} our definition takes both the velocity and the projected position of the star into account.
As an example, figure \ref{fig:IRS16CC_chi2} shows a cylindrical equal area projection of the $\chi^2$ probability (one degree of freedom) $p(\chi^2,1)$ that the stellar angular momentum points in the direction $(\theta,\phi)$ for the star IRS16~CC. 
The quantity $p(\chi^2,1) = 1 - \mathrm{Erf}(\sqrt{\chi^2/2})$ denotes  the probability to observe by chance a value of $\chi^2$ or larger, even for a correct model \citep{Yao2006}.

For comparison, figure \ref{fig:IRS16CC} shows the distribution of reconstructed angular momentum directions for the same star. The 
maximum probability contour in figure \ref{fig:IRS16CC_chi2} is identical in shape to the curve of the reconstructed angular momenta in figure \ref{fig:IRS16CC}. The $z$-prior determines the statistical weight of the curve in figure \ref{fig:IRS16CC}.

\begin{figure}[t!]
\begin{center}
\includegraphics[totalheight=7cm]{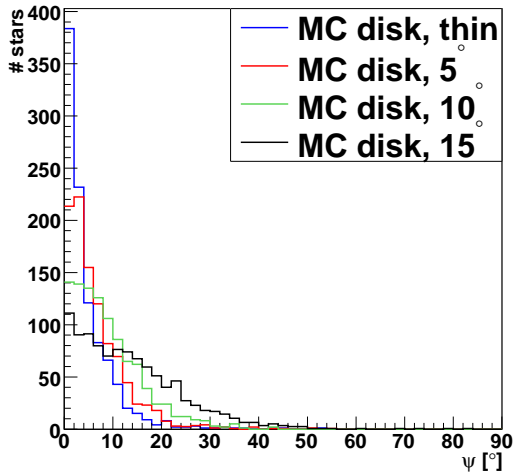}
\caption{\it \small Distribution of the reconstructed angular deviation $\psi$ of the disk stars from the common disk angular momentum direction for four MC generated disk of 1000 stars each, with $0^{\circ}$, $5^{\circ}$, $10^{\circ}$ and $15^{\circ}$ thickness.}
\label{fig:MC_angles}
\end{center}
\end{figure}

We define the deviation angle $\psi$ to be the minimum angle between the direction of the disk angular momentum and the $\chi^2=0$ contour. For small deviation angles, our definition is equivalent to the definition of \citet{Beloborodov2006}. However, in our definition angles up to $180^{\circ}$ are possible, while in the definition of \citet{Beloborodov2006} the deviation angle is between zero and $90^{\circ}$. Figure \ref{fig:MC_angles} shows the distribution of the angular deviation $\psi$ of the disk stars from the common disk angular momentum direction for four MC generated disk of 1000 stars each, with $0^{\circ}$, $5^{\circ}$, $10^{\circ}$ and $15^{\circ}$ thickness.

\section{Results}

\subsection{The Clockwise System}

\begin{figure}[t!]
\begin{center}
\includegraphics[totalheight=7cm]{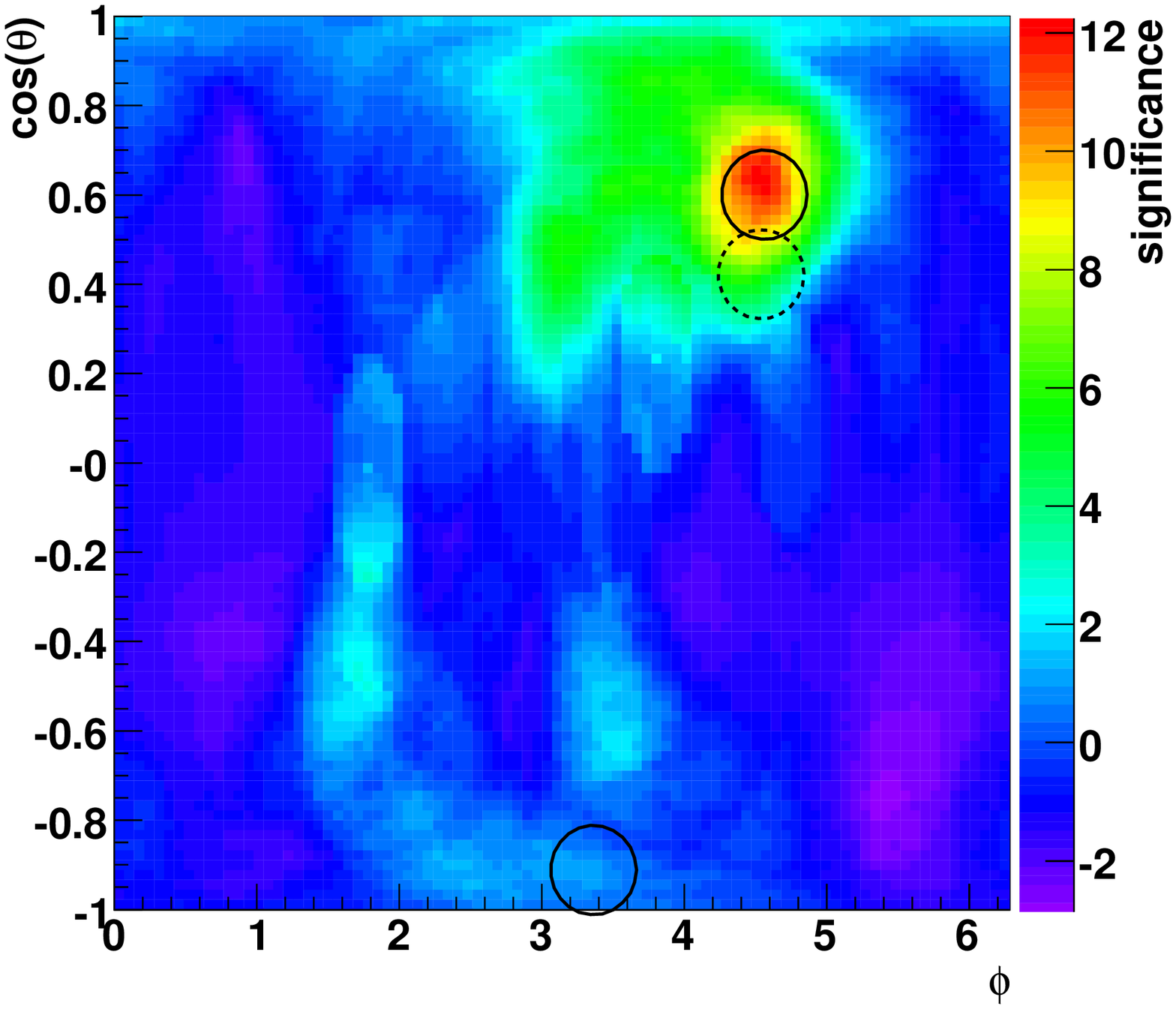}
\includegraphics[totalheight=7cm]{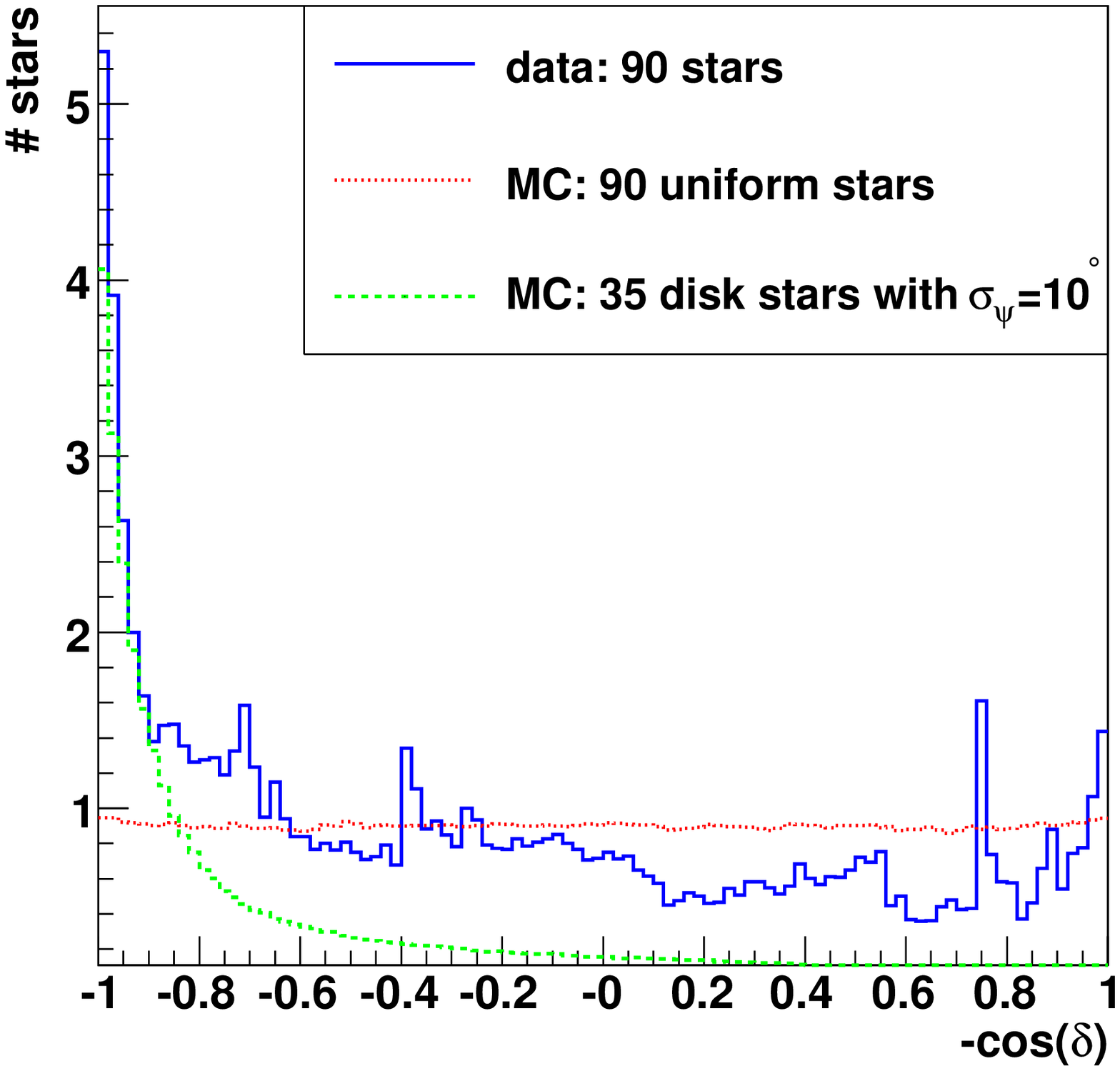}
\caption{\it \small (Left) cylindrical equal area projection of the distribution of significance in the sky for all 90 high quality WR/O stars with projected distances between 0.8'' and 12''. The disk positions of \citet{Paumard2006} are marked with full black circles and the position of the clockwise disk of \citet{Lu2008} is marked by a broken black circle. There is a maximum significance of $12.2\sigma$ at $(\phi,\theta)=(262^{\circ},51^{\circ})$, compatible with the clockwise system of \citet{Paumard2006}. The excess appears to have a narrow core and an extended tail to lower values of $\phi$. Moreover, there is an extended excess of counter-clockwise orbits. 
(Right) Distribution of the negative cosine of the angular distance between the individual reconstructed angular momentum directions for each of the 1000 generated $z$-values for all 90 WR/O stars with respect to the observed excess center $(\phi,\theta)=(262^{\circ},51^{\circ})$ (full blue line), the MC simulation expectation for 90 isotropic stars (red dotted line) and for a MC simulated disk of 35 stars with a two-dimensional sigma of $10^{\circ}$.
}
\label{fig:sky_sig_Fabrice}
\end{center}
\end{figure}

Figure \ref{fig:sky_sig_Fabrice} (left panel) shows a cylindrical equal area projection of the distribution of significances as a function of the sky coordinates  $\cos\theta_J$ vs. $\phi_J$ for the 90 selected high quality WR/O stars in the range $0.8'' \leq \sqrt{x^2+y^2} \leq 12''$. 
There is a global maximum excess significance at $(\phi,\theta)=(262^{\circ},51^{\circ})$ of $12.2 \sigma$. This corresponds to a reconstructed angular momentum density of 8.3 stars per $15^{\circ}$ radius aperture. 
In the case of 90 isotropic stars the MC simulations predict an average value of 1.6 stars per $15^{\circ}$ radius aperture, see figure \ref{fig:sky_simul_stars} (left panel) and an RMS of 0.55~stars per $15^{\circ}$ radius aperture.
The excess appears to have a narrow core. 
Figure \ref{fig:sky_sig_Fabrice} (right panel) shows the distribution of the negative cosine of the angular distance $\delta$ between the individual reconstructed angular momentum directions for each of the 1000 generated $z$-values for all 90 WR/O stars with respect to the observed excess center.
In addition it shows the respective distributions for MC simulated isotropic stars and a MC generated disk of stars with a two-dimensional sigma thickness of $10^{\circ}$. A flat distribution on a sphere is mapped to a flat distribution of the negative cosine of the angular difference.
We determine the half width at half maximum (HWHM) of the excess as the angular difference $\delta=18^{\circ}$, for which the distribution of the negative cosine of the angular distance with respect to the excess center falls to half the maximum value. 
The width of the excess core is compatible with the width of a disk with a two-dimensional Gaussian sigma thickness (for a definition see equation \ref{eq:inclin_2D}) of $10^{\circ}$.

The peak angular momentum density for a MC simulated disk of 90 stars with a two-dimensional sigma thickness of $10^{\circ}$ is 17.2 stars per $15^{\circ}$ radius aperture (see figure \ref{fig:sky_simul_stars} right panel). Comparing this value to the observed excess, we estimate that 35 stars contribute to this excess peak. Dividing the HWHM width of the excess peak by the square root of this number of stars, we estimate the statistical uncertainty of the peak position to be $3^{\circ}$.
We re-ran the same analysis steps with $R_0=7.5$~kpc, $M_{\mathrm{SgrA*}}=3.5\times10^6 M_{\odot}$ and $R_0=8.5$~kpc, $M_{\mathrm{SgrA*}}=4.5\times10^6 M_{\odot}$. This yielded maximum offsets of $2^{\circ}$ in $\phi$ and $\theta$ and $0.5\sigma$ in significance. Thus we estimate the systematic errors due to the uncertainties in $R_0$ and $M_{\mathrm{Sgr~A*}}$ to $2^{\circ}$ in $\phi$ and $\theta$ and $0.5\sigma$ in significance.
Within errors the excess peak position is compatible with the direction of the clockwise disk of \citet{Paumard2006}. 

The excess has -- in addition to the narrow core -- an extended tail to lower values of $\phi$. Moreover, in figure \ref{fig:sky_sig_Fabrice} shows an extended U-shaped excess of counter-clockwise orbits including the direction of the counter-clockwise system of \citet{Paumard2006}.


\begin{figure}[p]
\begin{center}
\includegraphics[totalheight=7cm]{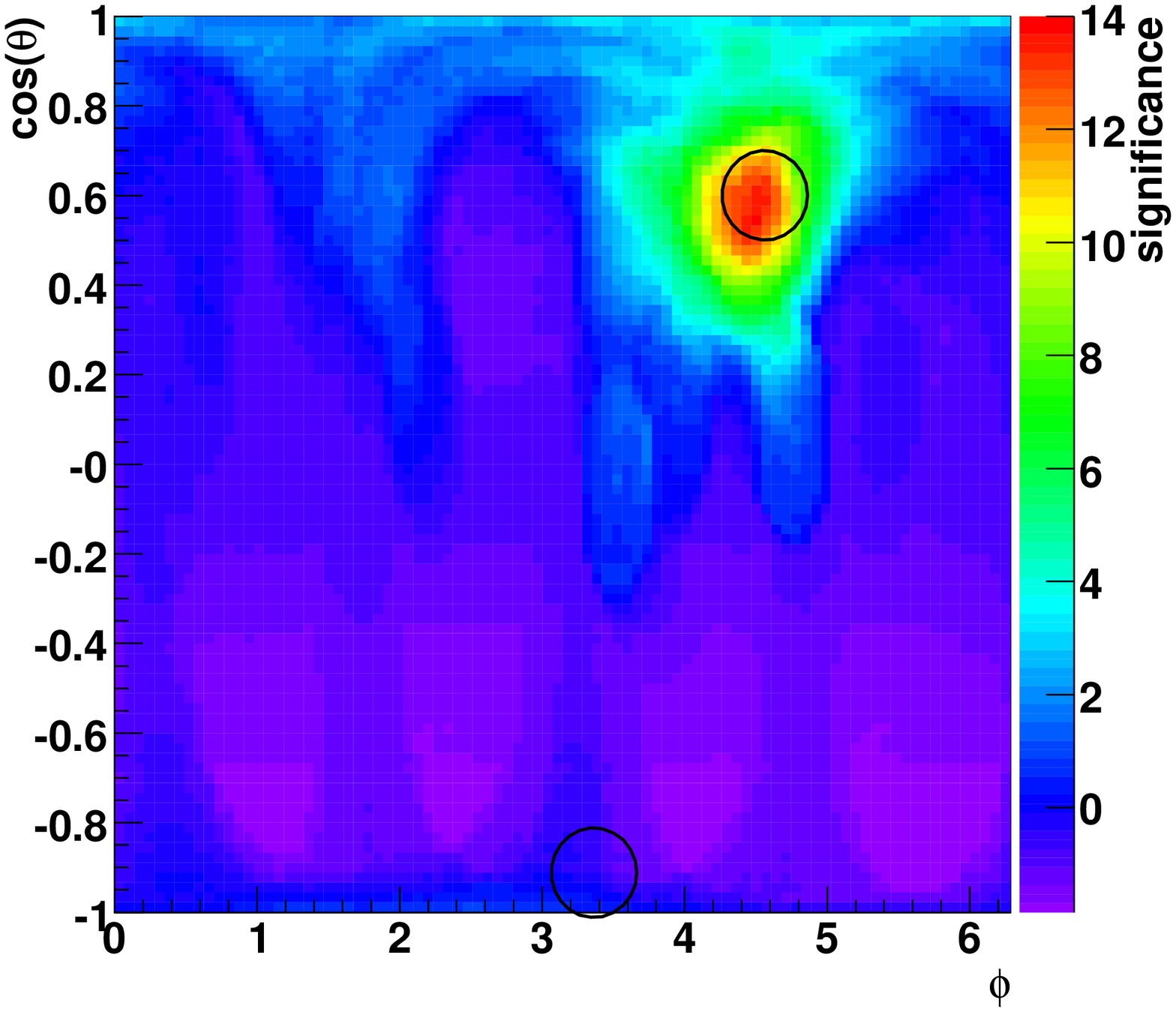}
\includegraphics[totalheight=7cm]{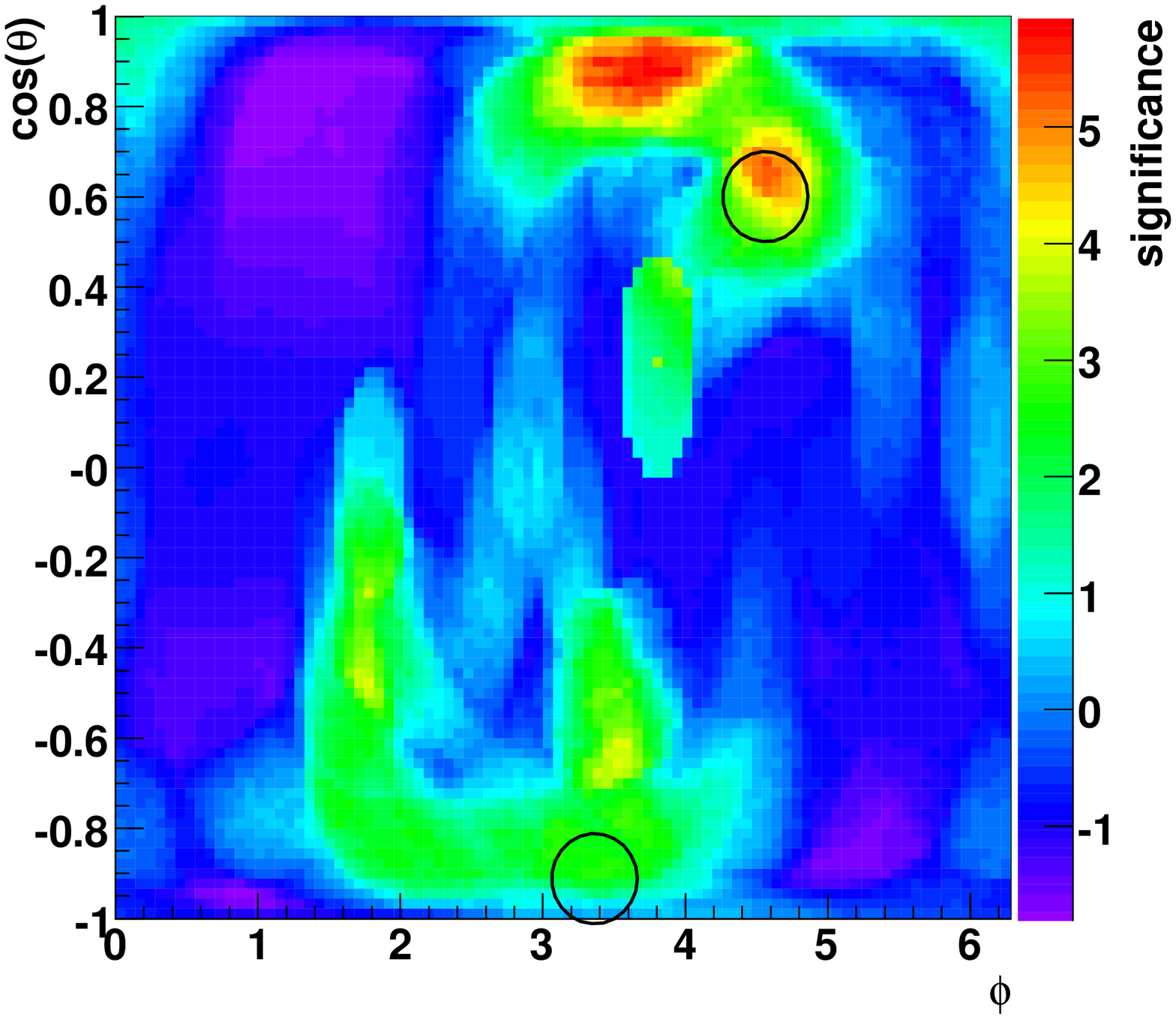}
\includegraphics[totalheight=7cm]{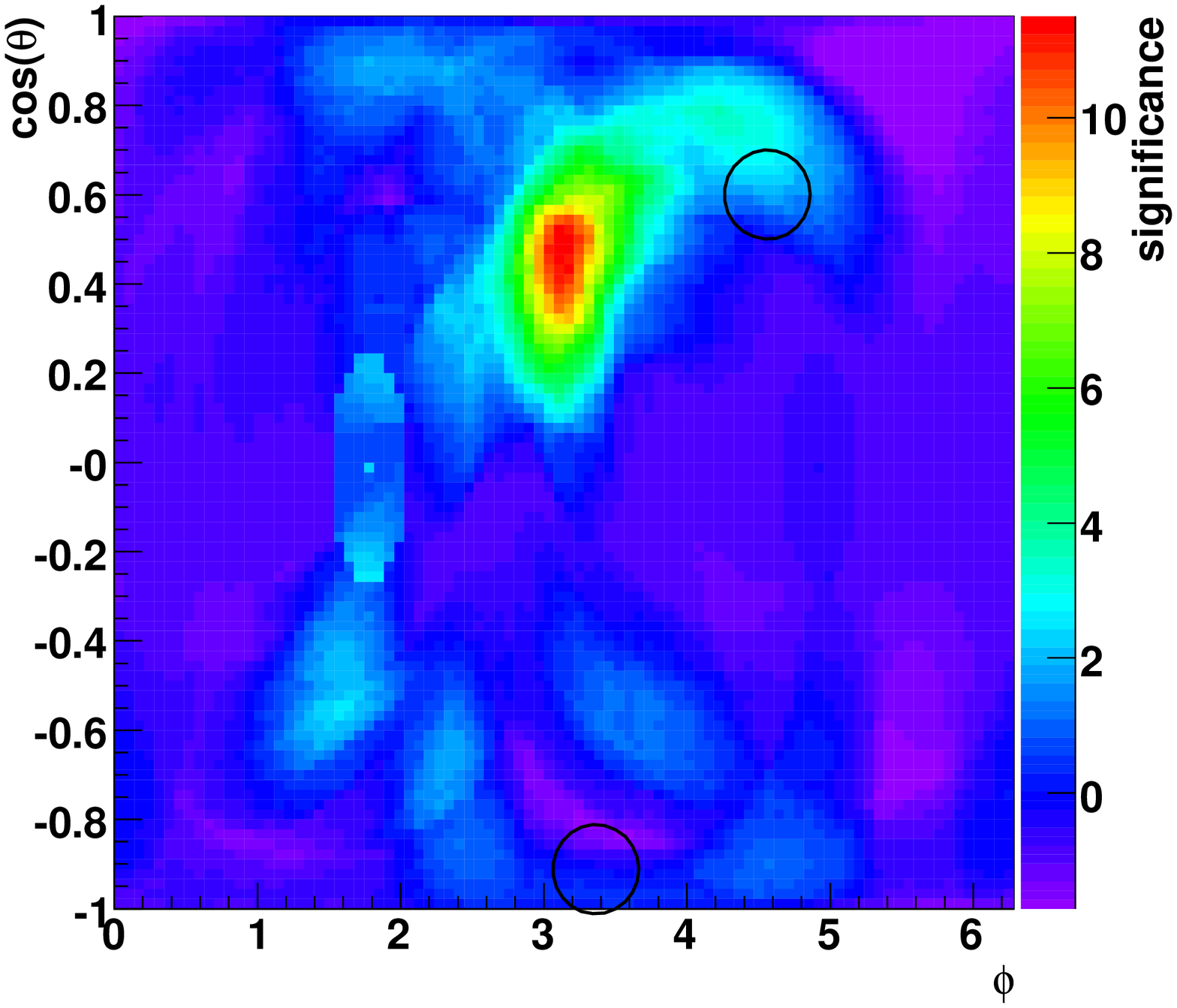}
\caption{\it \small cylindrical equal area projections of the distributions of significance in the sky for three radial bins: (upper left panel) 32 WR/O stars with projected distances in the bin 0.8''--3.5'' (upper right panel) 30 WR/O stars in the bin 3.5''--7'' and (lower panel) 28 WR/O stars in the bin 7''--12''.
In the inner bin there is a maximum excess significance of $13.9\sigma$ at $(\phi,\theta)=(256^{\circ},54^{\circ})$, compatible with the clockwise system of \citet{Paumard2006}. The significance map in the middle interval shows two extended excesses, one for clockwise and one for counter-clockwise orbits. The clockwise excess has a local maximum significance of $5.4\sigma$ at $(\phi,\theta)=(262^{\circ},48^{\circ})$ (compatible with the orientation of the clockwise system of \citet{Paumard2006}) but a global maximum significance of $5.9\sigma$ at a clearly offset position: $(\phi,\theta)=(215^{\circ},28^{\circ})$. The significance map in the outer bin shows a maximum excess significance of $11.5\sigma$ at yet another position $(\phi,\theta)=(179^{\circ},62^{\circ})$. The morphology of the excesses in the clockwise system may indicate a smooth transition of the excess center with projected radius. The disk positions \citep{Paumard2006} are marked with black circles.}
\label{fig:sky_sig_disk}
\end{center}
\end{figure}

\subsubsection{Radial Dependence of the Excess Position} \label{chi2_analysis}

In order to study a possible change in the excess features as a function of distance, we subdivided the sample of 90 WR/O stars into three radial intervals: 32 stars in the interval 0.8''--3.5'', 30 stars in the interval 3.5''--7'' and 28 stars in the interval 7''--12''. We chose the interval sizes such that each of the intervals contains approximately the same number of stars.
Figure \ref{fig:sky_sig_disk} shows cylindrical equal area projections of the significance sky distributions for these three intervals in projected distance to Sgr~A*. 
In the inner interval there is a maximum excess significance of $13.9\sigma$ at $(\phi,\theta)=(256^{\circ},54^{\circ})$, corresponding to a reconstructed angular momentum density of 5.5 stars per $15^{\circ}$ radius aperture. In case of 32 isotropic stars an average density of 0.63 stars and an RMS of 0.35 stars per  $15^{\circ}$ radius aperture are expected from MC simulations.
The map shows a well-defined peak with a HWHM of $16^{\circ}$. It is compatible with the excess density of angular momenta from a MC simulated disk of 25 stars with a $10^{\circ}$ thickness. We thus estimate the error of the peak position to be $3.2^{\circ}$ and find the peak position to be compatible with the clockwise system of \citet{Paumard2006}. 
Only 4 out of the 32 stars in the inner radial interval are on counter-clockwise orbits.

The significance map in the middle radial interval shows two extended excess structures, one for clockwise and one for counter-clockwise orbits. The clockwise excess structure has a local maximum significance of $5.4\sigma$ at $(\phi,\theta)=(262^{\circ},48^{\circ})$ (compatible with the orientation of the clockwise system of \citet{Paumard2006}) but a global maximum significance of $5.9\sigma$ at a clearly offset position: $(\phi,\theta)=(215^{\circ},28^{\circ})$. 15 out of the 30 stars in the middle radial interval are on counter-clockwise orbits. An extended U-shaped excess structure at the 3--4$\sigma$ significance level is visible. For a more detailed discussion about this counter-clockwise excess see section \ref{sec:counter_significance}. 

The significance map in the outer interval shows a maximum excess significance of $11.5\sigma$ at yet another position $(\phi,\theta)=(179^{\circ},62^{\circ})$, corresponding to an angular momentum density of 4.6 stars per $15^{\circ}$ radius aperture. For 28 isotropic stars, a density of 0.5 stars per  $15^{\circ}$ radius aperture is expected from MC simulations. 
The map shows a well-defined peak with a HWHM of $16^{\circ}$. It is compatible with the excess density of angular momenta from a MC simulated disk of 16 stars with a $10^{\circ}$ two-dimensional sigma thickness. We estimate the error of the peak position to be $4^{\circ}$.
10 of the 28 stars in the outer radial interval are on counter-clockwise orbits.

The angular distance between the significance peaks in the inner and outer intervals is $(64\pm6)^{\circ}$. 

The significance map of the inner interval shows a significance below $1 \sigma$ at the maximum position of the outer interval $(\phi,\theta)=(179^{\circ},62^{\circ})$. Moreover, the significance map of the outer interval shows a significance of only about $1.5\sigma$ at the maximum position of the inner interval $(\phi,\theta)=(256^{\circ},54^{\circ})$. We estimate the significance for a change of the maximum excess position as $>10\sigma$.
The morphology of the excesses in the clockwise system thus indicates that the excess center varies with projected radius.
Table \ref{tab:significances} summarizes the excess positions and significances for the full sample of stars and the three radial intervals.

\begin{table*}[t!]
\caption{Parameters of the clockwise system. The given position error is the statistical error only. We estimate the systematic error as $2^{\circ}$ in $\phi$ and $\theta$.} 
\begin{center}
\begin{tabular}{l | l | l | l | l | l} 
\hline 
\hline
radial interval &  \#stars & max. significance   & max. position                           & position error & HWHM \\

\hline
0.8''--12''    &  90     & $12.2\sigma$         & $(\phi,\theta)=(262^{\circ},51^{\circ})$ & $3^{\circ}$     & $18^{\circ}$   \\
                &         &                      & $(\Omega,i)=(98^{\circ},129^{\circ})$    &                &               \\
\hline
0.8''--3.5''   &  32     & $13.9\sigma$         & $(\phi,\theta)=(256^{\circ},54^{\circ})$ & $3.2^{\circ}$   & $16^{\circ}$  \\
                &         &                      & $(\Omega,i)=(104^{\circ},126^{\circ})$   &                &   \\ 
\hline
3.5''--7''     &  30     & $5.9\sigma$          & $(\phi,\theta)=(215^{\circ},28^{\circ})$ & extended       & extended \\ 
                &         &                      & $(\Omega,i)=(145^{\circ},152^{\circ})$   &                &   \\ 
\hline
7''--12''      &  28     & $11.5\sigma$         & $(\phi,\theta)=(181^{\circ},62^{\circ})$ & $3.8^{\circ}$   & $16^{\circ}$  \\
                &         &                      & $(\Omega,i)=(118^{\circ},126^{\circ})$   &                &   \\ 
\hline 
\hline
\end{tabular} 
\end{center}
\label{tab:significances} 
\end{table*}

\begin{figure}[t!]
\begin{center}
\includegraphics[totalheight=7cm]{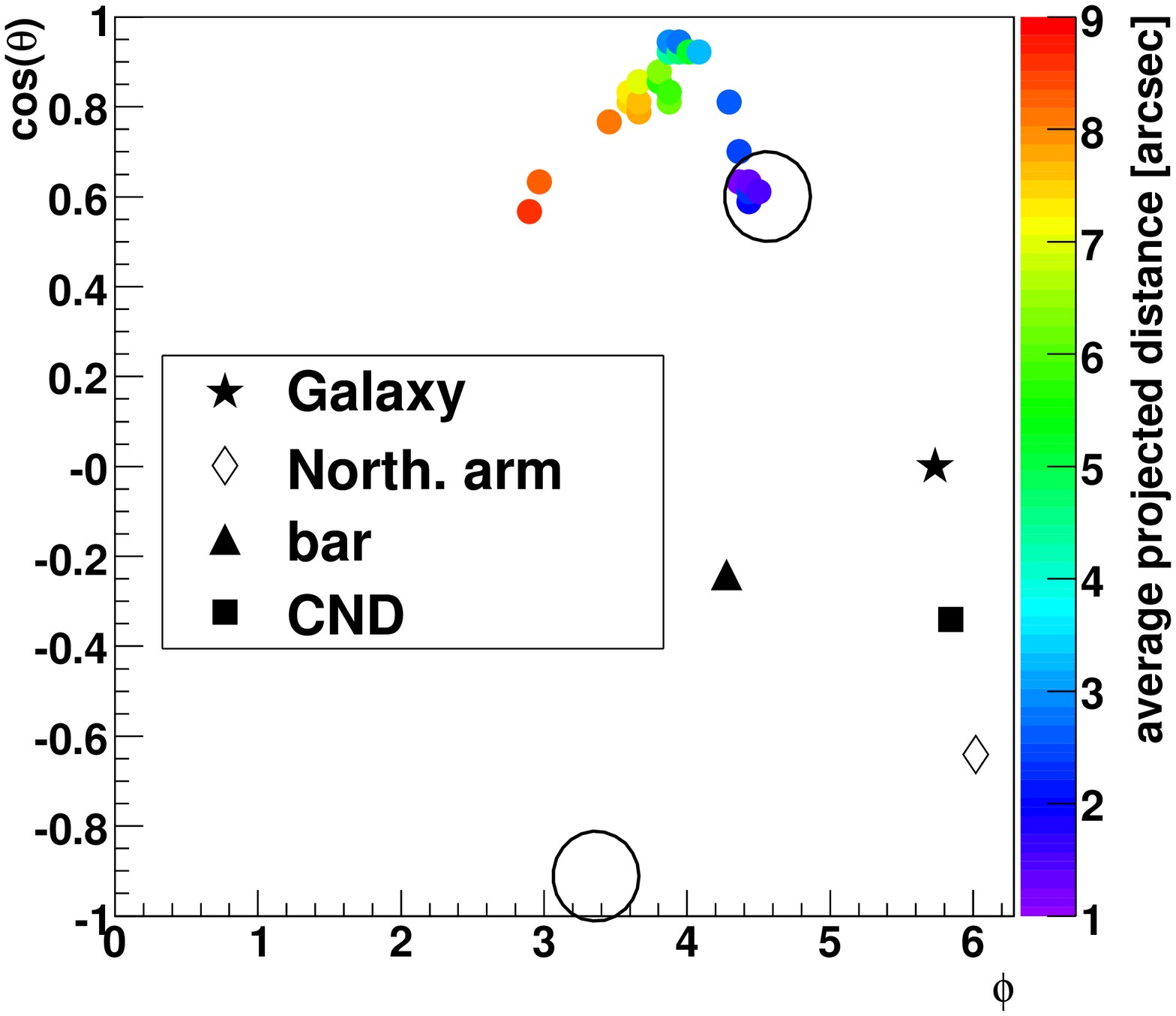}
\includegraphics[totalheight=6.5cm]{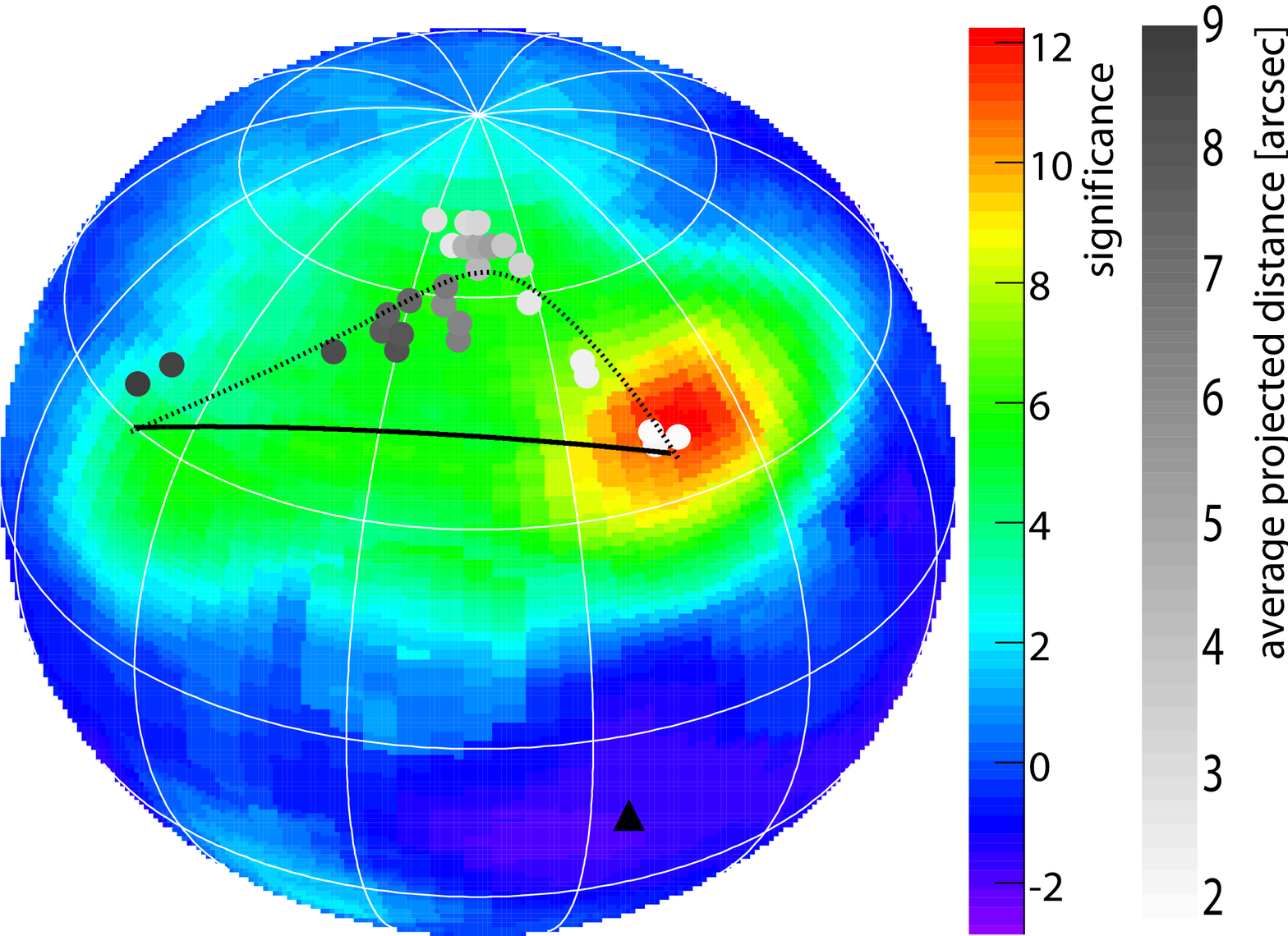}
\caption{\it \small 
(Left) Cylindrical equal area projection of the local average stellar angular momentum direction for the clockwise stars as a function of the average projected distance. The points are correlated, see text. The average angular momenta for the innermost stars agree well with the orientation of the clockwise disk of \citet{Paumard2006}, shown by the black circle. The asterisk shows the Galactic pole \citep{Reid2004}, the diamond indicates the normal vector to the northern arm of the minispiral \citep{Paumard2004}, the triangle indicates the normal vector to the bar of the minispiral \citep{Liszt2003}, the square shows the rotation axis of the circum-nuclear disk (CND) \citep{Jackson1993}.
(Right) 
Orthographic projection of the sky significance distribution of all WR/O stars (see figure \ref{fig:sky_sig_Fabrice}), seen from $\phi_0=226^{\circ}$, $\theta_0=54^{\circ}$ to highlight the clockwise system. The gray points show the average stellar angular momentum directions for the clockwise stars as a function of the average projected distance.
The full black line shows a great circle between the angular momentum directions of the inner and outer borders of the clockwise system. The broken black line shows a fit with quadratic polynomials to $\theta_J = \theta_J(R)$ and $\phi_J = \phi_J(R)$. It describes the observed change of the angular momentum direction with radius better than the great-circle.
}
\label{fig:disk_position_vs_distance_sphere}
\end{center}
\end{figure}

\begin{figure}[t!]
\begin{center}
\includegraphics[totalheight=7cm]{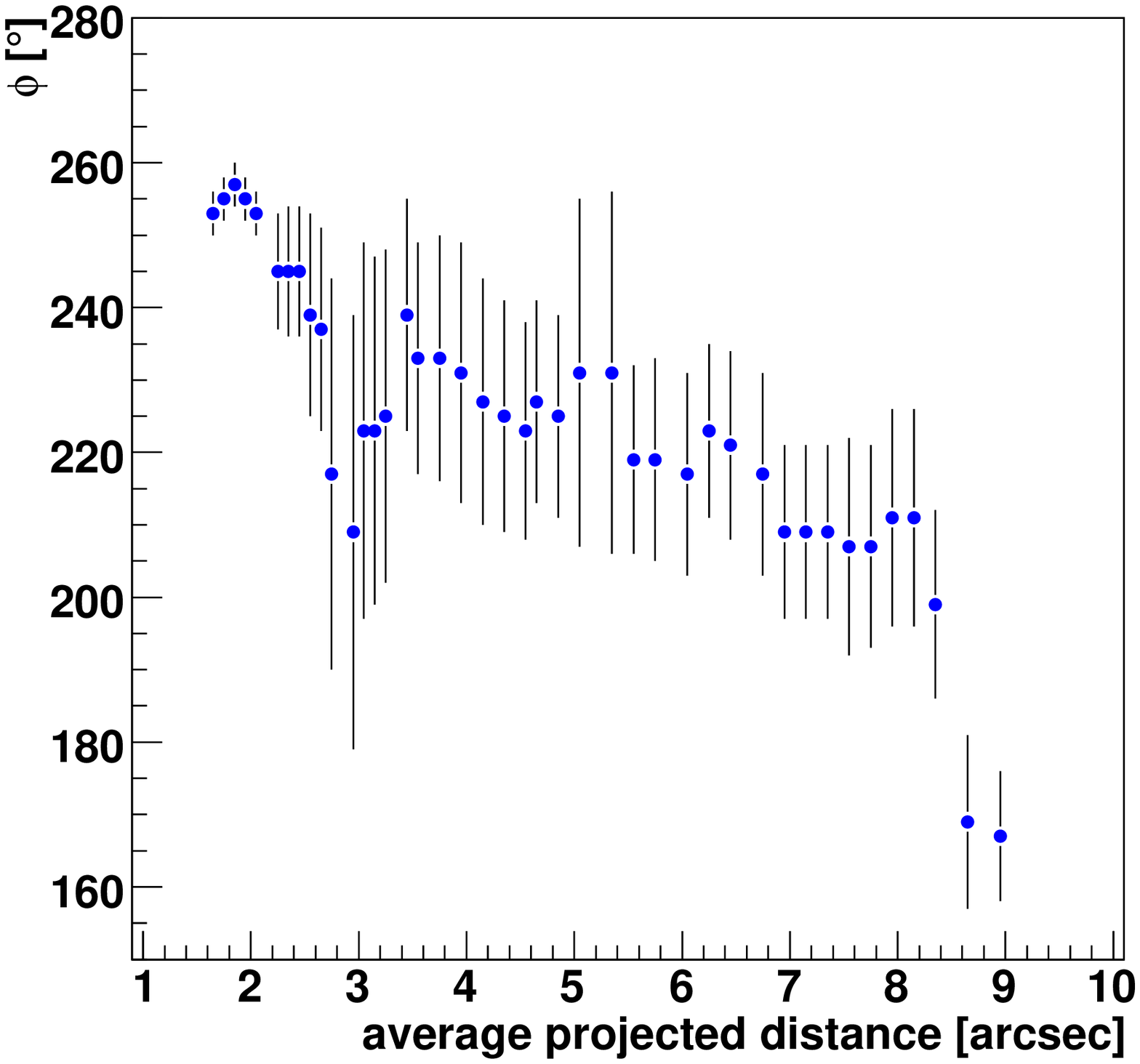}
\includegraphics[totalheight=7cm]{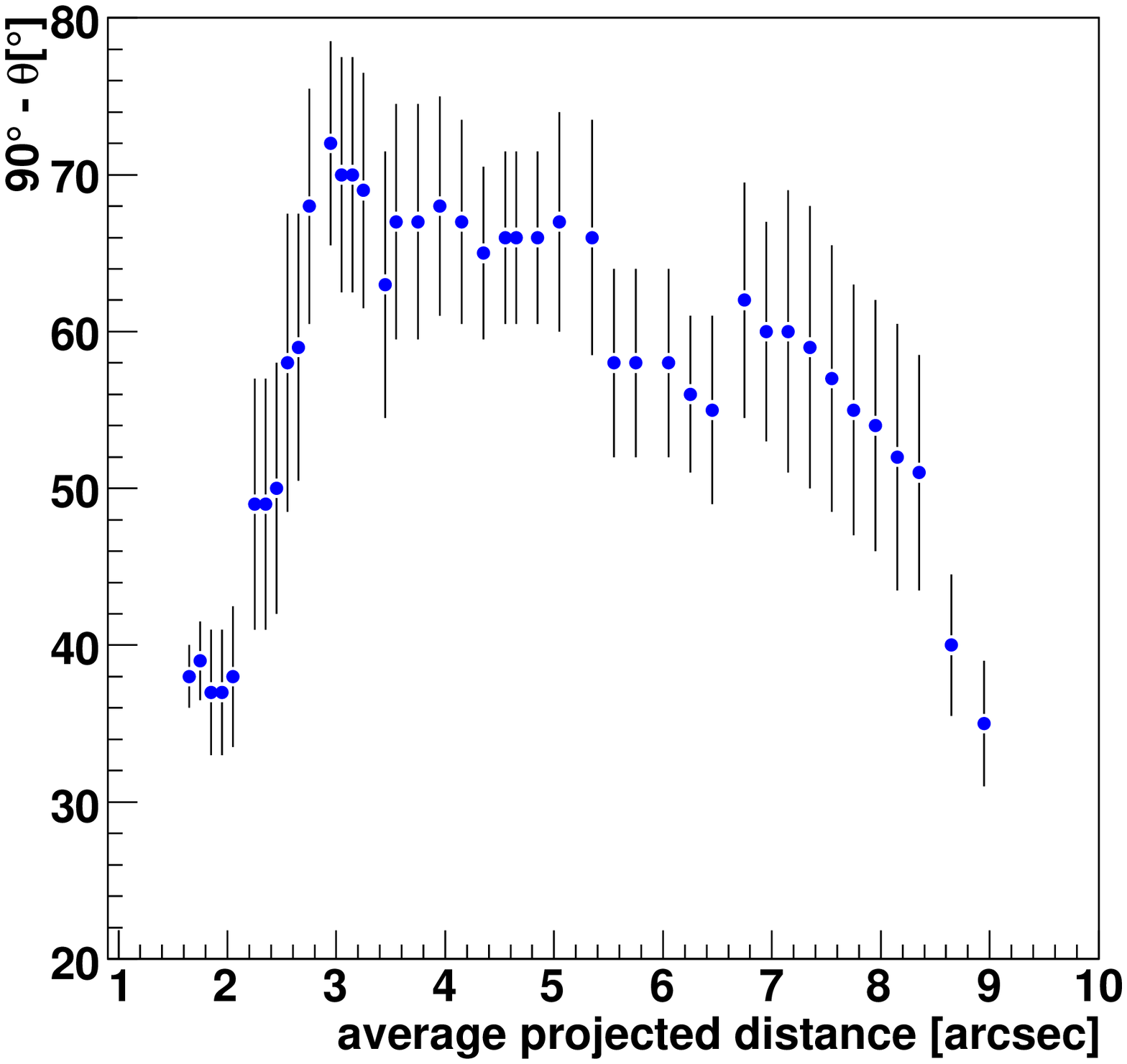}
\caption{\it \small 
Local average stellar angular momentum direction for the clockwise stars as a function of the average projected distance: (Left) $\phi_J = \phi_J(R)$ (Right) $\theta_J = \theta_J(R)$. The 42 points are correlated. They correspond to the 42 positions of a window of width 19 stars which is slid over the 61 clockwise moving WR/O stars ordered by their projected distances to Sgr A*.
}
\label{fig:disk_l_vs_p}
\end{center}
\end{figure}

We conclude that there is a significant change in the orientation of the clockwise system with projected distance. This change could reflect a large scale warp of a single disk, or the superposition of at least two disks or planar streamers located at different radii. The inner and outer radial interval can be well described by stars in $10^{\circ}$ thick disks with a relative inclination of $(64\pm6)^{\circ}$. The middle radial interval appears to represent a transition region. It also contains a significant fraction of stars on counter-clockwise orbits.
In order to explore this transition of the angular momentum direction of the clockwise system further, we ordered the 61 clockwise
 moving WR/O stars by their projected distances to Sgr~A*. We slid a window of width 19
 stars over the ordered clockwise moving stars, resulting in 42
 groups of clockwise
 moving stars. 
For each group of stars we calculated the bin-by-bin sum of the $\chi^2(\theta,\phi)$ sky plots for all the individual stars. We determined the sky position $(\theta_{\mathrm{min}},\phi_{\mathrm{min}})$ of the minimum of the summed $\chi^2$ sky plot. 
In the case of a warped system of stars, we would expect a smooth dependence of  $(\theta_{\mathrm{min}},\phi_{\mathrm{min}})$ with distance. Isotropic stars would result in random fluctuations of $(\theta_{\mathrm{min}},\phi_{\mathrm{min}})$.
In order not to be influenced by outliers, we iterate the determination of the minimum position. We calculated for all 19 stars in the window the angular distance to the sky position $(\theta_{\mathrm{min}},\phi_{\mathrm{min}})$. We excluded stars with a $\chi^2(\theta_{\mathrm{min}},\phi_{\mathrm{min}})>100$ and again computed the sky position with the minimum $\chi^2$, $(\theta_{\mathrm{min, iter}},\phi_{\mathrm{min, iter}})$. This position is the average angular momentum direction of the group of stars. 
We note that the window selection on the projected distance may introduce biases in the average angular momentum position. Better selection criteria might be the 3D distance, the semi-major axis, or the total energy of the star in the potential well of the supermassive black hole. However, the use of these selection criteria introduces a model dependence as the $z$-position of the stars is unknown, not all stars are members of the clockwise system, and the candidate members have non-zero eccentricities. This will be the subject of future work.

Figure \ref{fig:disk_position_vs_distance_sphere} (left) shows a cylindrical equal area projection of the local average angular momentum direction for the clockwise moving stars as a function of the average projected distance to Sgr~A*.
Figure \ref{fig:disk_position_vs_distance_sphere} (right) shows the same data as gray scale points in orthographic projections centered on the clockwise disk together with the significance sky distribution from figure \ref{fig:sky_sig_Fabrice}.
The average angular momentum direction of the clockwise stars is a 
function of the projected distance to Sgr~A*. For the innermost stars, the average angular momentum direction is compatible with the clockwise system of \citet{Paumard2006} but for higher projected distances the average angular momentum is clearly offset. This confirms the shift of the excess in the significance sky plots in figure \ref{fig:sky_sig_disk} as a function of projected distance to the Galactic Center. 
The full black line shows a great circle between the angular momentum directions of the inner and outer borders of the clockwise system. The broken black line shows a fit with quadratic polynomials to $\theta_J = \theta_J(R)$ and $\phi_J = \phi_J(R)$. It describes the observed change of the angular momentum direction with radius better than the great-circle.
Figure \ref{fig:disk_l_vs_p} shows the local average stellar angular momentum direction for the clockwise stars as a function of the average projected distance: $\phi_J = \phi_J(R)$ (left panel) and $\theta_J = \theta_J(R)$ (right panel). Within errors the angular momentum direction is compatible with being a smooth function of the average projected distance. We attribute the observed scatter to Poisson noise plus an additional local disk thickness.

The most likely explanation for the observed change in the angular momentum direction of the clockwise system with distance to Sgr~A* is a tilted warp in the clockwise disk (see the discussion section \ref{sec:discussion}), whose innermost part is the clockwise disk described in previous works. The angular difference between the innermost and outermost radii sampled  is $(64\pm6)^{\circ}$. The angular momentum direction as a function of the average projected distance does not follow a great circle, as expected for a simple warp. Instead, there is a significant offset from the great-circle, a tilt.

\subsubsection{Inclinations to the Clockwise System}

\begin{figure}[t!]
\begin{center}
\includegraphics[totalheight=7cm]{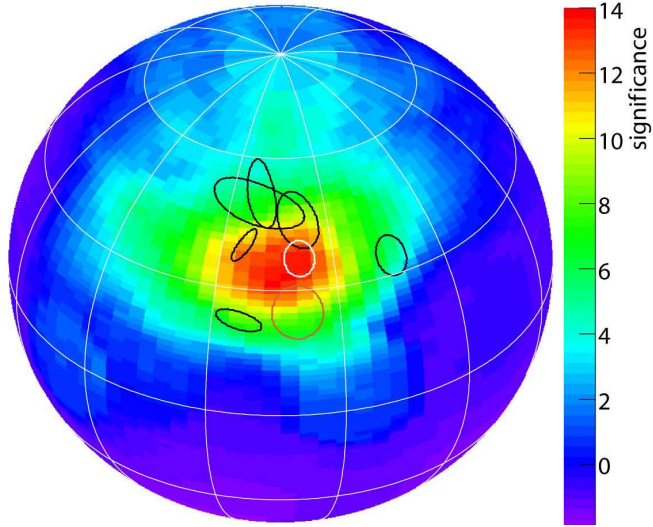} 
\caption{\it \small Orthographic projection (seen from $\phi_0=254^{\circ}$, $\theta_0=54^{\circ}$) of the significance sky map in the interval of projected distances 0.8''--3.5''. It is overlaid with the 2 sigma contours (black lines) for the direction of the orbital angular momentum vectors of the six early-type stars (S66, S67, S83, S87, S96 and S97) with $0.8''\leq R\leq1.4''$ for which \citet{Gillessen2008} were able to derive individual orbital solutions. All of these stars seem compatible with being members of the clockwise system. Still the orbital angular momenta of all of these stars are offset from the local angular momentum direction of the clockwise system at a confidence level beyond 90\%. The white ellipse shows the 2 sigma contour of the clockwise system as determined by \citet{Paumard2006} and the brown one the 2 sigma contour of \citet{Lu2008}. 
}
\label{fig:sky_zoom_Sstars}
\end{center}
\end{figure}

\begin{figure}[t!]
\begin{center}
\includegraphics[totalheight=7cm]{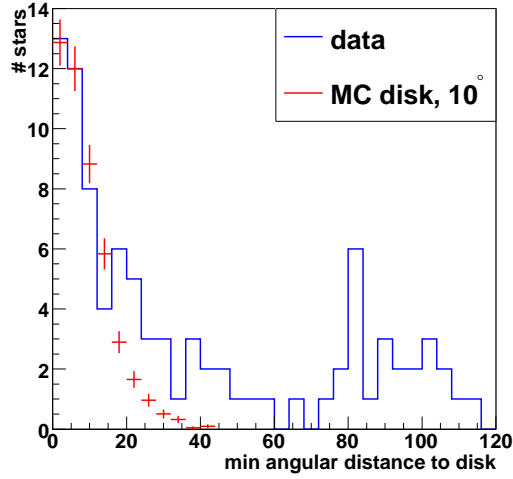}
\caption{\it \small Distribution of the reconstructed angular difference $\psi$ for all the 90 WR/O stars from the local average angular momentum direction of the clockwise system (blue histogram). The peak is well described by the expected distribution for a disk with a $10^{\circ}$ two-dimensional Gaussian sigma thickness and 46 disk members. In addition, there is a long tail to large angular distances towards to clockwise system.}
\label{fig:h_angle_clock}
\end{center}
\end{figure}

Figure \ref{fig:sky_zoom_Sstars} shows an orthographic projection (seen from $\phi_0 = 256^{\circ}$, $\theta_0=54^{\circ}$) of the significance sky map in the interval of projected distances 0.8''--3.5'' (upper left panel, figure \ref{fig:sky_sig_disk}). It is overlaid with the $2\sigma$ contours for the direction of the orbital angular momentum vector of the six early-type stars (S66, S67, S83, S87, S96 and S97) with $0.8''\leq R\leq1.4''$ for which \citet{Gillessen2008} were able to derive orbital solutions. All of these stars appear to be compatible with being members of the clockwise system. Still the orbital angular momenta of all of these stars are offset from the local angular momentum direction of the clockwise system at a confidence level beyond 90\%. The white ellipse shows the $2\sigma$ contour of the clockwise system as determined by \citet{Paumard2006}. The mean angular distance between the angular momenta of the six stars and the disk angular momentum direction is $<\psi>=14.5^{\circ}$ and the mean of the angular distances squared is $<\psi^2>^{1/2}=15.4^{\circ}$. The distribution of angular offsets can be fit with a two-dimensional Gaussian (see equation \ref{eq:inclin_2D}) with $\sigma_{\psi}=(11\pm2)^{\circ}$.

In a next step we want to determine the angular offsets of all WR/O stars from the local angular momentum direction of the clockwise system. We interpolated the points shown in figure \ref{fig:disk_position_vs_distance_sphere} (left) with a third degree spline function in projected distance to obtain the direction of the clockwise system as a function of average projected distance. 
Figure \ref{fig:h_angle_clock} shows the distribution of the reconstructed angular difference $\psi$ for all the 90 WR/O stars from the local average angular momentum direction of the clockwise system (blue histogram). The distribution has a peak at small inclinations, which is well described by the expected distribution for a disk with a $10^{\circ}$ two-dimensional Gaussian sigma thickness and 46 disk members. In addition, there is a long tail to large angular offsets from the clockwise system. 11 out of the 61 clockwise stars have inclinations larger than $30^{\circ}$ from the angular momentum direction of the clockwise system.
The clustering between $80^{\circ}$ and $100^{\circ}$ inclinations is due to the counter-clockwise rotating stars.

Within errors the inclinations of the stars to the clockwise system are independent of distance to Sgr~A*.

\subsubsection{Surface Density}

\begin{figure}[t!]
\begin{center}
\includegraphics[totalheight=7cm]{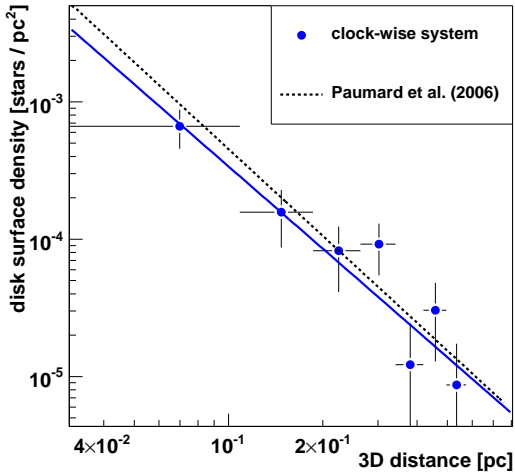}
\caption{\it \small Surface number density of the 30 candidate stars in the clockwise system (blue filled circles) as a function of the three-dimensional distance from Sgr~A* , which have a minimum angular distance below $10^{\circ}$ from the (local) average angular momentum direction of the clockwise system. The dashed line shows the $r^{-2.1}$ power-law of \citet{Paumard2006}. The full blue line shows the best fit power-law to the cock-wise disk: $\Sigma(r_{\mathrm{disk}}) \propto r^{-1.95\pm0.25}$.}
\label{fig:surface_density}
\end{center}
\end{figure}

Both the calculation of the surface density of the clockwise system as well as the calculation of the stellar orbital elements requires the determination of the unknown $z$-coordinate. 
In order to obtain the least biased results, we apply a hard criterion for the choice of candidate stars for the clockwise system. We only consider the 30 of the 61 clockwise moving WR/O stars which have angular distances below $10^{\circ}$ from the local angular momentum direction of the clockwise system. The fraction of stars which have angular distances below $10^{\circ}$ from the local angular momentum direction of the clockwise system of all clockwise moving stars is independent of distance to Sgr~A* within errors.
For each star we generate 1000 $z$ positions  taking into account the $\chi^2$ probability for the stellar angular momentum direction (see e.g. figure \ref{fig:IRS16CC_chi2}). Moreover we apply the prior that the local inclinations to the clockwise system follow a two-dimensional Gaussian distribution with a $\sigma_{\psi} = 10^{\circ}$.

We compute the surface number density of observed candidate stars of the clockwise system as: $\Sigma(r_{\mathrm{disk}}) = \Delta N_{\mathrm{stars}}(r_{\mathrm{disk}}) /  \Delta A_{\mathrm{disk}}(r_{\mathrm{disk}})$ with $\Delta A_{\mathrm{disk}}(r_{\mathrm{disk}}) = 2\pi r_{\mathrm{disk}} \Delta r_{\mathrm{disk}}$, where $r_{\mathrm{disk}}$ is the distance to Sgr~A* measured in the local angular momentum direction of the clockwise system. Figure \ref{fig:surface_density} shows the surface number density of candidate stars in the clockwise system as a function of the three-dimensional distance from Sgr~A* in the local angular momentum direction of the clockwise system.
The best fit power-law to the clockwise system is $\Sigma(r_{\mathrm{disk}}) \propto r_{\mathrm{disk}}^{-1.95\pm0.25}$. We estimate the systematic error of the power-law slope due to uncertainties in $R_0$ and $M_{\mathrm{Sgr~A*}}$ to $0.2$.
This result is compatible with the value found by \citet{Paumard2006}, $\Sigma(r_{\mathrm{disk}}) \propto r_{\mathrm{disk}}^{-2.1\pm0.2}$, and \citet{Lu2008}, $\Sigma(r_{\mathrm{disk}}) \propto r_{\mathrm{disk}}^{-2.3\pm0.66}$. Both groups did not take into account the change of the angular momentum direction with distance from Sgr~A*.

\subsubsection{Reconstruction of Orbital Elements}

\begin{figure}[t!]
\begin{center}
\includegraphics[totalheight=13cm]{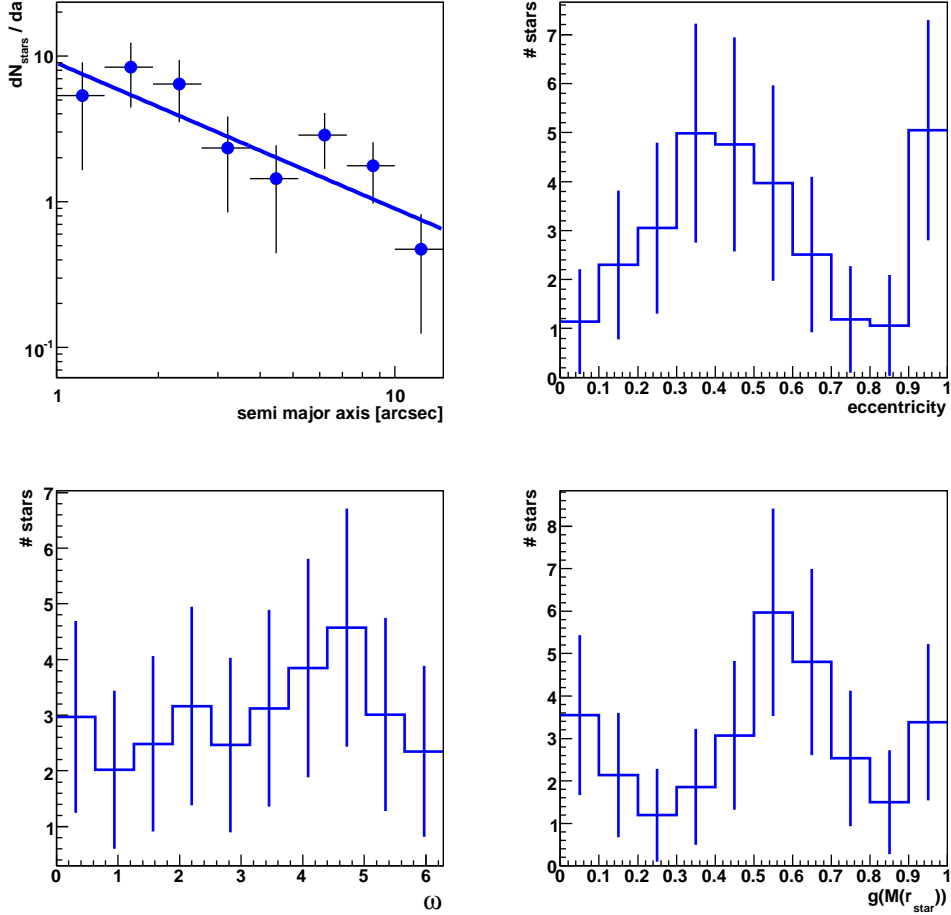}
\caption{\it \small Distributions of the orbital elements for the 30 clockwise moving WR/O stars in the radial bin 0.8''--12'', which have a minimum angular distance below $10^{\circ}$ from the (local) average angular momentum direction of the clockwise system.
(Upper left) $\mathrm{d}N_{\mathrm{stars}}/\mathrm{d}a$, the full line shows an $a^{-1}$ power law, (upper right) $\mathrm{d}N_{\mathrm{stars}}/\mathrm{d}\epsilon$, (lower left) $\mathrm{d}N_{\mathrm{stars}}/\mathrm{d}\omega$, (lower right) $\mathrm{d}N_{\mathrm{stars}}/\mathrm{d}g(M(r_{\mathrm{star}})$, $g=2\tau$ for $v_z>0$ and $g = 2(1-\tau)$ for $v_z<0$, see \citet{Beloborodov2004}.
}
\label{fig:elements_disk}
\end{center}
\end{figure}

\begin{figure}[t!]
\begin{center}
\includegraphics[totalheight=7cm]{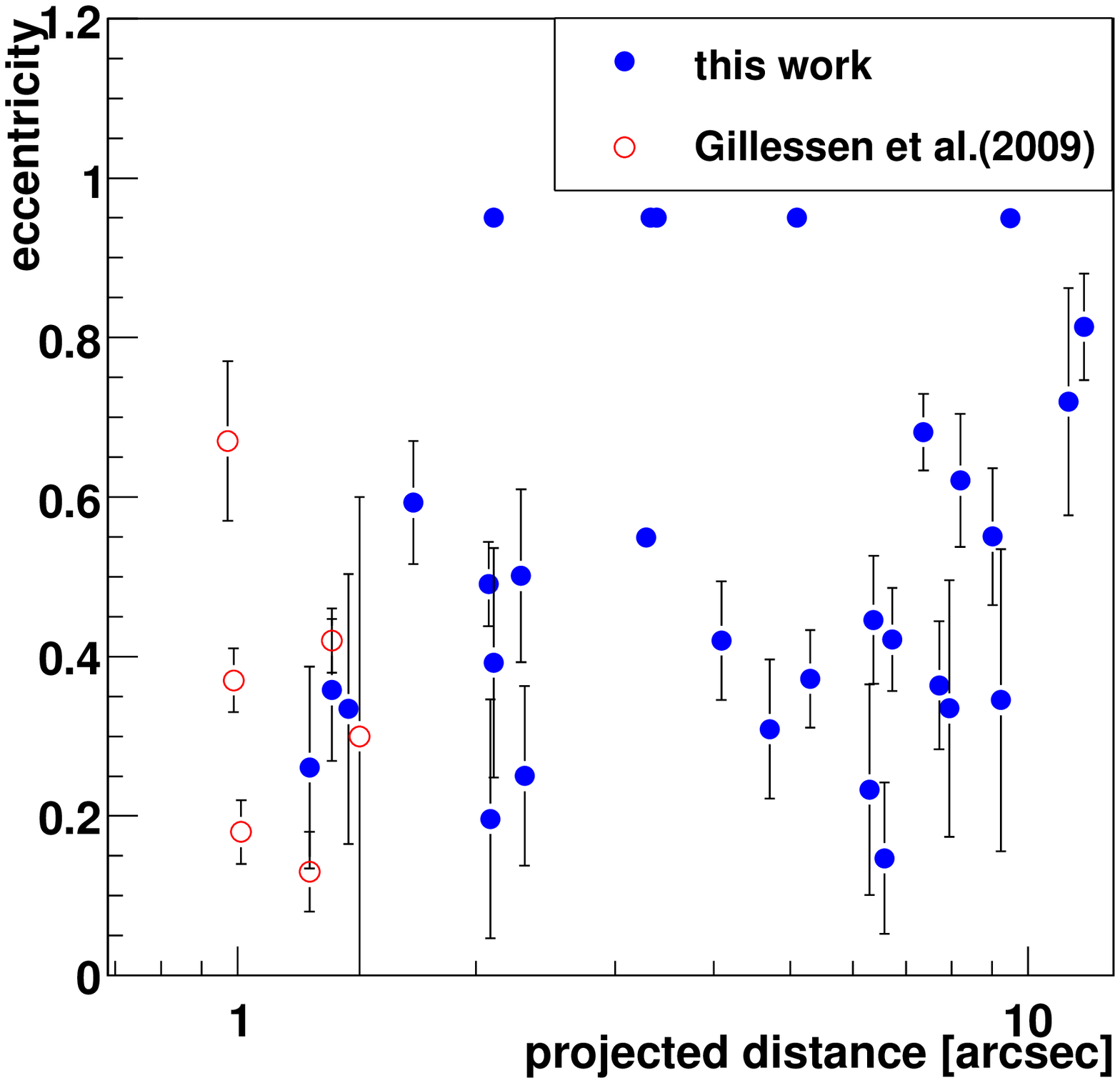}
\caption{\it \small Reconstructed eccentricity as a function of projected distance for the 30 clockwise moving WR/O stars (blue points), which have a minimum angular distance below $10^{\circ}$ from the (local) average angular momentum direction of the clockwise system. Red circles show the six early-type stars (S66, S67, S83, S87, S96 and S97) with $0.8''\leq R\leq1.4''$ for which \citet{Gillessen2008} were able to derive individual orbital solutions.
Error bars denote the RMS of the reconstructed eccentricities.}
\label{fig:eccentricity_vs_p}
\end{center}
\end{figure}

The analysis above was, to lowest order, independent of the enclosed mass. The potential only influenced the range of the generated $z$-values due to the assumption of bound orbits, see section \ref{sec:z_generation}. 
On the contrary, the determination of stellar orbital elements depends on the enclosed mass of the orbit, which is given by the sum of the mass of Sgr~A* plus the mass of the cluster of late-type stars. We used the following parameterization of the enclosed stellar mass $M(r_{\mathrm{star}})$ as a function of the 3D distance $r_{\mathrm{star}}$ to Sgr~A* \citep{Trippe2008}:
%
\begin{equation}
M(r_{\mathrm{star}}) = 4\times10^6M_{\odot} + \int_0^{r_{\mathrm{star}}}{\frac{2.1\times10^6M_{\odot}\mathrm{pc}^{-3}}{1 + (r/8.9 ``)^2} 4\pi r^2 \mathrm{d}r} \ .
\end{equation}
The presence of an extended mass induces retrograde pericenter shifts which result in open rosetta shaped orbits \citep{Alexander2005}. We approximate such an orbit locally by a Kepler ellipse and use the enclosed mass in a sphere given by the instantaneous position of the star to compute the orbital elements.

Figure \ref{fig:elements_disk} 
shows the distributions of the orbital elements $a,\epsilon,\omega,\tau$ for the 30 candidate stars of the clockwise system.
The distribution of semi major axis $a$ (upper left panel of figure \ref{fig:elements_disk}) is compatible with an $a^{-1}$ power law.
The distribution of the reconstructed eccentricities (upper right panel of figure \ref{fig:elements_disk}) has a mean of 0.51 and an RMS of 0.27. 
The eccentricity distribution shows a two-peak structure: a broad distribution between 0 and 0.8 and five stars between 0.9 and 1. Figure \ref{fig:eccentricity_vs_p} shows the eccentricity of the 30 candidate stars as a function of projected distance to Sgr~A*. The five stars with high eccentricities have very small RMS values for their reconstructed eccentricities. They could only be disk members in a very small region of their parameter space, limited to nearly radial orbits. This fine-tuning may be a sign that these five stars do not dynamically belong to the clockwise system, but rather are by chance compatible with being candidates. Still, we cannot exclude them at this point. Without these five stars, the mean eccentricity is $0.42\pm0.05$ and the RMS is $0.20\pm0.03$. We note that the eccentricity is a purely positive quantity. For bound orbits it is less than one. Therefore the eccentricities do not follow a Gaussian distribution. We use the observed mean and RMS eccentricities to quantify the first two moments of the a priory unknown eccentricity distribution.
Projection effects and the prior of a clockwise system with a Gaussian distribution of inclinations bias the distributions of reconstructed eccentricities to larger values. We have applied our reconstruction method to MC simulated stars in a disk with a $10^{\circ}$ two-dimensional Gaussian sigma thickness with a flat eccentricity distribution (mean eccentricity of 0.5 and RMS of 0.28). The reconstructed eccentricity distribution has a mean of 0.55 and an RMS of 0.26. We estimate the bias of the average reconstructed eccentricity to 0.05 and in addition quote a systematic error of 0.05. 
We estimated the systematic errors due to uncertainties in $R_0$ and $M_{\mathrm{Sgr~A*}}$ to be as small as 0.02.
The mean eccentricity of the candidate stars of the clockwise system is $0.37 \pm 0.05_{\mathrm{stat.}} \pm 0.05_{\mathrm{syst.}}$.

Moreover, figure \ref{fig:eccentricity_vs_p} includes the six early-type stars (S66, S67, S83, S87, S96 and S97) with $0.8''\leq R\leq1.4''$ for which \citet{Gillessen2008} were able to derive individual orbital solutions (see figure \ref{fig:sky_zoom_Sstars}).
Of these six so-called S-stars only two (S87 and S96) fulfill our strict selection criteria for disk candidate stars. The reconstructed eccentricities of S87 and S96 in this work agree within errors with the values of \citet{Gillessen2008}.
S67, S83, and S97 have inclinations of more than $10^{\circ}$ to the disk angular momentum direction and S66 is a B dwarf ($m_K = 14.8$). The six S-stars have a mean eccentricity of 0.36 and an RMS of 0.20, giving an error of the mean of 0.09.
In contrast the so-called S-stars have a different eccentricity distribution: $\mathrm{d}N_{\mathrm{stars}}/\mathrm{d}\epsilon \propto \epsilon^{2.6\pm0.9}$ \citep{Gillessen2008}.
We combine our result of the mean eccentricity with the mean eccentricity of the six stars from \citet{Gillessen2008} to a weighted average of $0.36 \pm 0.06$.


Figure \ref{fig:elements_disk}(lower left panel) shows the distribution of the reconstructed arguments of periapsis of the orbits $\omega$. For uniformly populated disks and a uniform azimuthal exposure the distribution of $\omega$ is expected to be flat. The observed distribution is compatible with a flat distribution within errors.

Figure \ref{fig:elements_disk}(lower right panel) shows the distribution of the time separating the stars from their nearest passage of their pericenter normalized to their half-periods $g(M(r_{\mathrm{star}}))$. For the true enclosed mass and a random snapshot time, the expected $g$ obeys Poisson statistics: it has a flat probability distribution between 0 and 1 with the mean expectation value $\overline{g}=0.5$ and the standard deviation $\Delta g= 12^{-1/2}\approx0.29$. $\overline{g}\rightarrow 0$ for small assumed masses and $\overline{g}\rightarrow1$ for too large assumed masses \citep{Beloborodov2004,Beloborodov2006}. In our data we determine $\overline{g}(M(r_{\mathrm{star}}))=0.52\pm0.05$ and $\Delta g(M(r_{\mathrm{star}}))= 0.27\pm0.04$. In case we adopt $R_0=7.5$~kpc and $M_{\mathrm{Sgr~A*}}=3.5\times10^6M_{\odot}$ we get $\overline{g}=0.48\pm0.05$ and for the parameters $R_0=8.5$~kpc and $M_{\mathrm{Sgr~A*}}=4.5\times10^6M_{\odot}$ we get $\overline{g}=0.53\pm0.05$.

\subsection{Hertzsprung-Russell Diagram}
\label{sec:HR}

\begin{figure}[t!]
\begin{center}
\includegraphics[totalheight=13cm]{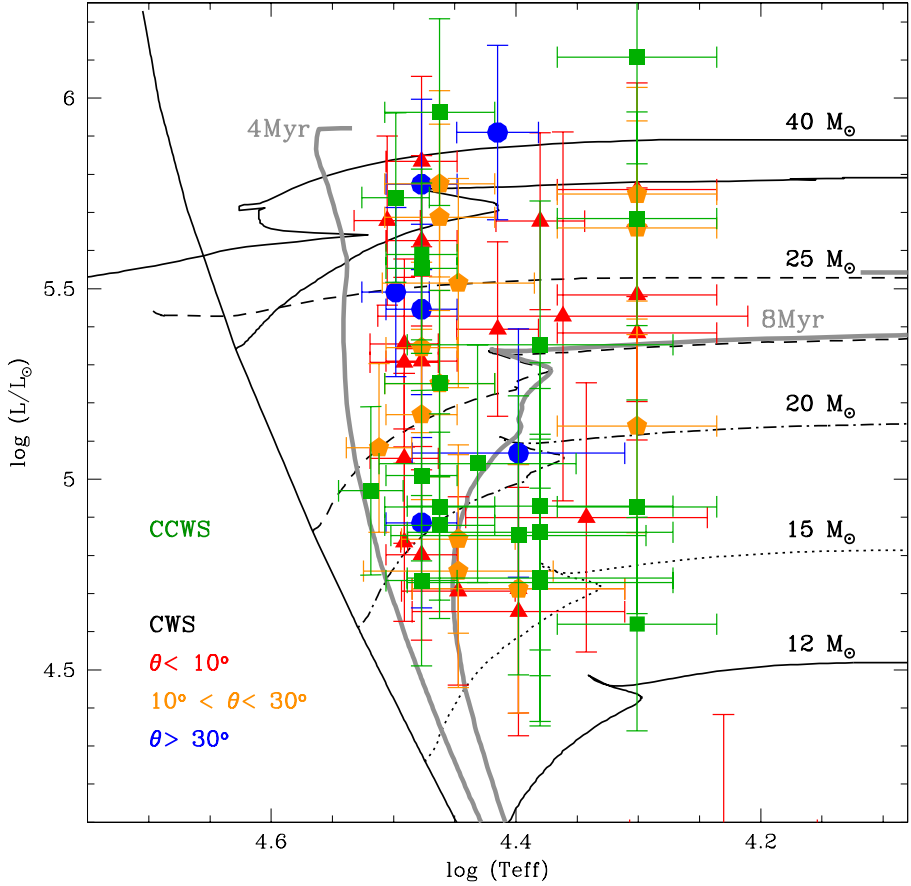}
\caption{\it \small Location of the O-type super giants, giants and dwarfs and B super giants in a Hertzsprung-Russel diagram. Red triangles denote the candidate stars with a maximum angular offset of $10^{\circ}$ from the local angular momentum direction of the clockwise system, the orange pentagons show the clockwise stars with off sets between $10^{\circ}$ and $30^{\circ}$ and the blue points show the clockwise rotating stars which have a minimum offset of $30^{\circ}$ from the clockwise system. The green squares represent the counter-clockwise stars.
 The numbers indicate the initial mass of the stars on the tracks starting from the zero age main sequence (solid black line). The two thick gray lines show two isochrones for 4 and 8 Myr. The tracks (various dotted/dashed lines) are based on Geneva models with rotation \citep{Meynet2005}.}
\label{fig:HR_diagram}
\end{center}
\end{figure}

Figure \ref{fig:HR_diagram} shows the location of the O-type super giants, giants and dwarfs and B super giants in a Hertzsprung-Russell (HR) diagram. Such a diagram can be used to estimate the age of the population by comparing of the stellar positions to theoretical isochrones. Such isochrones are displayed as solid grey lines. 
Our HR diagram excludes Wolf-Rayet stars since isochrones are ill-defined at the position occupied by these stars. For each star we assigned effective temperatures derived from spectral types using the effective temperature scales of \citet{Martins2005} for O stars and \citet{Crowther2006} for B supergiants. 
We estimated the luminosities from absolute K--band magnitudes and the bolometric corrections of \citet{Martins2006}. We assumed 
a distance of $(8.0\pm0.5)$~kpc and 
an extinction of $A_{\rm K}=2.8\pm0.5$ \citep{Scoville2003,Schoedel2007}. 
The errors in $T_{\mathrm{eff}}$ correspond
to uncertainties in spectral types. These errors propagate into the bolometric corrections, see equation~3 and table~1 of \citet{Martins2006}, 
and lead to the uncertainty in luminosities.
In figure \ref{fig:HR_diagram}, we differentiated four subgroups of stars: the candidate stars of the clockwise system with a maximum angular offset of $10^{\circ}$ from its local angular momentum direction, the clockwise rotating stars which have offsets between $10^{\circ}$ and $30^{\circ}$ and the one with offsets larger than $30^{\circ}$ from the clockwise system, as well as the counter-clockwise stars. 
Stars with $\log L/L_{\odot} < 5.0$ and $\log T_{\mathrm{eff}} < 4.45$ have a significant probability to be younger than 8 Myr. Indeed, these stars have the same probability to have  $\log T_{\mathrm{eff}}=30000$~K (corresponding to an age of 4-8 Myr) or $\log T_{\mathrm{eff}}=20000$~K (corresponding to older ages). This is due to the absence of  $T_{\mathrm{eff}}$ indicator in the K-band spectra of stars with $20000 \mathrm{K} < T_{\mathrm{eff}} < 30000$~K. Besides, the small number of red supergiants argues against an age much larger than 10 Myr \citep{Paumard2006,Davies2008}.

The main conclusion is that within errors these four groups of stars are coeval with an age between 4 and 8 Myrs, comparable to the earlier studies by \citet{Paumard2006}. There is no variation of average stellar age with projected distance to Sgr~A*.

\subsection{The Counter-Clockwise System}
\label{sec:counter_significance}

\begin{figure}[t!]
\begin{center}
\includegraphics[totalheight=7cm]{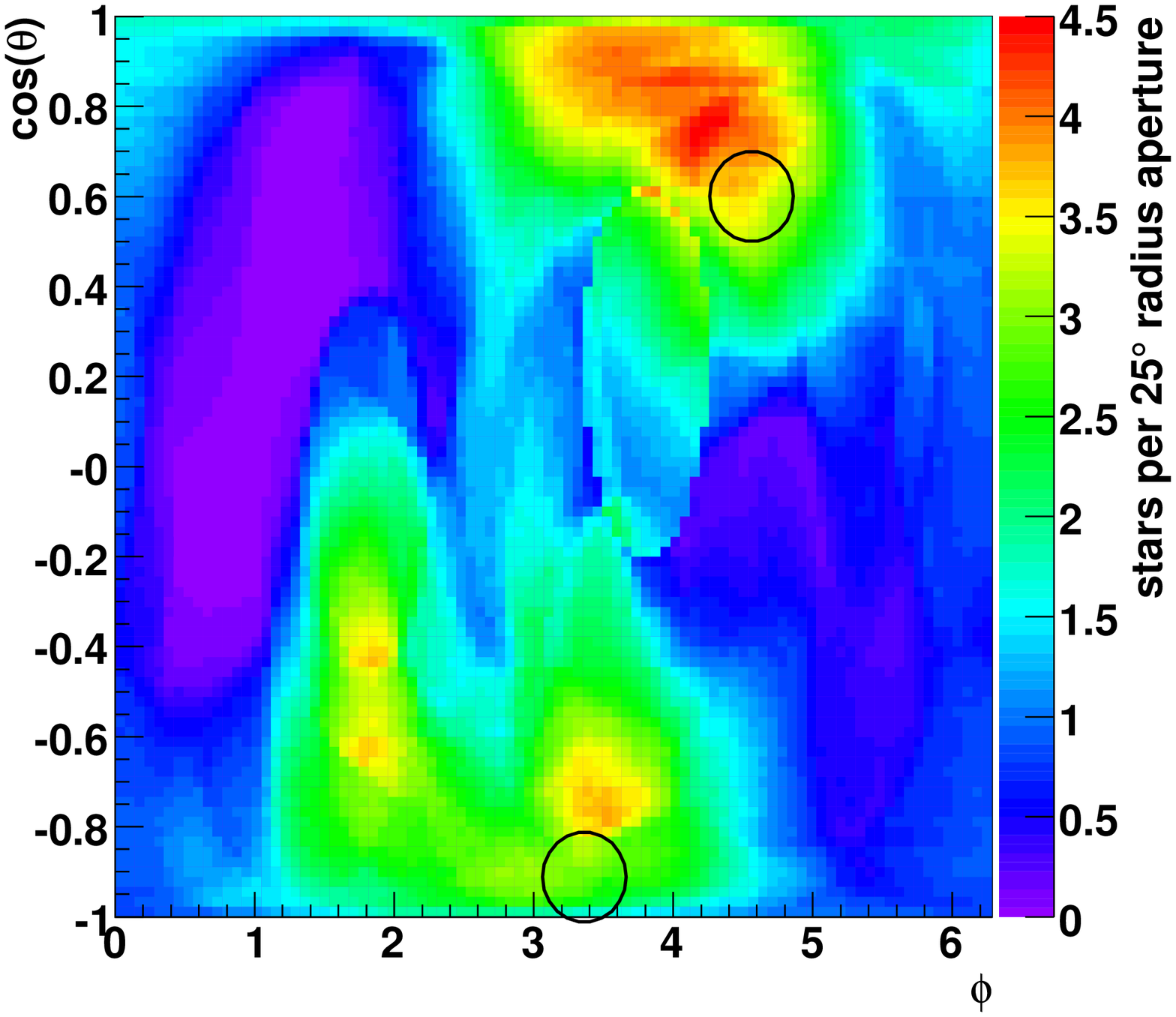}
\includegraphics[totalheight=7cm]{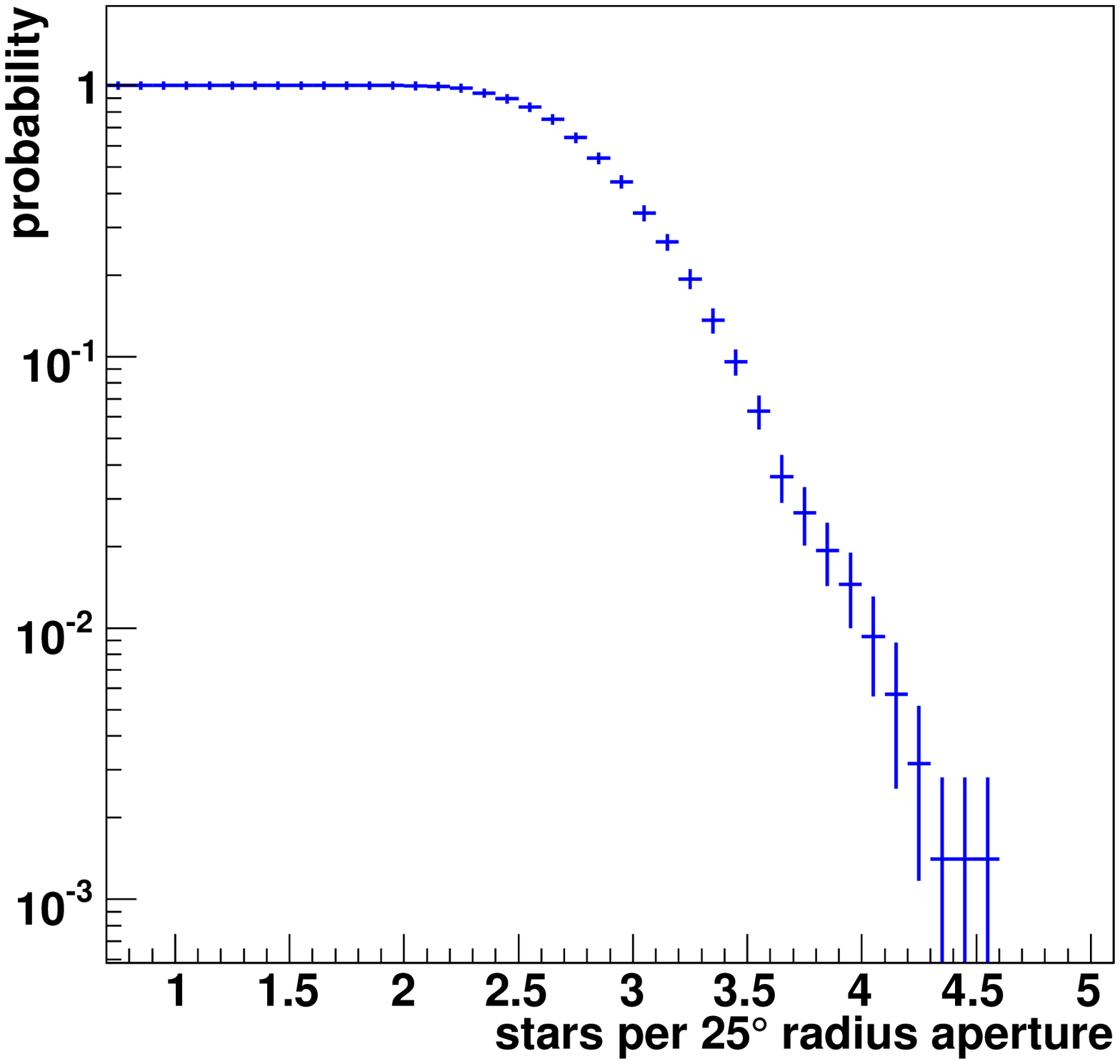}
\caption{\it \small (Left) cylindrical equal area projection of the density of reconstructed angular momenta in a fixed aperture of $25^{\circ}$ radius for 30 WR/O stars in the interval of projected distances to the Galactic Center 3.5''--7''. There is an extended U-shaped excess of counter-clockwise orbits with a global maximum angular momentum density of 3.8 stars per $25^{\circ}$ radius aperture at $(\phi,\theta)=(200^{\circ},142^{\circ})$, near the position of the counter-clockwise system of \citet{Paumard2006}.
The disk positions \citep{Paumard2006} are marked with black circles. 
(Right) Probability that an isotropic distribution of stars yields in at least one aperture of $25^{\circ}$ radius a certain density of reconstructed angular momenta. A maximum angular momentum density of 3.8 stars per $25^{\circ}$ radius aperture corresponds to a probability of 2\%.
}
\label{fig:sky_sig_counter_disk}
\end{center}
\end{figure}

\begin{figure}[t!]
\begin{center}
\includegraphics[totalheight=7cm]{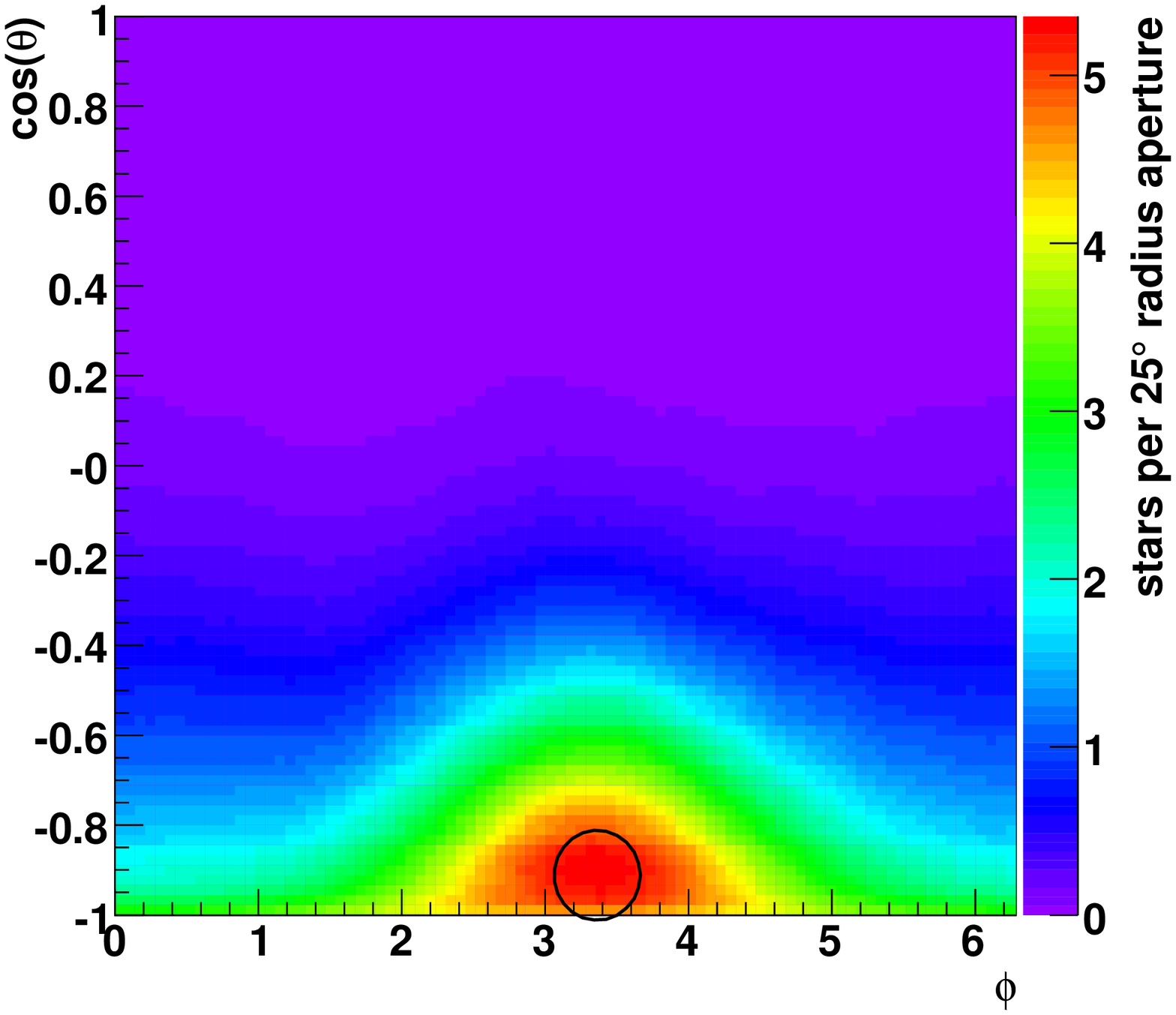}
\includegraphics[totalheight=7cm]{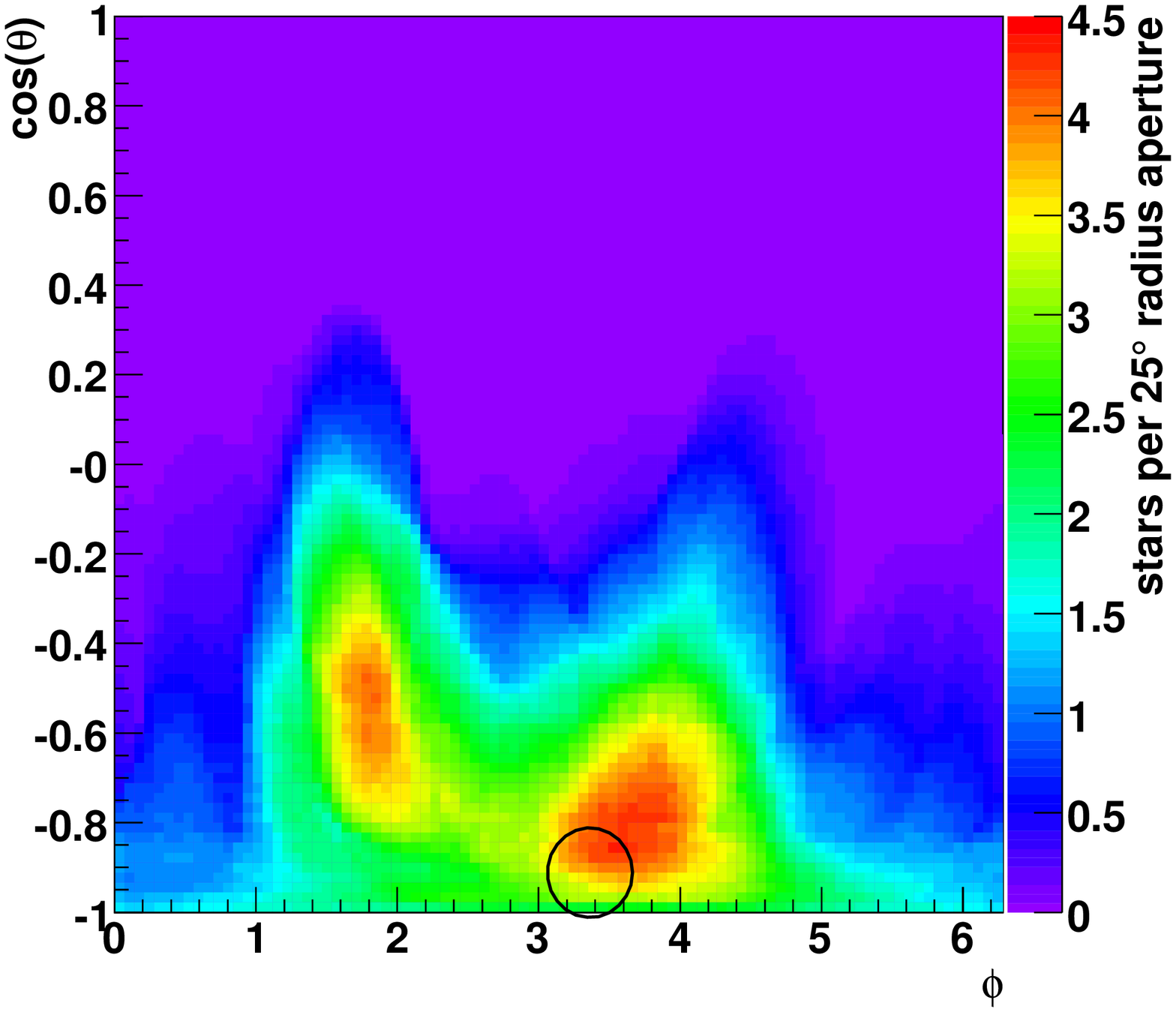}
\caption{\it \small  Cylindrical equal area projections of the density of reconstructed angular momenta in an aperture of $25^{\circ}$ radius (left) average for a MC simulated disk of many stars, scaled to 15 stars. The disk angular momentum points in the direction of the counter-clockwise disk of \citet{Paumard2006}. (right) 15 MC simulated stars with the same azimuthal distribution as the observed stars. All 15 stars are members of a disk in the direction of the counter-clockwise disk of \citet{Paumard2006}.}
\label{fig:counter_disk_simul}
\end{center}
\end{figure}

\begin{figure}[t!]
\begin{center}
\includegraphics[totalheight=7cm]{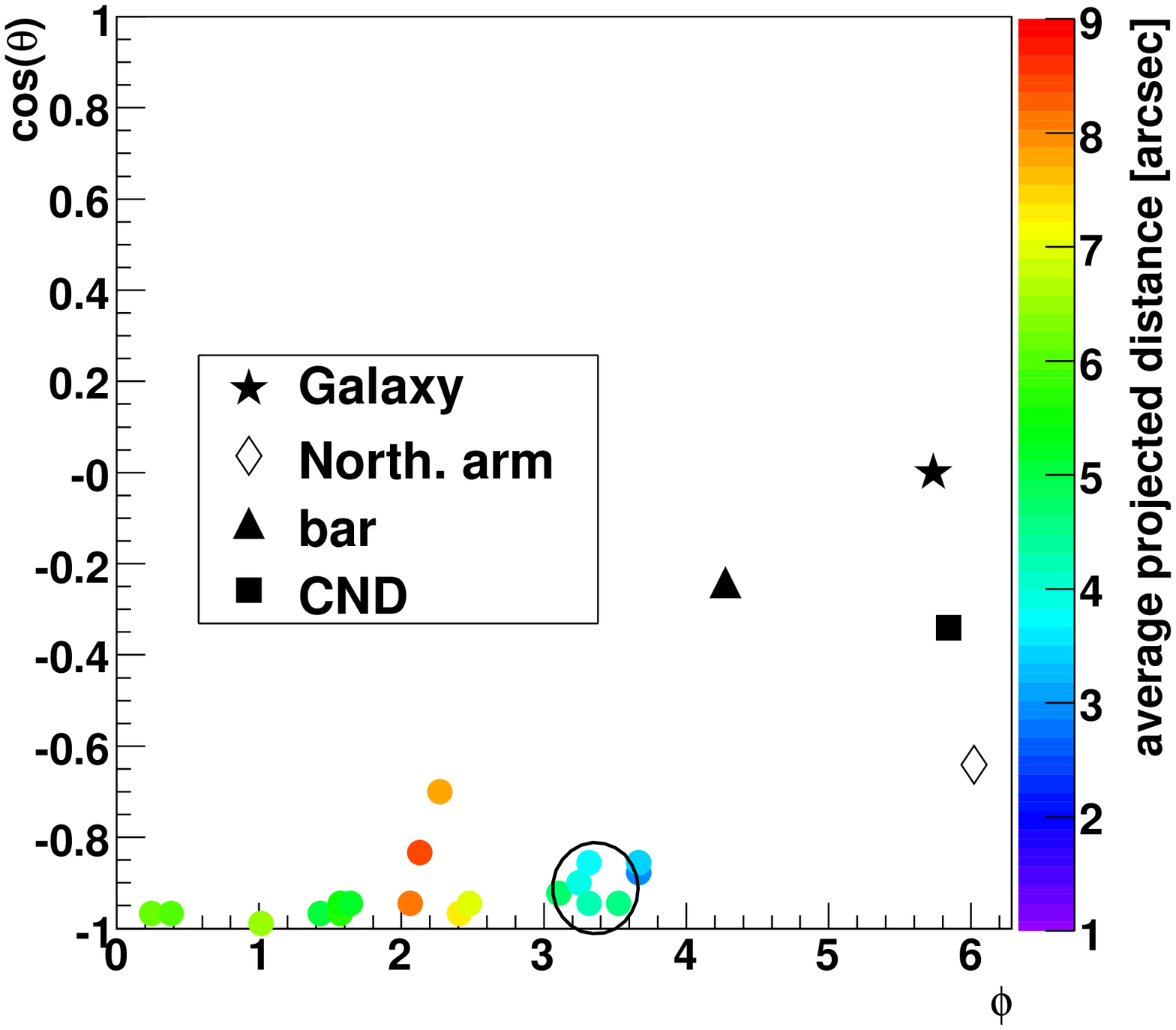}
\includegraphics[totalheight=6.5cm]{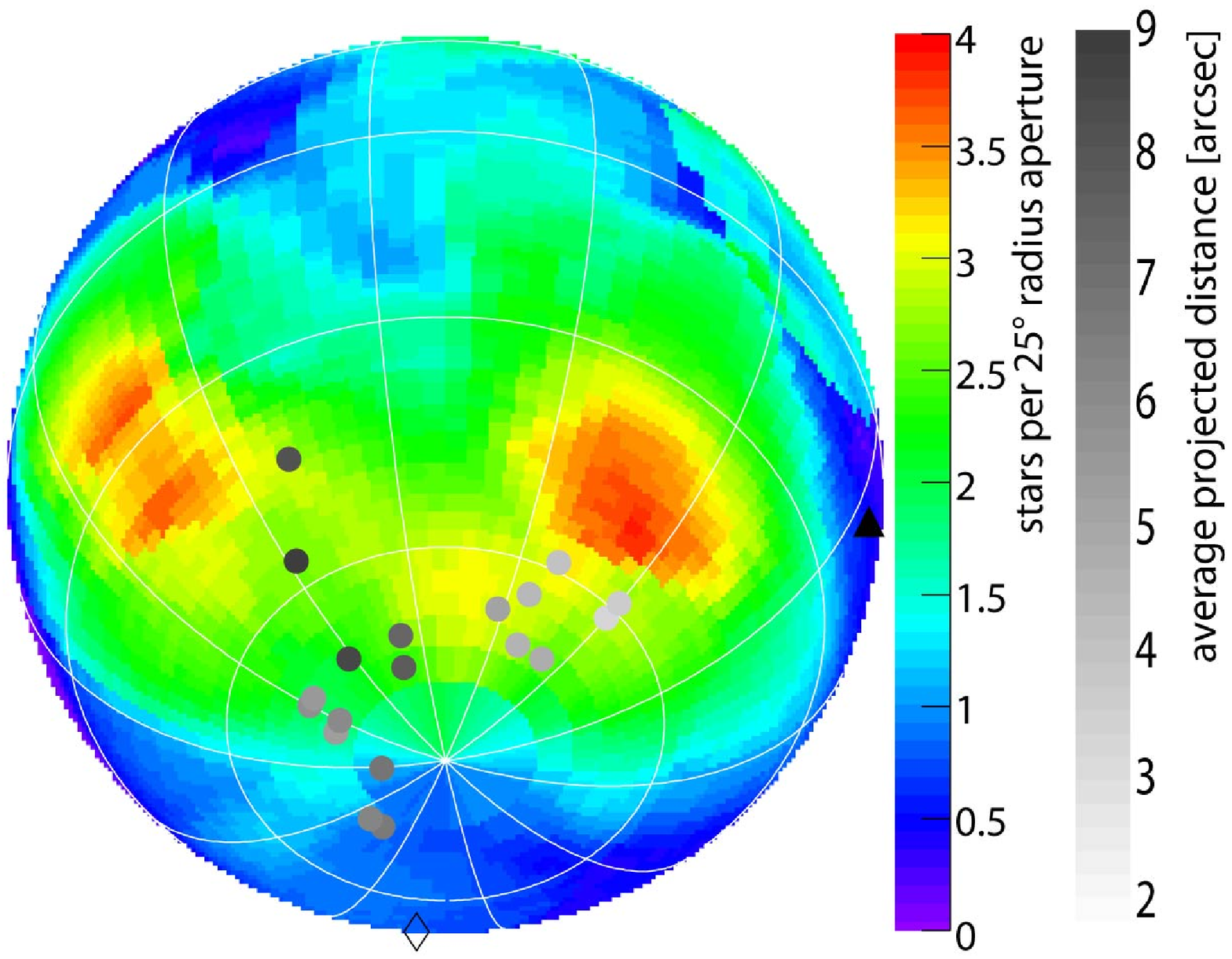}
\caption{\it \small 
(Left) Cylindrical equal area projection of the local average stellar angular momentum direction for the counter-clockwise stars as a function of the average projected distance. The points are correlated, see text. The average angular momenta for the innermost 16 stars (corresponding to the 7 points with smallest average projected distance) agree well with the orientations of the counter-clockwise disk of \citet{Paumard2006}, shown by the black circle. The asterisk shows the Galactic pole \citep{Reid2004}, the diamond indicates the normal vector to the northern arm of the minispiral \citep{Paumard2004}, the triangle indicates the normal vector to the bar of the minispiral \citep{Liszt2003}, the square shows the rotation axis of the CND \citep{Jackson1993}.
(Right) Orthographic projection of the density of reconstructed angular momenta in an aperture of $25^{\circ}$ radius for the 30 WR/O stars in the interval of projected distances 3.5''--7'' (see figure \ref{fig:sky_sig_counter_disk}), seen from $\phi_0=160^{\circ}$, $\theta_0=142^{\circ}$ to highlight the counter-clockwise stars. The gray points show the average stellar angular momentum direction for all counter-clockwise stars as a function of the average projected distance.
}
\label{fig:counter_disk_position_vs_distance_sphere}
\end{center}
\end{figure}

\begin{figure}[t!]
\begin{center}
\includegraphics[totalheight=7cm]{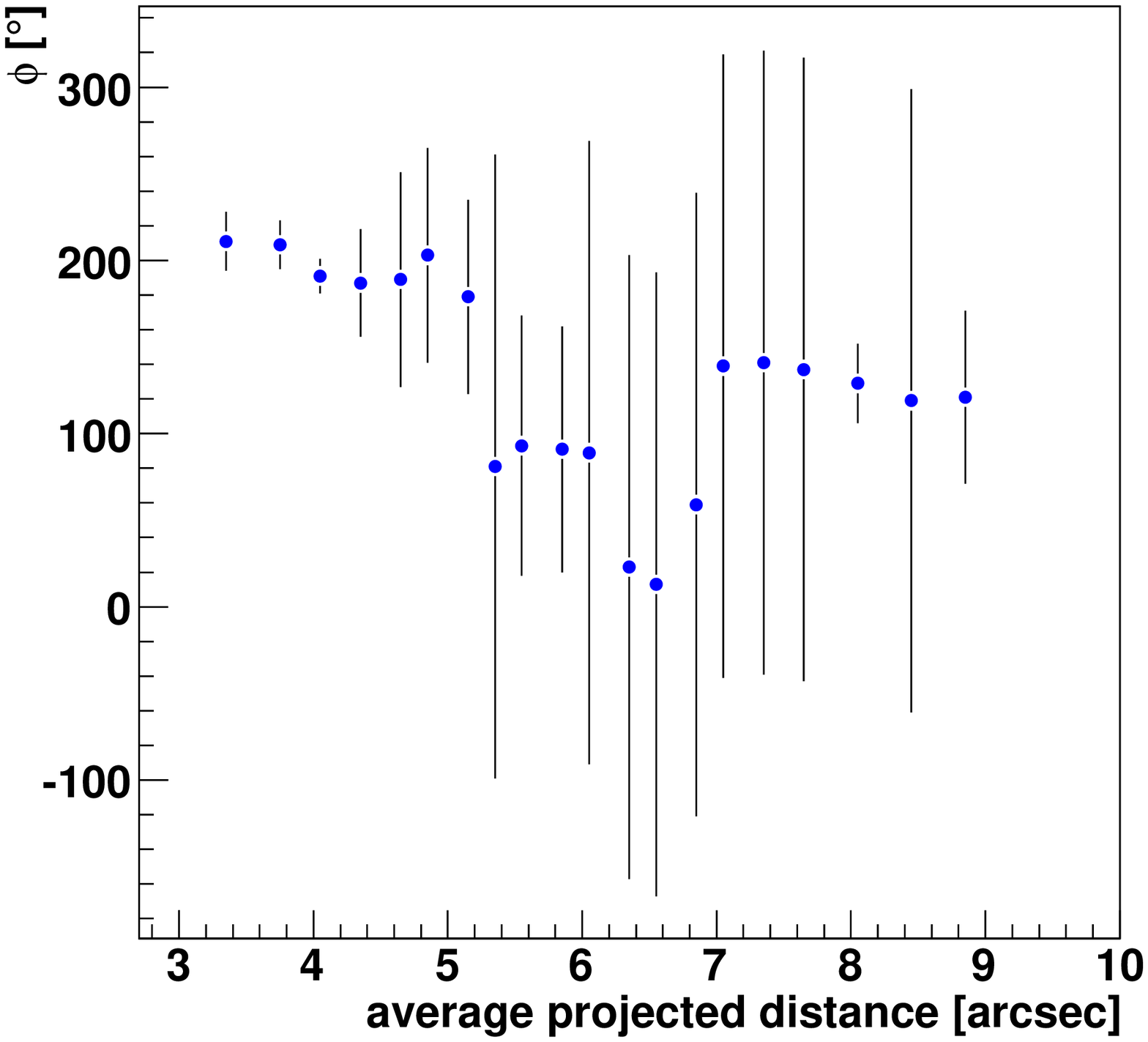}
\includegraphics[totalheight=7cm]{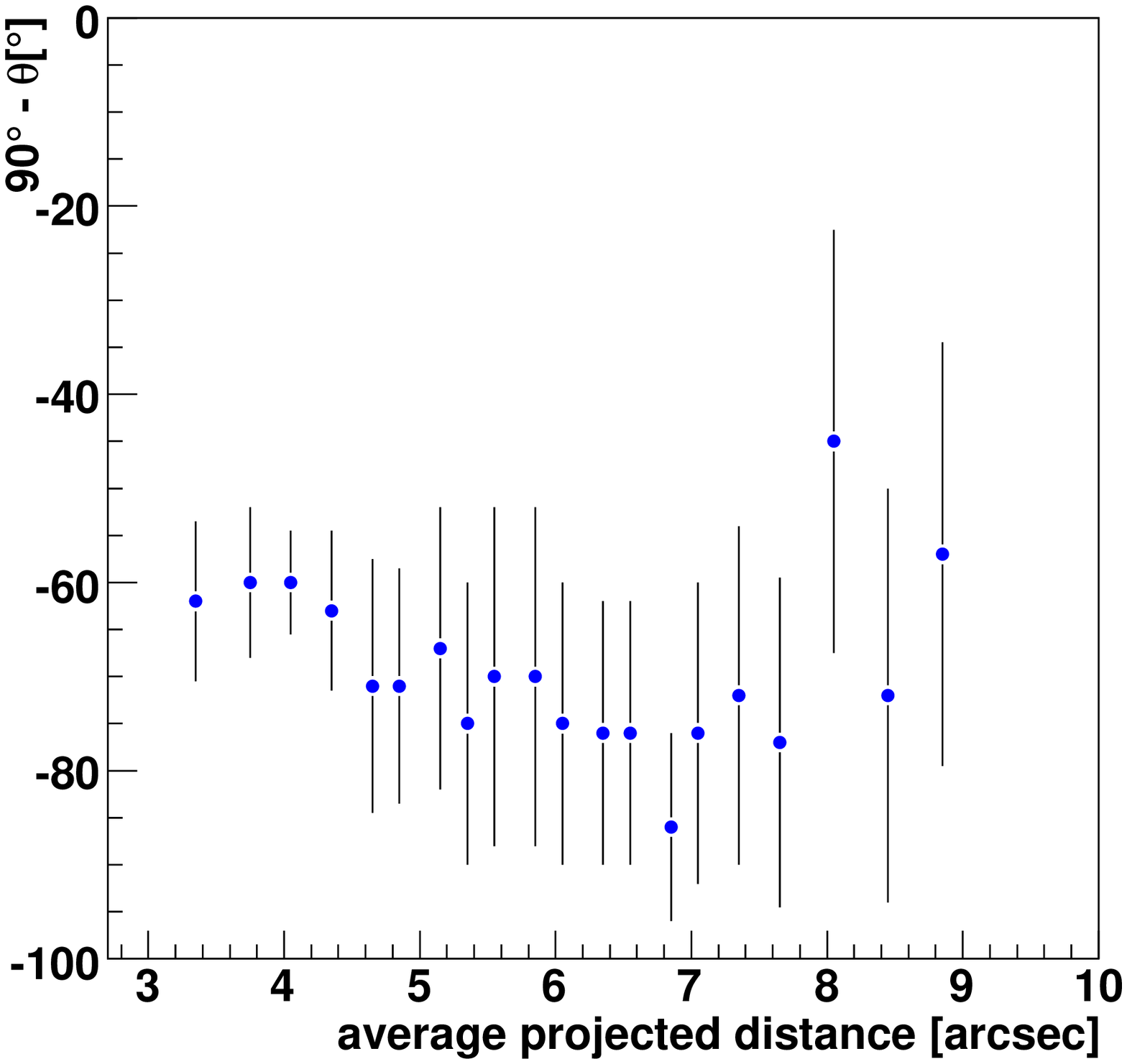}
\caption{\it \small 
Local average stellar angular momentum direction for the counter-clockwise stars as a function of the average projected distance: (Left) $\phi_J = \phi_J(R)$ (Right) $\theta_J = \theta_J(R)$. The 20 points are correlated. They correspond to the 20 positions of a window of width 10 stars which is slid over the 29 clockwise moving WR/O stars ordered by their projected distances to Sgr A*
}
\label{fig:counter_disk_l_vs_p}
\end{center}
\end{figure}

The existence of a counter-clockwise disk of early-type stars has been called into question by \citet{Lu2006,Lu2008}. 
It is undisputed that about 30\% of all WR/O stars (29/90) are on counter-clockwise orbits.
So the question is not whether such stars exist but whether their kinematic distribution is indicative of a second, well defined disk structure. The upper right panel (interval of projected distances 3.5''--7'') of  figure \ref{fig:sky_sig_disk} shows an extended U-shaped excess in the distribution of counter-clockwise stellar angular momenta. 
We calculated the significance for a sky density of angular momenta averaged over an aperture of $15^{\circ}$, chosen to maximize the significance of a moderately extended disk of $10^{\circ}$ two-dimensional sigma extension.
Figure \ref{fig:sky_sig_counter_disk} (left panel) shows for the same stars as in the upper right panel (interval of projected distances 3.5''--7'') of  figure \ref{fig:sky_sig_disk} the density of reconstructed angular momenta in a fixed aperture of $25^{\circ}$ radius, better suited for more extended features. There is an extended U-shaped excess of counter-clockwise orbits with a global maximum angular momentum density of 3.8 stars per $25^{\circ}$ radius aperture at $(\phi,\theta)=(200^{\circ},142^{\circ})$, near the position of the counter-clockwise system of \citet{Paumard2006}. In case of 15 isotropic distributed counter-clockwise stars we expect a mean density of 1.5 stars and an RMS of 0.52 stars per $25^{\circ}$ radius aperture centered at $(\phi,\theta)=(200^{\circ},142^{\circ})$, corresponding to an excess of 4.5 times the RMS.
In figure \ref{fig:sky_sig_counter_disk} (left panel) we did not use any additional $z$-position prior for the IRS 13E stars. If IRS 13E is a star cluster at $z\sim7"$ \citep{Paumard2004,Paumard2006}, the stars IRS 13E2 and E4 would not be candidate members of a counter-clockwise system. In this case, the maximum angular momentum density of counter-clockwise orbits would slightly be reduced to 3.7 stars per $25^{\circ}$ radius aperture at the same position as given above.

There are 15 stars with counter-clockwise orbits located at projected 
distances between 3.5" and 7". The location of the excess is not known a priori. What is the probability to find such an excess in an isotropic distribution of stars when searching all possible aperture directions?
We computed angular momentum density maps ($25^{\circ}$ radius aperture) for many groups of 15 simulated isotropic stars on counter-clockwise orbits. Figure \ref{fig:sky_sig_counter_disk} (right panel) shows the probability that an isotropic distribution of stars yields, in at least one aperture of $25^{\circ}$ radius, a certain density of reconstructed angular momenta. A maximum angular momentum density of 3.8 stars per $25^{\circ}$ radius aperture corresponds to a probability of 2\%. Taking the whole observed U-shaped excess structure into account the probability is even lower.
Hence we conservatively exclude at the 98\% confidence level that the observed excess of counter-clockwise stars is due to a fluctuation of isotropic stars.

In addition to the global maximum angular momentum density, the significance sky distribution (left panel of figure \ref{fig:sky_sig_counter_disk}) shows an extended U-like shape with local maxima up to 3.7 stars per $25^{\circ}$ radius aperture. This U-like shape may be due to

\begin{itemize}
\item{projection effects in the reconstruction of nearly face-on orbits ($\cos{\theta}\sim -1$)}
\item{a small number of stars with a possible non-isotropic distribution on the sky} 
\item{an intrinsic property of the counter-clockwise structure of WR/O stars like a large thickness, a warp, a state of dissolution or the presence of separate streamers etc.}
\end{itemize}

Figure \ref{fig:counter_disk_simul}(left) shows the  average density of reconstructed angular momenta in an aperture of $25^{\circ}$ radius for a MC simulated disk ($10^{\circ}$ thickness) of many stars, scaled to 15 stars.
The disk angular momentum points in the direction of the counter-clockwise disk of \citet{Paumard2006}.
 Although the simulated disk is nearly face-on ($\cos{\theta}\sim -1$) the angular momentum direction of the disk is not biased by the reconstruction procedure.
Figure \ref{fig:counter_disk_simul}(right) shows one representation of 15 MC simulated stars with the same azimuthal distribution as the observed counter-clockwise stars. All 15 stars are members of a disk ($10^{\circ}$ thickness) in the direction of the counter-clockwise disk of \citet{Paumard2006}. The excess has the same U-like structure as observed in the data with similar values for the angular momentum density.
Hence the combination of a small number of observed stars, which are members of a $10^{\circ}$ thick disk, at their observed positions on the sky can --- but does not necessarily has to --- produce the U-like shape of reconstructed angular momenta.


In an analogous study to the clockwise system, we  investigated the average angular momentum of the counter-clockwise stars as a function of projected distance, keeping in mind that  the small number of stars limits the significance of the result. We ordered the 29 counter-clockwise
 moving WR/O stars by their projected distances to Sgr~A*. We slid a window of 10
 stars width over the ordered counter-clockwise moving stars and computed the average angular momentum direction for these stars.
Figure \ref{fig:counter_disk_position_vs_distance_sphere} (left) shows a cylindrical equal area projection of the local average angular momentum direction for the counter-clockwise moving stars as a function of the average projected distance to Sgr~A*.
Figure \ref{fig:counter_disk_position_vs_distance_sphere} (right) shows the same data as gray scale points in an orthographic projection centered on the counter-clockwise excess together with the density of reconstructed angular momenta in an aperture of $25^{\circ}$ radius.
The average angular momentum direction of the innermost 16 counter-clockwise stars (corresponding to the 7 windows of 10 stars each with smallest average projected distances) clusters around the position of the counter-clockwise system of \citet{Paumard2006}. The average angular momentum direction of the counter-clockwise stars may be a function of the projected distance. Figure \ref{fig:counter_disk_l_vs_p} shows the local average stellar angular momentum direction ($\phi_J = \phi_J(R)$ and $\theta_J = \theta_J(R)$) for the counter-clockwise stars as a function of the average projected distance. Due to the small number of counter-clockwise stars, the spread is large and the significance of a possible shift of the average angular momentum direction is marginal. Still the plot $\phi_J = \phi_J(R)$ suggests an evolution of the $\phi$ angle of the angular momentum vector with distance to Sgr~A*. 
10 of the 29 counter-clockwise stars have inclinations larger than $30^{\circ}$ with respect to the local angular momentum direction of the counter-clockwise system.

Our orbital analysis is in good agreement with the results of \citet{Genzel2003} and \citet{Paumard2006} which were only based on velocity vectors. They found that there is probably a significant second, counter-clockwise system, at large angles with respect to the main clockwise system. 
The significance of this second system is moderate.
The observed maximum angular momentum density occurs with a 2\% chance in an isotropic distribution given a large number of tries ($45\time90$ sky bin for counter-clockwise orbits $\cos\theta<0$). 
We cannot exclude that the observed counter clockwise structure is due two separate streamers etc.
However, the significance of the counter-clockwise system is about as large as can be reasonably expected on the basis of the small numbers of stars contained in it, as well as its 
observed azimuthal and broader angular distribution of stars. 

\subsection{Summary and Comparison to Previous Results}

In this work we have increased the data set by 27 new reliably measured WR/O stars in the central 12'' of our Galaxy, to a total of 90, and reduced the proper motion uncertainties by a factor of four compared to the data set previously analyzed by \citet{Paumard2006}. The previous analysis by \citet{Levin2003,Genzel2003,Paumard2006} only included the 3D velocity information of the early-type stars to compute a $\chi^2$ for stellar disks. Here we have in addition taken into account the 2D positions on the sky. The results from \citet{Paumard2006} were based on 59 quality 1 and 2 early-type stars with projected distances between 0.8'' and 8''.

We have shown in this paper that the angular momentum direction of the clockwise system is a function of the projected distance from Sgr~A*. Its innermost edge is statistically compatible with the angular momentum direction of the clockwise disk determined by \citet{Paumard2006}. There does not appear to be any significant difference between an analysis with all velocities and spatial coordinates and an analysis based on velocities only.
In a reanalysis of the full \citet{Paumard2006} data set (all quality 2 stars with projected distances between 0.8'' and 12'') we see already a hint of the angular momentum direction change, albeit at lower significance. 
As \citet{Paumard2006} restricted their analysis to the range of projected distances between 0.8'' and 8'', they did not see any significant change of the average angular momentum.

The surface number density of observed stars is within errors compatible with the one given by \citet{Paumard2006}. 
Also the distribution of inclinations of the stellar orbits to the local angular momentum direction of the clockwise system agrees well with the disk thickness of $(14\pm4)^{\circ}$ determined by \citet{Paumard2006}. The average star in the clockwise system is not on a circular orbit.
Our analysis indicates somewhat larger eccentricities of the candidate stars of the clockwise system compared to \citet{Paumard2006}. Our results are compatible with the analysis of \citet{Beloborodov2006} and the results for the six early-type stars with $0.8''\leq R\leq1.4''$ by \citet{Gillessen2008}. 

\citet{Paumard2006} used a uniform $z$-prior to compute the eccentricities and noted a bias to too larges values, which they tried to correct by applying a weight of $1/(1-\exp(-\epsilon/0.35))$. 
Moreover, \citet{Paumard2006} used a smaller mass of Sgr~A* of $3.6\times10^6M_{\odot}$ and did not take the mass of the late-type cluster into account. A smaller mass favors smaller eccentricities.
Firmer conclusions on the eccentricity distribution will require a more thorough analysis of the systematic effects in the eccentricity reconstruction.

In addition, we confirm a non-isotropic distribution of the counter-clockwise stars at 
a confidence level beyond 98\%. 
The modest significance of this counter-clockwise structure is a combination of the small number of stars (Poisson noise) and the fact that it appears to be less well defined. The counter-clockwise system may be a second disrupted disk or streamer of stars.
For simplicity of analysis and visualization we have described the stellar structures in the central parsec in our discussion above in terms of flat disks or rings. We call deviations from such simple flat structures 'disk warps' or 'dissolving/disrupted disks'. This approach is based on our aim of distilling the simplest physical model required and justified by the data. The reality is undoubtedly more complex. The stellar distribution in the Galactic Center may consist of a connected but complex system of separate streamers at different radii and different orientations. 

\citet{Lu2008} independently performed a similar analysis searching for disk features amongst the early-type stars in the Galactic Center. 
%
They use two different data sets: a {\it primary} set of 32 stars at the projected distance bin 0.8''--3.5'', for which they measured positions, proper motion velocities and acceleration upper limits but use the radial velocities of \citet{Paumard2006}. In this radial distance bin they find that most stars are members of a fairly thin (RMS dispersion angle of $(7\pm2)^{\circ}$) clockwise disk at $\Omega=(100\pm3)^{\circ}$ and $i=(115\pm3)^{\circ}$. These numbers are in reasonable agreement with ours (above), as well as \citet{Paumard2006}. The slight offset in inclination may be due to different priors used for the distribution of the unknown $z$-coordinate of the young stars. \citet{Lu2008} do not use the volume density of the early-type stars to generate the $z$-coordinate (see section \ref{sec:z_generation}) but rather use a uniform acceleration prior. Moreover, the dispersion angle depends on the definition of candidate disk stars. 
\citet{Lu2008} show that some of the candidate stars of the clockwise system have eccentricities larger than 0.2 in agreement with our analysis.
\citet{Lu2008} also analyze a second ({\it extended}) data set of 73 stars: the 32 stars of their {\it primary} data set and in addition 41 quality 1 and 2 early-type stars from \citet{Paumard2006} at projected distances between 3.5'' and 12''. They concluded that the number probability of candidate stars being members of the clockwise disk decreases with increasing projected distance from Sgr~A*. Still they do not find this decrease to be significant at a level beyond $3\sigma$. Moreover, they do not find the pre-trial significance of the counter-clockwise system above $1 \sigma$. 
There are five main differences, why our analysis yields a higher significance of the counter-clockwise system compared to the analysis of \citet{Lu2008}:
\begin{itemize}
\item{the counter-clockwise disk has very few candidate members with projected distances below 3.5'', the limit of the {\it primary} sample of \citet{Lu2008}}
\item{analyzing the {\it extended} sample, \citet{Lu2008} do not make cuts on the projected distance}
\item{in this work we have reduced the proper motion uncertainties by a factor of four compared to the values published by \citet{Paumard2006}, which were used by \citet{Lu2008}}
\item{we added 27 new reliably measured WR/O stars to the data set, 15 of them on counter-clockwise orbits}
\item{we required a firm identification as a WR/O star with a maximum radial velocity uncertainty of 100 km/s. In their {\it extended sample} \citet{Lu2008} included 3 B stars and one O star affected by crowding from \citet{Paumard2006}, which we do not use in our analysis.}
\end{itemize}
Given the larger number of stars and smaller errors in our data set for projected distances beyond 3.5'', the results of \citet{Lu2008} are not in contradiction with our results.

\section{Discussion} \label{sec:discussion}

Our analysis confirms the existence of a relatively thin main disk-like structure of clockwise stars \citep{Levin2003,Paumard2006,Tanner2006,Lu2008}. 
This structure exhibits a significant change of its 
rotation axis at different radii. The main disk may be strongly warped, or consists of several, closely related streamers with radially varying orbital planes.
We also confirm at the 98\% confidence level the existence of a coherent structure of the counter-clockwise stars \citep{Paumard2006}. 
This second system may be a set of streamers at large angles relative to the primary clockwise system, or perhaps a second, albeit more disrupted, disk.

In the following we discuss the possible nature and origin of these disk-like structures in the framework of different models. Any such model can be tested against several observables:
%
\begin{itemize}
\item{number of disks, fractions of disk candidates and random stars, disk thickness}
\item{disk orientation, disk warps} 
\item{radial distribution of the early-type stars}
\item{eccentricity distribution of disk candidate stars}
\item{mass function.} 
\end{itemize}

We will shortly summarize our data on these five observables, and then discuss them in view of both the {\it infalling cluster} and the {\it in situ} star formation models, described below.

\subsection{Observables}

\subsubsection{Number of Stellar Structures and their Thickness}

55\% of the observed WR/O stars in the central 0.5~pc of the Galactic Center are candidate members of a warped clockwise system and 20\% of the stars are candidate members of a counter-clockwise system at an inclination of about $100^{\circ}$ with respect to the clockwise system. Both systems are not locally flat, the 
inclinations to the clockwise system are compatible with a two-dimensional Gaussian distribution with a sigma of $10^{\circ}$. 
Moreover, 25\% of the WR/O stars have angular separations from the disks too large to be considered candidate members. However, there is no further noticeable difference between the disk star candidates and the apparently random stars given their similar ages, see section \ref{sec:HR}. 
The data thus suggest the existence of two main structures; a main clockwise relatively thin disk-like structure and a somewhat 
thicker less massive counter-clockwise structure.

\subsubsection{Warps and Orientation}

\begin{figure}[t!]
\begin{center}
\includegraphics[totalheight=7cm]{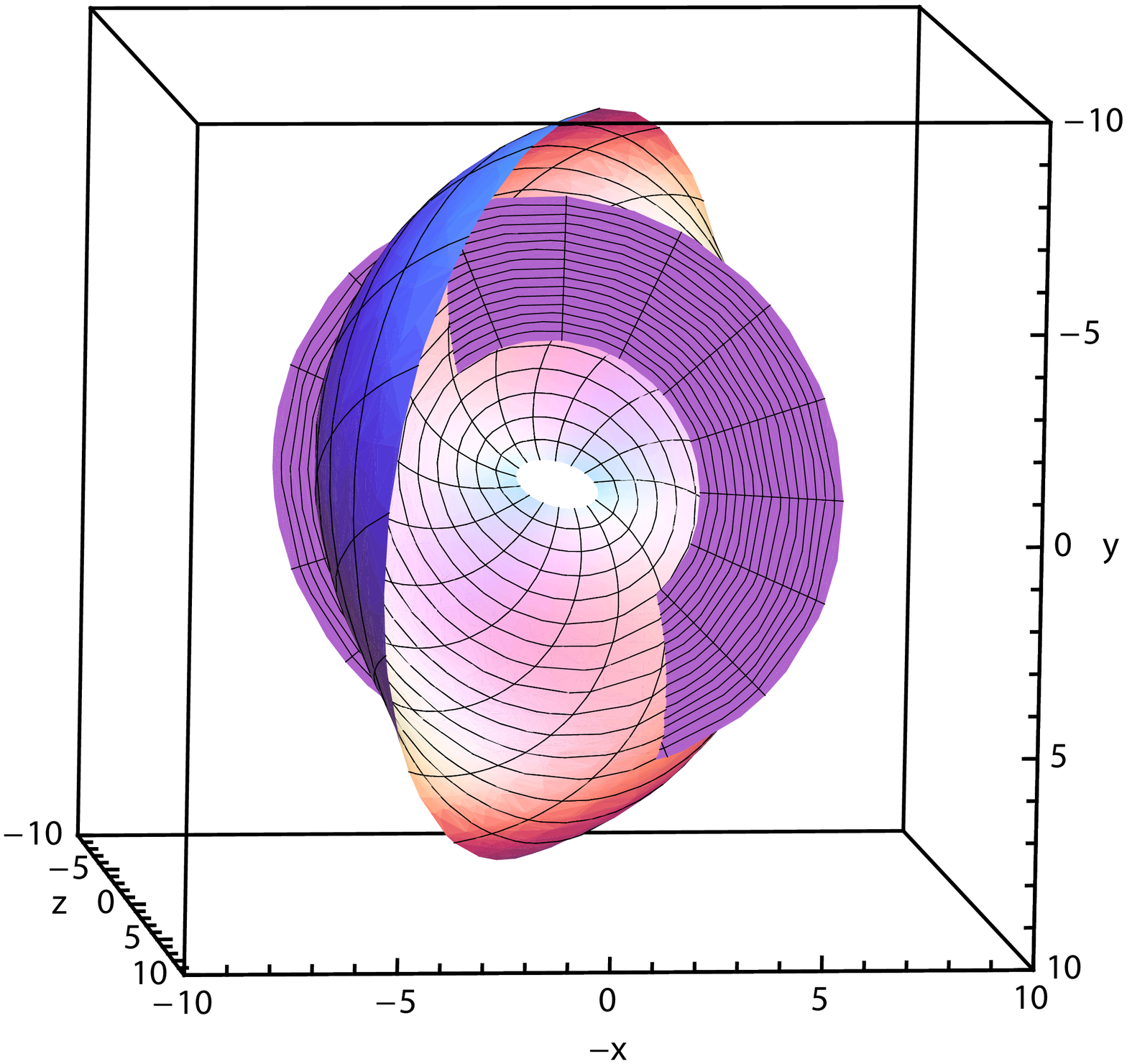}
\includegraphics[totalheight=7cm]{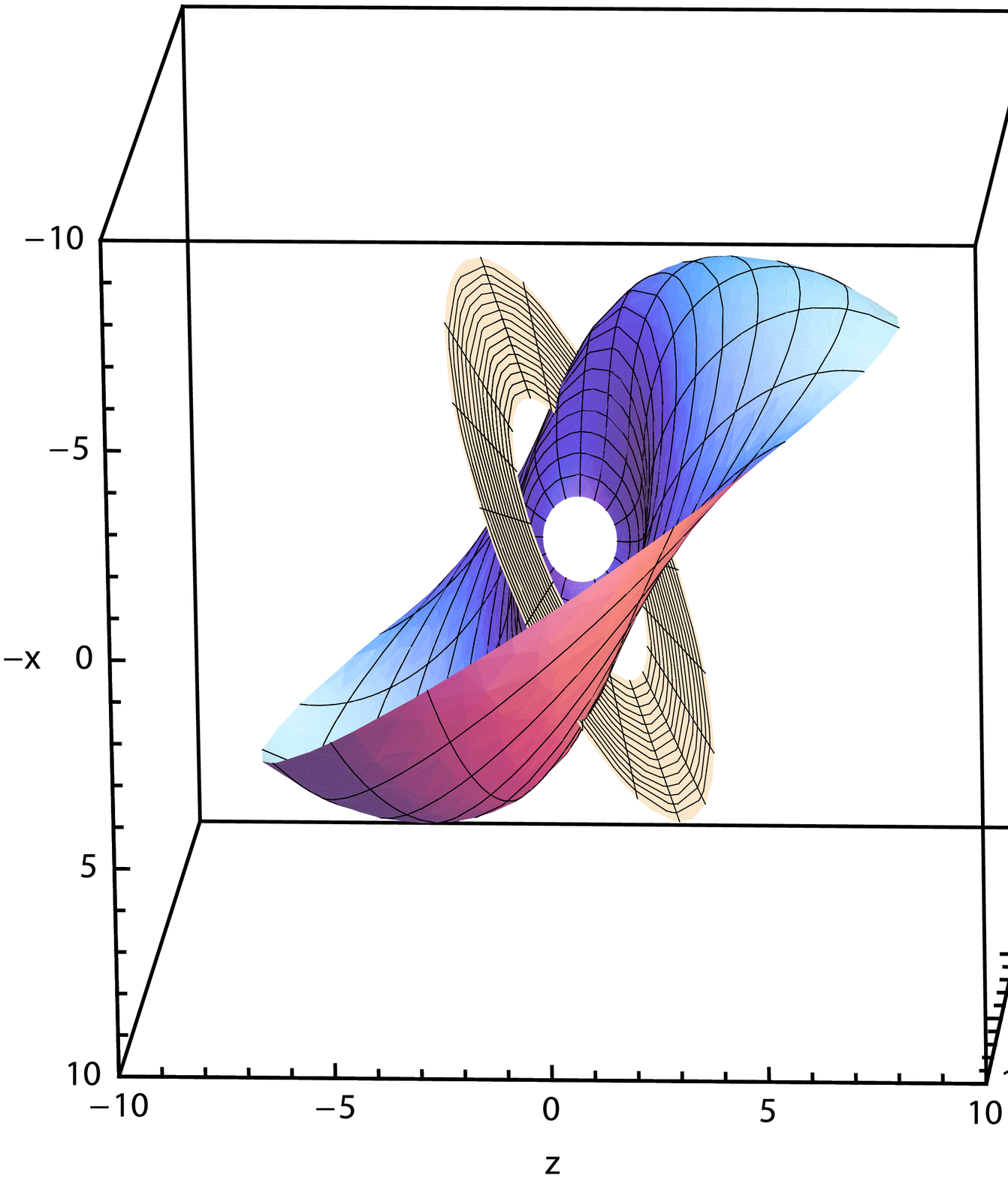}
\caption{\it \small Three-dimensional visualization of a warped and tilted clockwise disk.
It is shown assuming locally flat circular orbits (a = 1''--10''). The angular momentum vector is a smooth function of semi-major axis, the broken line in figure \ref{fig:disk_position_vs_distance_sphere} (right panel).
For comparison, the flat, non-warped ring in the range a= 3.5''--7'' at large inclination to the clockwise warped and tilted disk indicates the average angular momentum direction of the innermost 16 counter-clockwise stars.}
\label{fig:disk_warp_3D}
\end{center}
\end{figure}

The rotation axis of the observed clockwise system changes as a function of the projected distance from Sgr~A*. 
Its direction changes by $\Delta_\theta=26^{\circ}$ and $\Delta_\phi=41^{\circ}$ between the inner region and the
intermediate one; and then changes again by $\Delta_\theta=-34^{\circ}$ and $\Delta_\phi=36^{\circ}$ in the transition to
 the outer region (see table \ref{tab:significances}). 
The angular momentum direction as a function of the average projected distance does not follow a great circle as expected for a simple warp. Instead, there is a significant offset from the great-circle, a tilt. The radial dependence of $\theta_J$ and $\phi_J$ can reasonably well be fitted by quadratic polynomials, see figure \ref{fig:disk_position_vs_distance_sphere} (right panel).   
The rotation axis of the counter-clockwise structure keeps a constant inclination ($\theta$) of approximately $100^{\circ}$ from the
main clockwise disk but has a broad distribution in $\phi$ throughout the intermediate region where it is observed.  

Figure \ref{fig:disk_warp_3D} shows a three-dimensional visualization of the warped and tilted clockwise disk.
It is shown assuming locally flat circular orbits (a = 1''--10''). The angular momentum vector is a smooth function of semi-major axis, given by the quadratic fits to the direction of the angular momentum of the clock-wise system as a function of distance to Sgr~A*. 
For comparison the flat, non-warped ring in the range a= 3.5''--7'' at large inclination to the clockwise warped and tilted disk indicates the average angular momentum direction of the counter-clockwise stars with projected distances between 3.5'' and 7''.

\subsubsection{Eccentricity Distribution of Disk Candidate Stars}
The main clockwise structure is very coherent and is most likely a single cohesive disk of stars. The eccentricity of most likely disk candidate stars can therefore be estimated under the assumption that they indeed belong to such a disk. 
The data suggest non-zero, moderate eccentricities. The a priory unknown eccentricity distribution has a mean of $0.36 \pm 0.06$ and an RMS of 0.2.
We can exclude at a confidence level beyond 5 sigma that all disk stars are on circular orbits. There are four candidate disk stars with eccentricities above 0.9, which may belong to a different population of stars (e.g. they are no disk stars).

\subsubsection{Mass Function (MF) and Radial Distribution} 
\citet{Paumard2006} showed that the MF of the disk stars is significantly flatter than a Salpeter MF. Also the observed X-ray emission constraints of \citet{NayakshinSunyaev2005} suggest a top heavy MF.
The surface number density of the disk stars is a steep function of distance from Sgr~A*: $\Sigma(r_{\mathrm{disk}}) \propto r^{-1.95\pm0.25}$. 
The current limited data on the mass function of the WR/O stars suggest the same distribution of the MF 
throughout the main clockwise system at different radii. It also seems to be similar to the MF of stars in the counter-clockwise system. However, to better constrain the MF of these stars, the identification of B dwarfs throughout the regions where the young stars are observed is required, in addition to the dynamical analysis of their kinematics. Moreover, a thorough completeness correction is required for a better understating of the MF. These caveats are beyond the scope of this paper and will be addressed in a forthcoming paper by \citet{Martins2009}.

\subsection{Models for the origin of the young GC WR/O stars}

The two main scenarios discussed in the literature for the origin of the disk-like structure of young stars in the GC are the {\it infalling cluster} model and the {\it in situ} star formation model.
In the {\it infalling cluster} scenario, a cluster of young stars formed far from the GC (a few parsecs) and in-spirals to Sgr~A*. During the in-spiral the stars are stripped from the young cluster, forming a disk-like structure. In a mass segregated cluster the most massive stars are positioned in the cluster core and are the last to be stripped at the innermost regions of the GC, close to the MBH. The final stripped stars are suggested to be the young O and WR stars observed in the GC.   

In the in-situ model, a clump or clumps of gas fall into the GC, where they form a disk-like structure. The gaseous disk then fragments to form new low mass stars in a stellar disk, that grow, through accretion, into the more massive stars currently observed in the GC \citep{Levin2003,Genzel2003,Goodman2003,Milosljevic2004,Nayakshin2005,Paumard2006,Bonnell2008,Mapelli2008,Hobbs2008}.

Any valid model can be constrained by the observations and should be able to explain the observed data described above. 
In the following we confront the current models with the data.

\subsubsection{In situ formation}

The tidal forces near the super-massive black hole Sgr~A* are so strong that a cloud cannot be gravitationally bound unless its density exceeds $6\cdot10^9 M_{\mathrm{Sgr A*}}/(4\cdot 10^6M_{\odot}) (R/7")^{-3} \mathrm{cm}^{-3}$,  
where $R$ is the distance to Sgr~A* 
, orders of magnitude denser than currently observed \citep{Morris1993}. The tidal shear can be overcome if the mass accretion was large enough at some point in the past causing a gravitationally unstable massive ($\sim10^4M_{\odot}$) disk to form. The stars formed directly out of the fragmenting disk.
Given the fact that the star formation event 6 Myrs ago seems to be isolated, with  no comparable activity for several tens of Myrs before and not much after \citep{Krabbe1995}, Occam's razor would suggest that only one disk of WR/O stars was present initially.
The $10^{\circ}$ sigma thickness of the clockwise system may be explained by disk heating due to interactions with cusp stars \citep{Perets2008}, massive disk stars \citep{Alexander2007,Cuadra2008} and binaries \citep{Perets2008a}. But such heating cannot excite and explain the large fraction of young stars at very high inclinations from the main clockwise disk structure, and/or the coherent counter-clockwise structure \citep{Cuadra2008}.
Moreover, even more massive and less likely perturbers such as in-spiraling intermediate mass black holes \citep{Yu2007,LoeckmannBaumgardt2008} have difficulties explaining both the clockwise and the inclined counter-clockwise systems. In addition they require a fine tuned scenario, where the in-spiral occurs shortly after the stellar disk formation. It is therefore more likely that the clockwise and counter-clockwise configurations formed together and were not due to some dynamical evolution of a single disk-like structure. 

In situ star formation following the infall of a single molecular cloud was recently studied in simulations \citep{Bonnell2008}. In such a scenario a gaseous disk is formed and then fragments to form a single 
disk of stars \citep{Nayakshin2005,Alexander2008,Bonnell2008}. Efficient circularization in the original infalling gas leads to low orbital eccentricities of the massive stars \citep{Sanders1998,Vollmer2001}. 
A significant average eccentricity of the disk stars of $0.36\pm0.06$ probably requires that the stars form on an orbital timescale, following the initial compression \citep{Bonnell2008}. However, the formation of warps in the disk and/or highly inclined stellar structures or even just a large fraction of stars outside the disk have not been observed in these simulations.
     
Recent simulations by \citet{Hobbs2008} suggest that strongly warped disks could form in-situ following cloud-cloud collisions in the GC. Moreover, highly inclined structures such as stellar filaments, streams and small clusters could be observed in such simulations. Such scenarios could, in principle, naturally explain our observations, showing both a main warped disk and another highly inclined stellar structure, possibly with other less pronounced smaller structures and stars with various angular momenta outside the main structures. The second main structure could also be evidence of a second disk, possibly in a state of dissolution, or another coherent structure as seen in cloud-cloud collision simulations \citep{Hobbs2008}. We caution, however, that such simulations, can currently explore only a small fraction of the possible phase space for cloud collisions, and do not include radiative transfer effects. 
The present data set does not yet allow us to distinguish between a disk in a 
state of dissolution and stellar filaments, streams and small clusters

The eccentricities of the stars in the observed structures and the thickness of these structures could be explained both by the in-situ formation of eccentric disks/structures \citep{Alexander2008,Bonnell2008,Hobbs2008,Wardle2008} and/or through the dynamical heating of more circular configurations mainly by cusp stars and stellar mass black holes which excite the mean eccentricities to be as high as 0.4. The latter possibility, however, is not likely to excite eccentricities as high as 0.9, as possibly observed for some of the candidate disk stars (if they indeed belong to the disk). 

Another possibility that has been discussed in the literature is nuclear disk warping and eccentricity excitations, due to interaction with other massive non-spherical structures. Any configuration of two non-spherical symmetric populations causes the structures to precess. Differential precession at different speeds are observed as warps. There is a coherent rotation of the cluster of late-type stars in the Galactic Center but their distribution is spherically symmetric \citep{Trippe2008} and hence does not cause any precession. Other planar structures in the Galactic Center such as the minispiral and the CND have too low masses \citep{Jackson1993,Liszt2003} to cause significant warps.

On the other hand, an {\it in situ} formation of the WR/O stars in the Galactic Center in two disks, as possibly suggested by our observations, provides two massive inclined structures. Such a configuration has been suggested to induce warping and 
eccentricity excitations through differential precession and a Kozai resonance \citep{Nayakshin2006,Loeckmann2009,Subr2008}. 
The expected magnitude of a warp due to differential precession induced by two disks can be estimated analytically. \citet{Nayakshin2005a} assumes one massive ring (mass $M_1$) at radius $R_1$ and a less massive test particle ring (mass $M_2$) at radius $R_2$. Let $\alpha$ be the angle between the angular momentum vectors of the two rings. The precession frequency $\omega_P$ of the less massive ring around the more massive one is than given by

\begin{equation}
\frac{\omega(R_2)}{\Omega(R_2)} \approx -\frac{3}{4} \frac{M_1}{M_{\bullet}} \cos\alpha \frac{R_2^3 R_1^2}{(R_2^2+R_1^2)^{5/2}} \ ,
\end{equation}
where $\Omega = \frac{2\pi}{T} = \sqrt{\frac{G_{\mathrm{N}} M_{\bullet}}{R_2^3}}$ is the angular velocity of the less massive ring.
For fixed masses the precession frequency has a maximum at $R_{2,\mathrm{max}} = \sqrt{3/7} R_1$. Therefore the precession time $T_P$ is
\begin{equation}
T_P \geq  \frac{4}{3} \frac{M_{\bullet}}{M_1} \frac{1}{\cos\alpha} \frac{(R_{2,\mathrm{max}}^2+R_1^2)^{5/2}}{R_{2,\mathrm{max}}^3 R_1^2}2\pi\sqrt{\frac{R_{2,\mathrm{max}}^3}{G_{\mathrm{N}} M_{\bullet}}} \ .
\end{equation}
With $M_{\bullet} = 4\times10^6_{\odot}$ we get
\begin{equation}
T_P \geq 10^{7} \mathrm{years} \frac{(R_1/3'')^{3/2}}{M_1/(5000 M_{\odot})} \ .
\end{equation}

\citet{Paumard2006} estimate the total mass of the counter-clockwise system to be $5000 M_{\odot}$. A counter-clockwise ring of about 5000 solar masses at a distance of about 3'' from the Galactic Center, causes precession periods in the clockwise system of about $10^7$~years. This is compatible with a warp due to differential precession in the clockwise disk of about $65^{\circ}$. This estimate is approximate. It is significantly modified for radially extended disks, or for stellar disks with eccentric orbits.
A counter-clockwise ring with a radius of about 3'' would cause a maximum precession of the clockwise disk at a radius of 3'' and less precession for smaller and larger distances. In case of an initially flat clockwise disk, the innermost and the outermost border of that disk would have similar rotation axis, contrary to the data. This disagreement is intriguing but needs to be followed up with future more detailed models in the spirit of \citet{Nayakshin2006}  with updated constraints on the stellar orbits.
Moreover, the above estimation of the warp amplitude did not take into account the influence of the cusp stars in which the disks are embedded. 
Analytical studies suggest that the stellar cusp around the disks would quench their strong interaction and would not allow
for the Kozai resonance to be effective \citep{Ivanov2005,Karas2007,Chang2008}.
N-body simulations show that the cusp
stars stabilize the disks and the eccentricity excitation is quenched (Perets et al., in prep.). 
If theoretical models with an initially flat distribution cannot match the configuration seen in the data through warping, it would imply that the disks were born warped. This would provide an additional and very important constraint to the models.
 

It is also interesting to note a well known example of another warped nuclear disk surrounding a supermassive black hole in another galaxy; the maser disk observation of NGC~4258 \citep{Miyoshi1995,Herrnstein1996}. The NGC~4258 disk extends from 0.13 to 0.26~pc from the nucleus and shows a moderate warp of $\sim8^{\circ}$ 
\citep{Herrnstein1996}.
%
Although interesting by itself, the observed nuclear maser disk shows much weaker warping than observed in the clockwise disk in the GC. 
It is likely to be of a different origin of the GC disk, and various suggestions have been discussed in the literature 
\citep[e.g.][]{Arnaboldi1993,Caproni2006,Ferreira2008}.
\citet{Ayse2007} found that there are stable configurations of single massive nuclear disks in galaxies with pronounced warps for a wide mass range of the nuclear disk.

The in-situ formation model requires that the gas densities in the infalling clouds are sufficiently high so that star formation occurs rapidly after the infall (but not before), so that the massive stars end up in the central parsec where they are observed and yet the gas does not have time to dissipate to one plane.
Such a scenario produces a surface number density of stars in the disk of $\Sigma(r) \propto r^{-2}$ \citep{Lin1987,Hobbs2008}, compatible with the observations. The top heavy initial MF expected from such a scenario \citep[e.g.][]{Bonnell2008} is also consistent with 
our observations. We note, however, that in some of the simulations by 
\citet{Hobbs2008} they find that different regions in the main disk or in the other structures have a different mass function. The observation of more candidate member stars of the clockwise system might be required to confirm or refute this.  

\subsubsection{Cluster Infall}
In most scenarios for the {\it infall of a cluster}, the stars remain close to the in-spiral plane \citep{Berukoff2006} and form a single flat star distribution with a thickness comparable to or larger than the original cluster. \citet{Levin2005} have shown that after the infall of a cluster of young stars around an intermediate mass black hole on an initially eccentric orbit a significant fraction of stars get pushed out of the in-spiral plane and end up with large inclinations. However, the stars with large inclinations do not have a preferred orbital orientation;
thus it is difficult for this scenario to form a second inclined coherent structure.
Therefore, in the framework of the infalling cluster scenario, the observation of two disks of young stars requires two cluster infall events in the last $6 \times 10^6$ years, rendering this scenario even more unlikely.

In the infalling cluster scenario the stars which are stripped from the cluster as it spirals in have similar inclinations and eccentricities as the cluster itself requiring an initial eccentricity of the cluster of $0.36\pm0.06$. An in-spiral through dynamical friction
would suggest a more moderate eccentricity. Nevertheless, observations of the GC Arches cluster show that it is on a non-circular orbit \citep{Stolte2008}; although at a large distance from the Sgr~A*.  

The observed warps give strong evidence against the {\it infalling cluster} scenario. In the framework of the infalling cluster scenario, the stars will keep the mean angular momentum orientation of the initial cluster if the cluster sinks in rapidly resulting in a flat, non-warped disk of stars.
To form a warp of the observed magnitude through differential precession, the stars need to have been in a relatively thin disk for a long time. But in the infalling cluster scenario, the stars barely had the time to get to their sub-pc orbits, and thus they could not have spent much time as a disk.

The infalling cluster scenario predicts surface density profiles as shallow as $\Sigma(r) \propto r^{-0.75}$ \citep[see e.g.][]{Hansen2003,Berukoff2006}. This is considerably shallower than that observed. \citet{Guerkan2005} argue that this discrepancy can, in principle, be overcome by initial mass segregation in the cluster. The lower mass stars of the cluster are lost at larger distances and the most massive stars inside. This proposition will be tested by a future search for lower mass B stars at distances beyond 12'' from Sgr~A* \citep{Martins2009}.

\subsection{Summary}

To sum up, our observations are compatible with an {\it in situ} formation of these WR/O stars in a clockwise rotating disk and another highly inclined counter-clockwise structure, possibly a disk. 
Although the infall of two clusters may possibly explain the two systems of WR/O stars, the stars would have remained in a disk for too short to develop the observed warp.

\section{Conclusions and Outlook}

Our latest sample of early-type stars between 0.8'' and 12'' from the Galactic Center contains 90 WR/O stars. The most important results are:

\begin{itemize}
\item 55\% of the WR/O stars are candidate members of a clockwise disk
\item the clockwise disk shows a large warp (in total $(64\pm6)^{\circ}$, at a significance level of $>10$ sigma), 
the disk angular momentum direction is a function of the
projected distance.
\item the clockwise disk is not locally flat, the inclinations of the clockwise system are compatible with a two-dimensional Gaussian distribution with a sigma of $10^{\circ}$
\item the clockwise system is not circular; it has a mean eccentricity of $0.36\pm0.06$.
\item 20\% of the WR/O stars are candidate members of  a coherent feature amongst the counter-clockwise stars (pre-trial significance of 4.5 sigma for the interval of projected distances between 3.5'' and 7'', corresponding to a 2\% chance that this feature is a statistical fluctuation in an isotropic cusp star distribution).
\item the counter-clockwise system may be a disk in a dissolving state: 10 out of 29 counter-clockwise rotating stars have angular separations larger than $30^{\circ}$ from the counter-clockwise system, while only 11 out of 61 clockwise moving stars have a separation larger than $30^{\circ}$ from the clockwise disk.
\end{itemize}

The observation of a warped disk of WR/O stars in the Galactic Center has substantially strengthened our previous conclusion that the population of 100 young, massive stars arranged mainly in two coherent structures between 0.8'' and 12'' from Sgr~A* is due to in situ star formation (as opposed to an infalling cluster scenario) from gas that fell into the central region about $6\times10^6$ years ago and became dense enough to overcome the tidal forces.



\appendix

\section{Coordinate systems} \label{sec:coordinate}

As in \citep{Paumard2006} we used the usual Cartesian coordinate system in offsets from Sgr~A$^*$: $x=\cos \delta \mathrm{d} \alpha$ increased eastward, $y=\mathrm{d} \delta$ increased northward and $z=\mathrm{d}D$ increased along the line of sight away from the observer. We used a spherical coordinate system $(x,y,z) \Leftrightarrow (r,\theta,\phi)$:

\begin{eqnarray}
x &=& r \sin{\theta} \cos{\phi} \\
y &=& r \sin{\theta} \sin{\phi} \\
z &=& r \cos{\theta} \ .
\end{eqnarray}

\begin{figure}[t!]
\begin{center}
\includegraphics[totalheight=12cm]{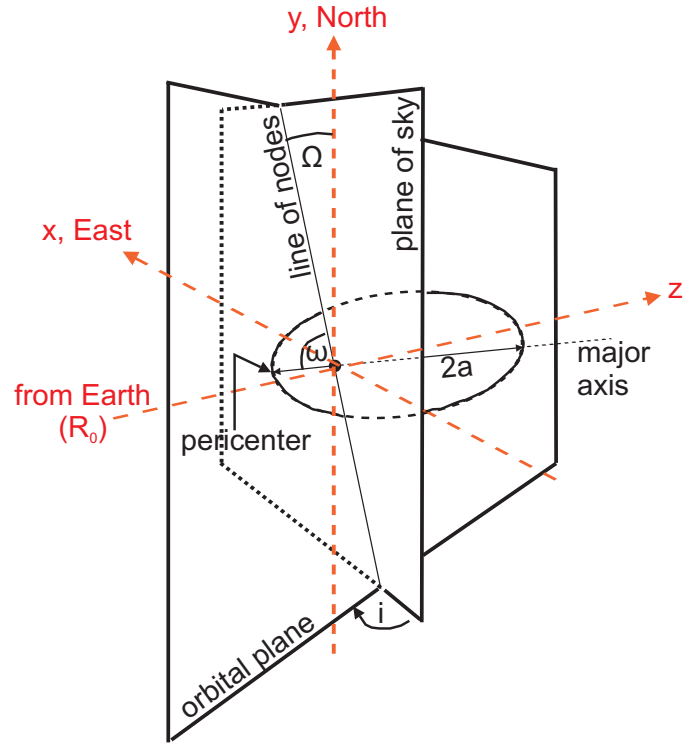}
\caption{\it \small The definition of the classical orbital elements used in this work in the framework of the chosen coordinate system in this work. The angular momentum $\vec{J}$ is orthogonal to the orbital plane. $\theta_J$ is the angle between the $z$-axis and $\vec{J}$, $\phi_J$ is the angle between the $x$-axis and the projection of $\vec{J}$ onto the plane of the sky ($xy$-plane).}
\label{fig:coordinates}
\end{center}
\end{figure}

Figure \ref{fig:coordinates} illustrates the definition of the classical orbital elements used in this work in the framework of the chosen coordinate system in this work.

\section{Acknowledgements}

We thank J. Cuadra for his useful comments. TA is supported by ISF grant 968/06, Minerva grant 8563 and ERC Starting Grant 202996.

\bibliographystyle{apj}

\end{document}